\documentclass[preprint, times]{elsarticle}

\usepackage{amssymb}
\usepackage{subfigure} 
\usepackage{graphicx}
\usepackage{amsmath}
\usepackage{algorithmic}
\usepackage{algorithm}
\usepackage{bm}
\usepackage{makecell}
\usepackage{siunitx}
\usepackage{geometry}
\usepackage{ntheorem}
\usepackage{graphicx}
\usepackage{array}
\usepackage{hyperref}
\RequirePackage{CJKnumb}

\journal{elsevier}

\begin{document}

\begin{frontmatter}



\title{Deep learning accelerated solutions of incompressible Navier-Stokes equations on non-uniform Cartesian grids}


\author[inst1]{Heming Bai}
\author[inst2]{Dong Zhang}
\author[inst3]{Shengze Cai}
\author[inst1]{Xin Bian\corref{cor1}}
\ead{bianx@zju.edu.cn}

\affiliation[inst1]{organization={State Key Laboratory of Fluid Power and Mechatronic Systems, Department of Engineering Mechanics, Zhejiang University},
            city={Hangzhou},
            postcode={310027}, 
            country={China}}
\affiliation[inst2]{organization={China Ship Scientific Research Center}, city={Wuxi}, postcode={214082}, country={China}}
\affiliation[inst3]{organization={Institute of Cyber-Systems and Control, College of Control Science and Engineering, Zhejiang University},
           city={Hangzhou},
           postcode={310027},
           country={China}}

\cortext[cor1]{Corresponding author}

\begin{abstract}
In incompressible flow simulations, non-uniform grids efficiently capture localized flow features; however, their spatially varying resolutions severely exacerbate computational complexity. The pressure Poisson equation (PPE) formulated on these grids yields highly complex linear systems, forming the primary computational bottleneck in fractional step method. To address this, we develop an extended hybrid framework tailored for non-uniform Cartesian grids, integrating deep learning with classical iterative solvers to accelerate PPE solutions. Specifically, the framework employs a deep operator network with a U-Net-based branch network. To effectively capture spatially varying resolutions, we propose a multi-level distance vector map construction strategy that computes discrete grid-spacing information corresponding to each hierarchical level of the U-Net. This grid-spacing information is explicitly fused into feature maps prior to convolution operations. Empowered by this grid-spacing-aware architecture, the framework seamlessly extends to simulate flows interacting with solid structures using a decoupled immersed boundary projection method. By training exclusively on \textit{fabricated} linear systems rather than conventional flow-dependent datasets, the model generalizes effortlessly across diverse immersed obstacle geometries with \textit{fixed network weights}. Benchmark results demonstrate that the framework significantly outperforms standalone preconditioned conjugate gradient methods and its standard convolution counterpart, underscoring its exceptional potential for real-world computational fluid dynamics applications.

\end{abstract}



\begin{keyword}
   Non-uniform Cartesian grid \sep Incompressible flows \sep Hybrid method \sep Pressure Poisson equation \sep Immersed boundary method 

\end{keyword}

\end{frontmatter}


\section{Introduction}
\label{introduction}

Computational fluid dynamics~(CFD) has become an indispensable tool for both fundamental fluid dynamics research and engineering applications, primarily due to its ability to provide detailed flow field information via numerical solutions of the Navier-Stokes~(NS) equations~\cite{anderson1995computational}. Nevertheless, CFD continues to face severe computational bottlenecks in iterative engineering applications, such as shape design optimization~\cite{viquerat2021direct} and active flow control~\cite{zheng2024transformerDRL}. These applications typically require not only numerous simulations but also non-uniform spatial discretizations to accurately resolve multi-scale flow features, leading to prohibitively high computational costs. This intrinsic limitation underscores an urgent need for innovative methodologies to accelerate classical numerical solutions.

The recent surge of artificial intelligence~(AI) has sparked a paradigm shift within computational science~\cite{lu2021deepxde}.  Within the realm of AI-driven CFD, existing research efforts have largely converged into two prominent directions: physics-informed neural networks~\cite{2019PINNs, meng2020ppinn, cai2021PINNscup, Cai2021_PINNs,jin2021nsfnets,zhu2024physics, qiu2025direct_wangyw, wang2025machinePINNJFM, zhu2026physics} and purely data-driven surrogate models.
The latter category encompasses a diverse range of approaches, including standard neural network-based methods~\cite{nazvanova2023data, han2019novel, zuo2023fastTransformer}, neural operator learning frameworks~\cite{lu2021DeepONet, li2020FNO, mao2021DeepMMNet, Lin_Maxey_Li_Karniadakis_2021, li2023longIUFNO, bai2024DON}, and pre-training and fine-tuning strategies~\cite{goswami2022, zhang2023Metalearning, xu2024self, zhang2025omnifluids}. Despite their theoretical appeal and rapid development, both types of frameworks frequently encounter bottlenecks regarding computational accuracy, efficiency, and generalizability, and thus currently remain incapable of completely supplanting traditional CFD solvers. To bridge this gap, \textit{hybrid methods} have emerged as a highly promising new direction. As a pragmatic alternative, rather than replacing traditional solvers entirely, this approach preserves the well-established advantages of CFD while strategically integrating deep learning techniques to enhance specific computational components. 

Within the research front of hybrid methods, a representative example is the application of neural networks to accelerate solutions of the pressure Poisson equation~(PPE)~\cite{yang2016data_poisson, tompson2017accelerating,xiao2018novel,2022chenJinLi} in fractional-step methods~\cite{chorin1967,perot1993_fractional_step_method} for incompressible flows. While these early approaches have shown success, they typically rely on flow-specific training datasets, which severely limits their generalization to out-of-distribution flow scenarios. To circumvent the specific-data dependency, the deep conjugate direction method~\cite{kaneda2023DCDM} is trained on datasets generated directly from the coefficient matrix of the PPE. However, this approach relies exclusively on neural networks to update its solutions and thus inevitably suffers from \textit{spectral bias}~\cite{rahaman2019spectral,Xuzhi2020frequency,xu2024overview}. 
More specifically, neural networks tend to be highly effective in rapidly reducing global, low-frequency errors, which is in sharp contrast to classical iterative methods that are generally better suited for attenuating local, high-frequency error components.
Recognizing the inherent complementarity between neural networks and classical iterative methods has motivated a new horizon in the development of hybrid methods.  By strategically harnessing their respective strengths, recent studies have successfully accelerated numerical convergence in solving partial differential equations~(PDEs)~\cite{zhang2024blending, kahana2023geometryHINTs,dong2024pinnMG,kopanivcakova2025deeponetHybrid, kopanivcakova2025leveraging, cui2025hybrid, lee2025fastmeta_solver}. 
Along this promising research line, our previous work introduced HyDEA~(Hybrid Deep lEarning line-search directions and iterative methods for Accelerated solutions)~\cite{bai2026hybrid} to accelerate PPE solutions within the fractional step method for incompressible flows. Specifically, HyDEA leverages a deep learning line-search method~(DLSM) to rapidly eliminate low-frequency errors, while alternating it with the conjugate gradient~(CG)-type methods, which serve as a critical complement by refining the remaining solution components. Despite the success, its current implementation remains confined to flow scenarios discretized on uniform Cartesian grids, as the core architecture is based on convolutional neural networks~(CNNs). In engineering applications such as fluid–structure interaction~(FSI) problems, local grid refinement near solid boundaries is essential to resolve fine-scale flow features. However, such non-uniform spatial discretizations severely exacerbate the computational complexity of the resulting linear systems. Consequently, extending HyDEA to accommodate non-uniform grid discretizations is of significant practical and scientific value. 

Standard convolution operators rely on spatially shared kernel parameters, implicitly assuming that data are on uniform grids. Consequently, when processing flow fields defined on non-uniform grids, a prevalent workaround is to interpolate the data onto a uniform grid before employing convolution operators to learn the fluid dynamics~\cite{han2019novel, bai2024DON, 2019LeeGAN}. In the presence of inner geometric obstacles, this interpolation inevitably distorts critical flow features around the boundaries. To circumvent this local distortion, a natural alternative is to employ network architectures tailored for non-uniformly distributed data. By conceptualizing grid points and their connectivity as nodes and edges, graph neural networks~(GNNs) emerge as an intuitive candidate. Nevertheless, extracting the multi-scale features inherent to fluid flows presents a profound challenge, as classical GNNs (e.g., Graph convolutional neural network~\cite{kipf2016GCN} and GraphSAGE~\cite{hamilton2017GraphSAGE}) suffer from severe over-smoothing and high computational overhead, preventing the straightforward layer-stacking strategy typically used in CNNs to expand the receptive field. Although advanced multi-scale GNNs, such as the architecture proposed by Lino et al.~\cite{lino2022MSGNN}, achieve remarkable predictive accuracy in resolving complex flows, their U-Net-like hierarchical structure incurs prohibitive computational complexity, largely dominated by the dynamic construction of graph levels and the exhaustive intra- and inter-level message-passing operations required to update node and edge features. Furthermore, treating discretized grid points as an ordered sequence of tokens enables the application of Transformer architectures~\cite{vaswani2017attention} to model flow evolution~\cite{xu2024self}. Despite their exceptional feature extraction capabilities, the substantial computational cost of the self-attention mechanisms renders them computationally intractable for large-scale, high-resolution flow simulations. Given the steep computational expenses inherent to both GNNs and Transformers when processing massive non-uniform datasets, adapting the highly efficient CNN paradigm to accommodate non-uniform discretizations remains a highly compelling research frontier. To this end, Hu et al.~\cite{hu2022meshconv} proposed the Mesh-Conv operator, which elegantly integrates local spatial metrics of grid spacing and angular information directly into the convolution operation. This approach significantly enhances the feature extraction on non-uniform structured grids while strictly preserving the computational efficiency of standard convolution operators.

In this work, we extend the HyDEA framework to incompressible flow simulations on non-uniform Cartesian grids. Inspired by the Mesh-Conv operator~\cite{hu2022meshconv}, we develop a multi-level distance vector map construction strategy to evaluate discrete local grid-spacing information corresponding to each hierarchical level of the U-Net branch. By explicitly fusing these grid-spacing representations with the feature maps prior to convolution operations at each stage, we successfully generalize the core philosophy of Mesh-Conv operator into a robust, multi-level network architecture.

Equipped with this multi-level grid-spacing awareness, the extended HyDEA framework preserves the following key advantages:
\begin{itemize}
 \item Reduces the number of iterations required for PPE solution across a wide range of flow scenarios on non-uniform Cartesian grids, compared with standalone CG-type methods.
 \item Achieves superior iterative convergence for solving PPE on non-uniform Cartesian grids with substantial variations in grid spacing, outperforming the standard convolution-based HyDEA.
 \item Generalizes seamlessly across diverse inner obstacle geometries in flow simulations without requiring network retraining.
\end{itemize}

The rest of this work is organized as follows. Section~\ref{method} elucidates the numerical methodology, beginning with an introduction to the decoupled immersed boundary projection method~\cite{li2016DIBPM}, followed by a comprehensive formulation of HyDEA. This section further describes the construction of the training dataset, the Mesh-Conv operator, the multi-level distance vector map construction strategy, the neural network architecture, and the overall training procedure. Section~\ref{result} systematically evaluates the performance of the extended HyDEA through multiple benchmark cases. Finally, Section~\ref{conclusion} summarizes the principal findings and outlines directions for future work.

\section{Methodology}
\label{method}

\subsection{Decoupled immersed boundary projection method}
\label{DIBPM}

The immersed boundary method~(IBM)~\cite{peskin1972flow} is employed to model FSI. By introducing a forcing term $\mathbf{f}$ into the momentum equation, this approach modifies the non-dimensional incompressible Navier-Stokes~(NS) equations to the following form:
\begin{eqnarray}
\label{momentum equation IBM}
 \frac{\partial{\mathbf{u}}}{\partial{t}} + \mathbf{u}\cdot\nabla_{}\mathbf{u} &=& -\nabla_{}p + \frac{1}{Re}\nabla_{}^{2}\mathbf{u}+\mathbf{f},
\\
\label{con equation}
  \nabla_{}\cdot\mathbf{u} &=& 0,
\end{eqnarray}
where $p$, $\mathbf{u}$ and $Re$ denote the pressure, velocity vector and Reynolds number, respectively.

Taira et al.~\cite{Taira_2007_IBPM} and Li et al.~\cite{li2016DIBPM} developed the immersed boundary projection method~(IBPM) and the decoupled IBPM~(DIBPM), respectively, based on the fractional step method of Perot et al.~\cite{perot1993_fractional_step_method}. The DIBPM effectively decouples the pressure, velocity and forcing term, resulting in a reduced computational expenditure compared to the IBPM in practice. The DIBPM is adopted in this work and its complete computational procedure is presented as follows:
\begin{eqnarray}
\label{1step_fr_secondLU_IBM_summary}
  A\mathbf{u}^{\ast\ast} &=& \mathbf{r}^{n} + bc_{1},
\\
\label{2step_fr_secondLU_IBM_summary}
  \Delta tEH\delta \mathbf{F} &=& \mathbf{U_{B}}-E\mathbf{u}^{\ast\ast},
\\
\label{3step_fr_secondLU_IBM_summary}
  \mathbf{u}^{\ast} &=& \mathbf{u}^{\ast\ast} + \Delta tH\delta \mathbf{F},
\\
\label{4step_fr_secondLU_IBM_summary}
 \Delta tDG\delta p &=& D\mathbf{u}^{\ast}-bc_{2},
\\
\label{5step_fr_secondLU_IBM_summary}
 \mathbf{u}^{n+1} &=& \mathbf{u}^{\ast} - \Delta tG\delta p,
\\
\label{6step_fr_secondLU_IBM_summary}
 p^{n+1} &=& p^{n} + \delta p, 
\\
\label{7step_fr_secondLU_IBM_summary}
 \mathbf{F}^{n+1} &=& \mathbf{F}^{n} + \delta \mathbf{F}.
\end{eqnarray}
Here $A$ denotes the implicit operator matrix for the advection-diffusion component, $\mathbf{r}^n$ is the explicit terms of the discrete momentum equation. The structure of $A$ and $\mathbf{r}^n$ depends on the specific mesh configuration and time advancement scheme. In this work, the finite difference method is applied to discretize the NS equations on a staggered-grid, utilizing the implicit second-order Crank-Nicolson method for the time integration of diffusion term and the explicit second-order Adams-Bashforth scheme for convective term. The remaining variables and operators are defined as follows:
\begin{itemize}
 \item $\mathbf{u}^{\ast}$ and $\mathbf{u}^{\ast\ast}$: The intermediate velocity vectors.
 \item $bc_1$ and $bc_2$: The boundary condition vectors for the momentum equation and the incompressibility constraint, respectively.
 \item $\Delta t$: CFD time step size.
\item $H$ and $E$: The discrete regularization and interpolation operator matrices, respectively. The former converts the variables defined at Lagrangian points to Eulerian coordinates utilizing a delta function, while the latter performs the reverse transformation. In this work, the discrete delta function of Roma et al.~\cite{roma1999adaptive_delta} is employed.
\item $\mathbf{U_B}$ and $\delta \mathbf{F}=\mathbf{F}^{n+1}-\mathbf{F}^{n}$: The Lagrangian velocity and the time increment of the momentum force on the immersed boundary.
\item $D$ and $G$: The divergence and gradient operator matrices, respectively.
\end{itemize}

Eq.~(\ref{4step_fr_secondLU_IBM_summary}) is the pressure Poisson equation~(PPE). The solution of Eq.~(\ref{4step_fr_secondLU_IBM_summary}), typically achieved through classical iterative methods, represents the most computationally demanding component of the entire solution process. $\delta p=p^{n+1}-p^{n}$ is the time increment of the pressure, and this formulation serves to mitigate the splitting error of the fractional step method~\cite{dukowicz1992_lowspliterror}. The detailed derivation of the DIBPM can be found in~\cite{li2016DIBPM}. The DIBPM simulations in this work are carried out using the open-source solver $\mathtt{PetIBM}$~\cite{chuang2018petibm}.

\subsection{Hybrid Deep lEarning line-search directions and iterative methods for Accelerated solutions}
\label{HyDEA}

The PPE given in~(\ref{4step_fr_secondLU_IBM_summary}) can be recast into a compact linear system:
\begin{eqnarray}
\label{discretLinearEquation}
 M\delta p = S,
\end{eqnarray}
where matrix $M \in \mathbf{R}^{a \times a}$ represents the discrete approximation of the Laplace operator. $\delta p \in \mathbf{R}^a$ and $S \in \mathbf{R}^a$ are the unknown and source term vectors, respectively. 
Our previous work~\cite{bai2026hybrid} introduced HyDEA~(Hybrid Deep lEarning line-search directions and iterative methods for Accelerated solutions),
which is a hybrid framework that seamlessly integrates a Deep learning Line-Search Method~(DLSM) into classical iterative methods, such as CG-type methods, leveraging their respective strengths to accelerate the convergence of solving Eq.~(\ref{discretLinearEquation}). 

Within the HyDEA framework, DLSM functions as a specialized line-search method. During the training phase, a deep neural network~(DNN) is employed to learn the underlying mapping from the iterative residual vector ($r$) to the corresponding iterative error vector ($e$). These two vectors at the $k$th iteration step are governed by the linear system:
\begin{eqnarray}
\label{discretLinearEquation_delta}
 Me_{k} = r_{k},
\end{eqnarray}
where $e_{k}= \Delta (\delta p) = \delta p_{\text{exact}}-\delta p_{k}$ and $r_{k}=S-M\delta p_{k}$ represent the error and residual vectors at the $k$th iteration, respectively. In the inference phase, at the $k$th iteration step, the trained DNN utilizes the residual vector $r_{k}$ to predict an approximate error vector $e_{k}^{NN}$, which serves as the line-search direction to iteratively update the solution for Eq.~(\ref{discretLinearEquation}): 
\begin{eqnarray}
\label{alpha_lineResearch}
 \alpha_{k} &=& \frac{r_{k}^{T}e_{k}^{NN}}{(e_{k}^{NN})^T M e_{k}^{NN}},
 \\
\label{lineResea_update}
 \delta p_{k+1} &=& \delta p_{k} + \alpha_{k}e_{k}^{NN}.
\end{eqnarray}
The detailed derivation and implementation of the DLSM can be found in our previous work~\cite{bai2026hybrid}.

\begin{figure}[htbp]
\centering
  \includegraphics[scale=0.55]{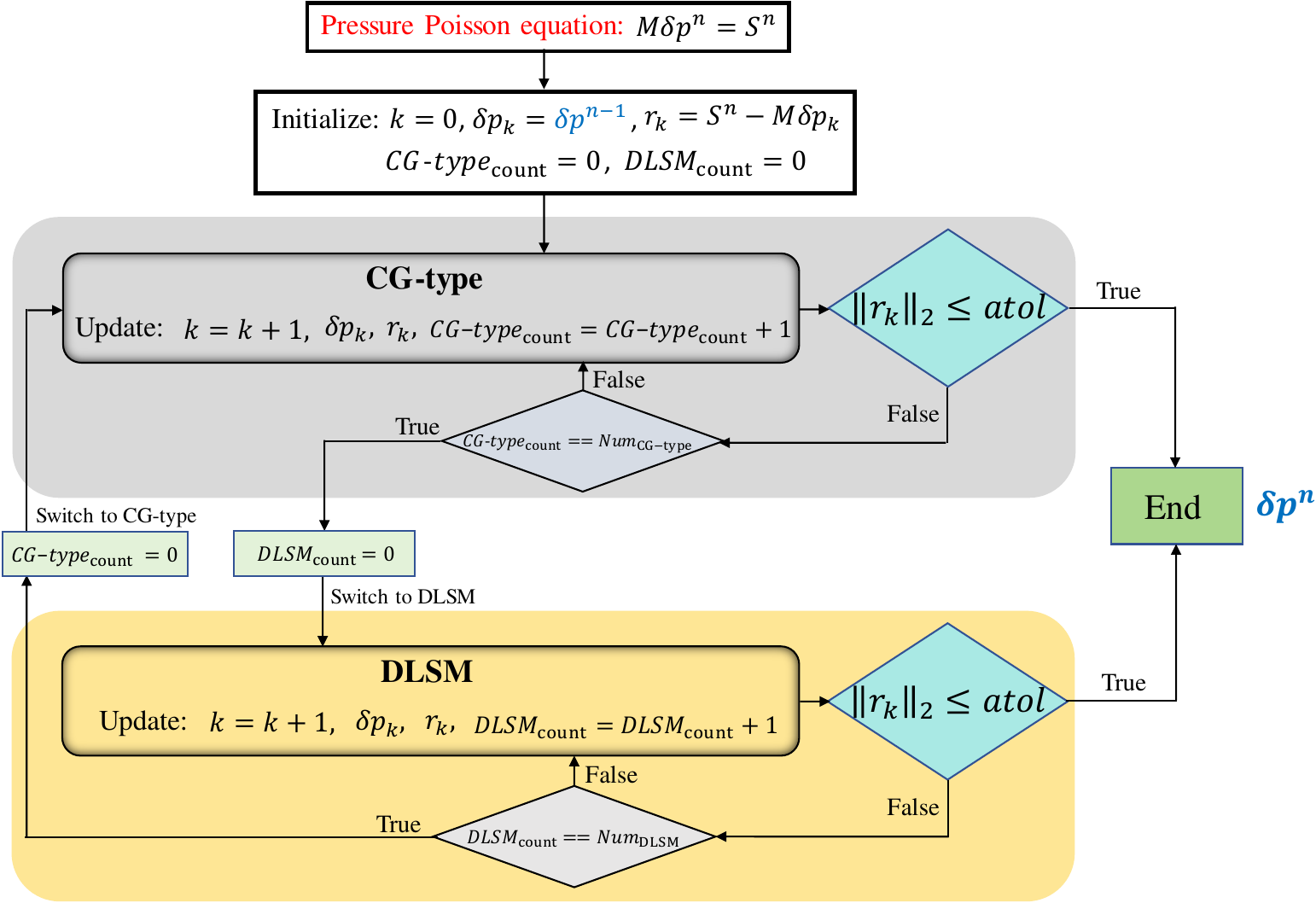}
  \caption{The workflow of HyDEA.}\label{HyDEAworkflow}
\end{figure}

The workflow of HyDEA for a single CFD time step is illustrated in Fig.~\ref{HyDEAworkflow}. Here, $n$ represents the time step counter for the CFD simulation, and $atol$ denotes the prescribed absolute residual tolerance. To manage the dynamic alternation between solvers, $Num_{\mathrm{CG-type}}$ and $Num_{\mathrm{DLSM}}$ are defined as the maximum allowed consecutive iterations for the CG-type method and DLSM within each alternating cycle, respectively. Their execution is monitored by the runtime counters $\mathit{CG-type}_{\mathrm{count}}$ and $\mathit{DLSM}_{\mathrm{count}}$. The entire hybrid iteration procedure continues, progressively refining the solution $\delta p_{k}$, until the $L2$-norm of the residual vector $r_k$ satisfies the convergence criterion. 

\subsection{Dataset construction}
\label{Dataset_Construction}

Following the methodology in Refs~\cite{kaneda2023DCDM,bai2026hybrid}, we construct random vectors biased toward the lower end of the spectrum of $M$ as the training dataset. More specifically, the crucial step in dataset construction lies in generating approximate eigenvectors $Q_{m}=[\mathbf{q}_{0},\mathbf{q}_{1},\cdots,\mathbf{q}_{m-1}] \in \mathbf{R}^{a \times m}$~(where $m < a$) that effectively represent the spectrum of $M$.

First, $M$ is converted to a low-dimensional tridiagonal matrix $T_{m}\in \mathbf{R}^{m \times m}$ through Lanczos iteration~\cite{lanczos1950iteration}, resulting in 
\begin{eqnarray}
\label{Lanczos}
 T_{m} = V^{T}_{m}MV_{m},
\end{eqnarray}
where $V_{m}\in \mathbf{R}^{a \times m}$ represents the orthogonal matrix containing the Lanczos iteration-generated orthogonal vectors. Subsequently, a matrix $U_{m}\in \mathbf{R}^{m \times m}$ is formed by the eigenvectors of $T_{m}$. Thereafter, the Ritz vectors of $M$ are obtained through $Q_{m}=V_{m} U_{m}$. Finally, the dataset is constructed through the following formulation:
\begin{eqnarray}
\label{b_construct}
 R^{i} &=& \frac{\sum_{j=0}^{m-1} c_{j}^{i} \mathbf{q}_{j}}{||\sum_{j=0}^{m-1} c_{j}^{i} \mathbf{q}_{j}||_{2}},
\\
\label{c_construct}
 c_{j}^i &=& \begin{cases}
 9 \cdot N(0,1), & \text{if } 0 \leq j \leq b \cdot m \\
  N(0,1), & otherwise
 \end{cases},
\end{eqnarray}
where $N(0,1)$ denotes a random variable drawn from the standard normal distribution, and $i$ denotes the index identifying individual samples within the dataset. The hyper-parameter $m$ is the number of Lanczos iterations, thereby directly controlling the number of approximate eigenvalues and eigenvectors obtained for $M$. The hyper-parameter $b$ is utilized to modulate the frequency characteristics of vectors $R$ in the dataset.

\subsection{Neural Network architecture and model training}
\label{NNs and training}

In our previous work~\cite{bai2026hybrid}, the branch network of the DeepONet in HyDEA employed the classic CNN-based U-Net~\cite{2015U-Net} as its backbone to process $r_k$ discretized on uniform Cartesian grids. However, standard convolution operators inherently assume uniform spatial distributions, rendering them inadequate for the non-uniform grids required to resolve fine-scale flow features near solid boundaries in FSI applications. To overcome this limitation, the present work is dedicated to extending the HyDEA framework specifically for incompressible flow simulations on non-uniform Cartesian grids.

\begin{figure}[htbp]
\centering
  \includegraphics[scale=1.15]{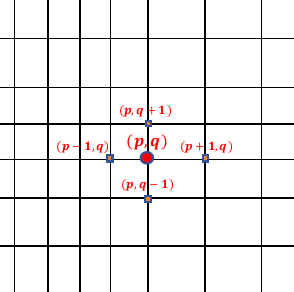}
  \caption{Schematic illustration of the distribution of target and source nodes within the non-uniform local Cartesian grid.}\label{Distance_4}
\end{figure}

To enhance the adaptability of CNNs to general non-uniform structured grids, Hu et al.~\cite{hu2022meshconv} introduced the Mesh-Conv operator, which incorporates a local weight function based on the spatial distribution of neighboring nodes. The original Mesh-Conv operator evaluates an eight-node neighborhood to encode both distance and angular information. However, because the Cartesian grids employed in the present work possess strict inherent orthogonality, such angular metrics become redundant. Consequently, the local weight calculation can be reduced to depend solely on the spatial distances to the four directly adjacent orthogonal nodes. Tailored to this geometric feature, we define the modified local weight function as follows:
\begin{eqnarray}
\label{Hu_local_weight}
 \bar{\eta}_{p,q} = \mathbf{exp}(-\Phi_{\Delta}(\Delta_{p,q})^2),
\end{eqnarray}
where $\bar{\eta}_{p,q}$ denotes the local weight at the target node, with $(p,q)$ representing its discrete spatial indices on the two-dimensional grid, as illustrated in Fig.~\ref{Distance_4}. $\Phi_{\Delta}$, implemented as a neural network, is the weight function designed to encode the distance vector $\Delta_{p,q}$. Here, $\Delta_{p,q}$ consists of the Euclidean distances between the target node $(p,q)$ and its four directly adjacent source nodes.

\begin{figure}[htbp]
\centering
  \includegraphics[scale=0.5]{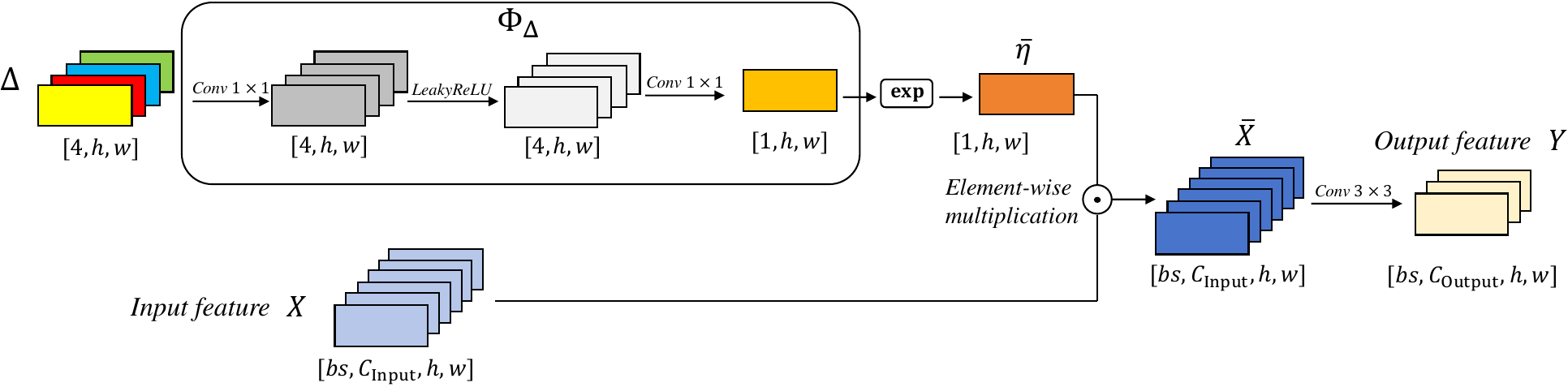}
  \caption{The overall computational workflow of Mesh-Conv operator based on the modified local weight. The tensor shape $[bs,c,h,w]$ represents the feature map dimensions, corresponding to batch size, the number of channel, height, and width.}\label{MConv}
\end{figure}

By incorporating $\bar{\eta}_{p,q}$, the input feature $X_{p,q}$ is transformed into a weighted feature $\bar{X}_{p,q}$, which explicitly encodes the local grid-spacing information:
\begin{eqnarray}
\label{Weighted_feature}
 \bar{X}_{p,q} = \bar{\eta}_{p,q} \cdot X_{p,q}.
\end{eqnarray}
Subsequently, the weighted feature map $\bar{X}$ is processed by a standard convolution operator to yield the final output feature map $Y$.
The overall computational workflow of the Mesh-Conv operator based on this modified local weight is illustrated in Fig.~\ref{MConv}.

The original implementation of the Mesh-Conv operator in~\cite{hu2022meshconv} was restricted to scenarios where the spatial resolutions of the feature maps remain constant. In contrast, the U-Net architecture builds a hierarchical multi-level representation through successive downsampling and upsampling, enabling progressively larger receptive fields to facilitate the capture of global features. Consequently, the feature maps exhibit varying spatial resolutions at different levels. Therefore, successfully integrating Mesh-Conv operator into the U-Net framework requires the distance vectors to be dynamically adapted to the specific grid resolution at each hierarchical level.

\begin{figure}[htbp]
\centering
  \includegraphics[scale=0.5]{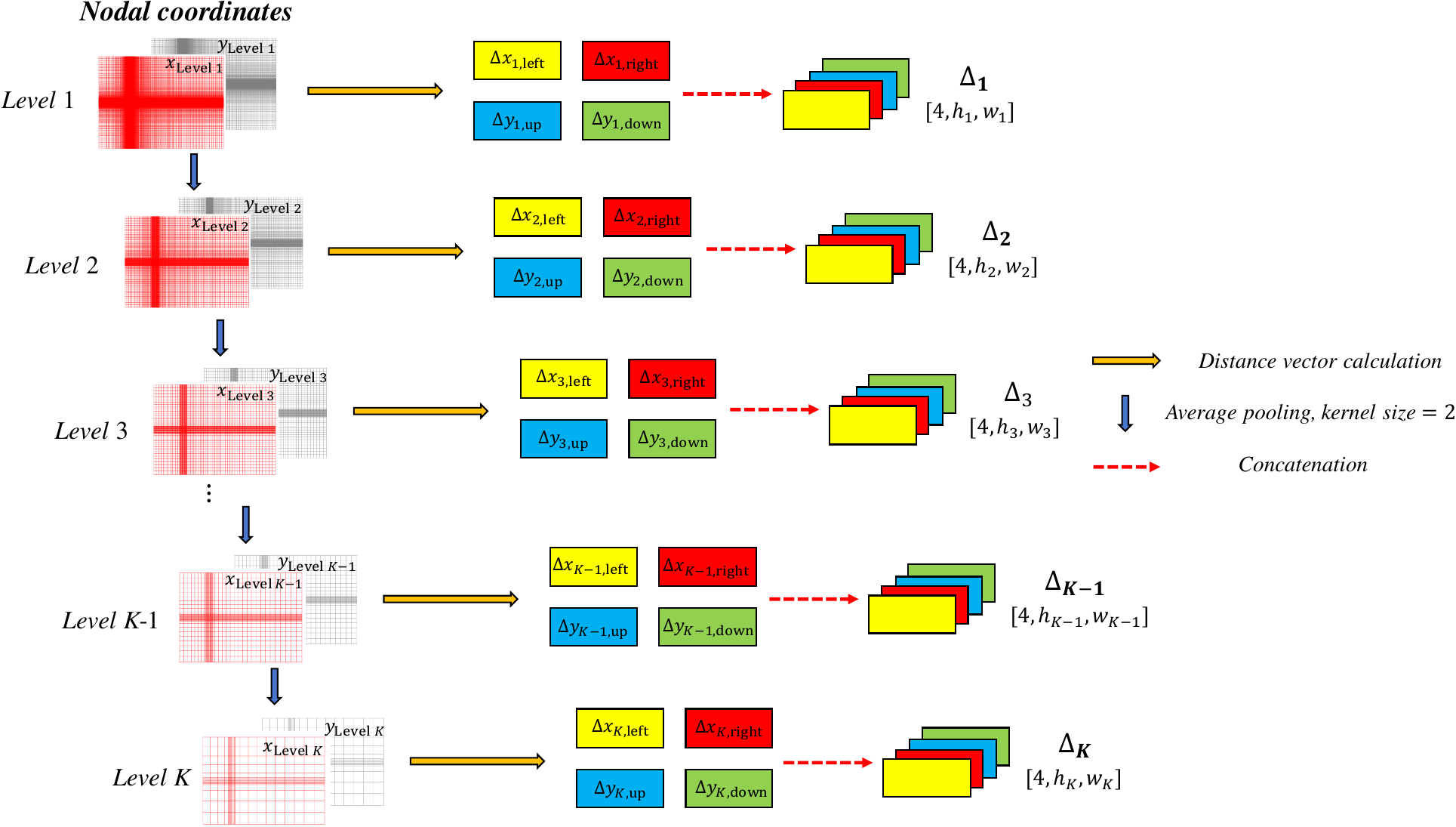}
  \caption{Schematic illustration of the multi-level distance vector map construction strategy.}\label{Level_distancevector_Cal}
\end{figure}

To enable this dynamic adaptation, we designed a multi-level distance vector map construction strategy, as explicitly detailed in Fig.~\ref{Level_distancevector_Cal}. Initially, an average pooling operation is applied to compute the physical nodal coordinates corresponding to each of the $K$ levels of the U-Net. Based on these pooled coordinates, the level-specific distance vector maps are calculated. Specifically, for any given target node, we compute the absolute coordinate differences between the target node and its four directly adjacent source nodes: the left and right neighbors along the $x$-axis, and the upper and lower neighbors along the $y$-axis. By computing these four directional distance components across all target nodes at a specific level $K$ and arranging them spatially, we construct four distinct grid-spacing maps~($\Delta x_{K,\text{left}}$, $\Delta x_{K, \text{right}}$, $\Delta y_{K, \text{up}}$ and $\Delta y_{K, \text{down}}$). Each map corresponds to one direction and possesses a spatial resolution of $h_{K} \times w_{K}$. Ultimately, these four maps are concatenated along the channel dimension to construct the complete distance vector map $\Delta_{K}$ with a shape of $[4,h_{K},w_{K}]$. $\Delta_{K}$ comprehensively encodes the local grid-spacing information of the input feature map at level $K$ of the U-Net, and is subsequently utilized in the level-specific Mesh-Conv operator. Furthermore, regarding the handling of grid-spacing information at the domain boundaries, we employ a nearest-neighbor padding strategy, wherein the virtual grid spacing exterior to a boundary target node is set equal to equal the grid spacing of its immediate internal neighbor.

\begin{figure}[htbp]
\centering
  \includegraphics[scale=0.5]{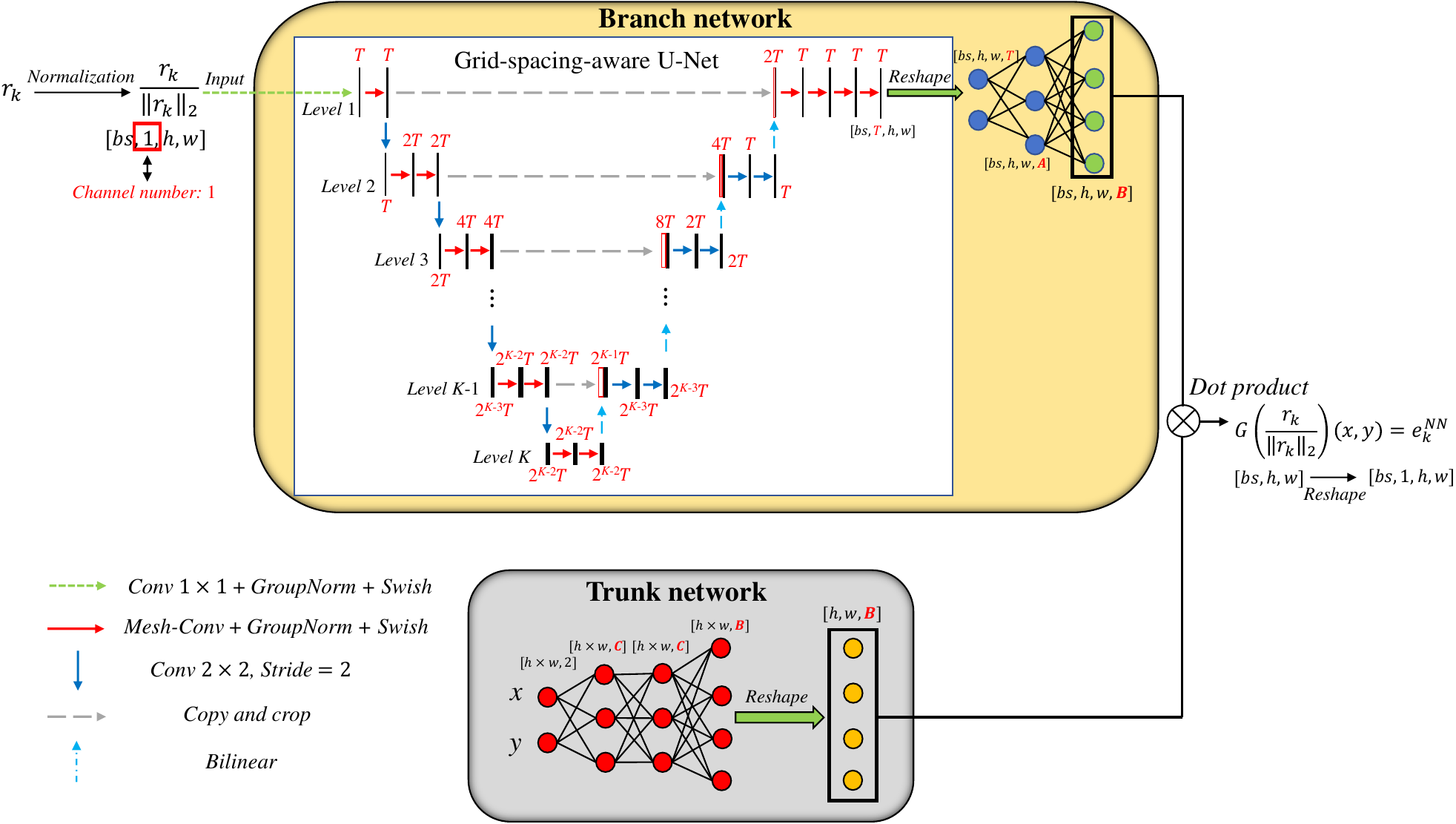}
  \caption{Architecture of deep operator network.}\label{DeepONet_MUNet}
\end{figure}

The resulting grid-spacing-aware U-Net is employed as the branch network within the DeepONet framework, and the overall architecture is illustrated in Fig.~\ref{DeepONet_MUNet}. Here, $T$, $2T$, $4T$, etc. denote the numbers of feature map channels at the corresponding network layers, while $A$, $B$ and $C$ represent the numbers of neurons in feedforward neural networks. Following our previous work~\cite{bai2026hybrid}, we set $K=5$, $T=40$, $A=100$, $B=200$, and $C=100$.

Any vector $R$ constructed according to Eq.~(\ref{b_construct}) is fed into the DeepONet as input, yielding a predicted output vector $NN(R)$. The loss function is defined as
\begin{eqnarray}
\label{Loss}
 Loss &=& \frac{1}{N} \sum_{i=1}^{N}[R_{i}-(M \cdot NN(R))_{i}]^{2},
\end{eqnarray}
where $N$ represents the total number of grid points. The random mini-batch~\cite{wandel2022spline} method is employed with a $batchsize$~($bs$) of 8, comprising 1000 iterations per epoch for a total of 1000 epochs. The neural network parameters are optimized using the Adam optimizer enhanced with Sharpness-Aware Minimization (SAM)~\cite{foret2020SAM}. The learning rate~($lr$) follows a cosine decay schedule with an initial value of 0.0002, and the perturbation radius in SAM is set to 0.0002~\cite{bai2024DON}. In all cases, the training dataset contains $54{,}000$ samples.

\subsection{Technical details}
\label{Technical_details}

Table~\ref{tab:components} summarizes components, number of grid points, maximum-to-minimum grid-spacing ratio~($\Delta_{\max}/\Delta_{\min}$) and applications of HyDEA. Each implementation of HyDEA incorporates a CG-type method and DLSM. We employ the vanilla conjugate gradient~(CG) method together with three preconditioned variants, namely, the incomplete Cholesky decomposition PCG~(ICPCG) method, the Jacobi PCG~(JPCG) method, and a 4-level multigrid PCG~(MGPCG-4) method as the CG-type method. For example:
\begin{itemize}
 \item HyDEA~(ICPCG + DLSM-1-Conv) denotes the hybridization of the ICPCG method and DLSM for Case 1, where the branch network of the DLSM employs the standard convolution operator. 
 \item HyDEA~(ICPCG + DLSM-2-MConv) denotes the hybridization of the ICPCG method and DLSM for Case 2, where the branch network of the DLSM employs the modified Mesh-Conv operator.
\end{itemize}

The multigrid preconditioner is based on the smoothed aggregation algebraic multigrid utilizing a V-cycle. On the finer levels of the multigrid hierarchy, a Gauss-Seidel-preconditioned Chebyshev polynomial smoother is applied with two pre- and post-smoothing iterations per level, whereas the system on the coarsest grid is solved exactly using a direct LU factorization. Furthermore, since the coefficient matrix of the PPE remains constant throughout the simulation, the preconditioner is constructed and stored exclusively at the initial time step and is then directly reused in all subsequent time steps to avoid redundant computational overhead during the setup phase.

DLSM and DeepONet are implemented utilizing $\mathtt{Python}$ and $\mathtt{PyTorch}$. CG-type methods are sourced from the $\mathtt{PETSc}$ library~\cite{balay2019petsc}, which is implemented in $\mathtt{C}$, with its $\mathtt{Python}$ interface $\mathtt{petsc4py}$.

The $\mathtt{Python}$-$\mathtt{PetIBM}$ interface is built using the $\mathtt{pybind11}$ library~\cite{Pybind11}.

The DeepONet models are trained and deployed exclusively on a single NVIDIA GeForce RTX4090, while remaining computations are executed on a single Intel Xeon Silver 4210R CPU.

The source code for HyDEA is available on Github at \url{https://github.com/HMB9666/HyDEA}.

\begin{table}
    \renewcommand{\arraystretch}{1.5}
    \setlength{\tabcolsep}{3pt}
    \normalsize
    \centering
    \caption{Components and associated number of grid points of HyDEA~(CG-type + DLSM).}
    \begin{tabular}{p{8.5cm}p{1.7cm}p{1.7cm}p{1.7cm}}
    
    \hline
      HyDEA(CG-type+Deep learning) &  DLSM-1  &  DLSM-2 & DLSM-3\\
    \hline
      Number of grid points &  $50{,}625$  & $107{,}016$ & $87{,}320$ \\
      CG    & \checkmark &  -  &  -   \\     
      ICPCG  & \checkmark &  \checkmark  & \checkmark \\
      JPCG    & \checkmark &  -  &  -   \\
      MGPCG-4  & \checkmark  &  \checkmark  &  -    \\
      $\Delta_{\max}/\Delta_{\min}$   &  2.3 & 41 & 28 \\
      Standard convolution operator in branch network~(-Conv)   &  \checkmark & \checkmark & - \\
      Modified Mesh-Conv operator in branch network~(-MConv)   &  \checkmark & \checkmark & \checkmark \\
    \hline
    \end{tabular}
    \label{tab:components}
\end{table}

\section{Results and discussions}
\label{result}

We systematically evaluate the performance of the grid-spacing-aware HyDEA through three benchmark cases governed by viscous incompressible fluid dynamics, and provide comprehensive comparisons against both the standalone CG-type method and standard convolution-based HyDEA in Ref.~\cite{bai2026hybrid}.

Case 1: In Section~\ref{Case1_cavity_cylinder}, HyDEA is evaluated for two-dimensional~(2D) lid-driven cavity flow with an embedded stationary circular cylinder. The computational grid is locally refined near the cylinder region, with $\Delta_{\max}/\Delta_{\min} \approx 2.3$. 

Case 2: Section~\ref{flow_around_obstacle} investigates more 2D flows past an obstacle. The computational grid is locally refined near the obstacle region, achieving $\Delta_{\max}/\Delta_{\min} \approx 41$. More specifically:
\begin{itemize}
 \item in Section~\ref{Re100Cylinder}, HyDEA is evaluated for classical 2D flow past a circular cylinder at $Re=100$.
 \item in Section~\ref{elliptical_cylinder}, HyDEA is evaluated for 2D flow past an elliptical cylinder.
 \item in Section~\ref{flow past suboff}, HyDEA is evaluated for 2D flow past the DARPA SUBOFF profile.
 \item in Section~\ref{flow past Oscylinder}, HyDEA is evaluated for 2D flow past an inline oscillating cylinder.
\end{itemize}

Case 3: Section~\ref{Flapwing} evaluates the performance of HyDEA for 2D flapping elliptical wing at $Re=75$. The computational grid is locally refined near the elliptical wing region, achieving $\Delta_{\max}/\Delta_{\min} \approx 28$.

For the training dataset preparation, we employ the parameter values $m=7000$ and $b=0.6$. These predetermined parameter values have been used in Ref.~\cite{bai2026hybrid} and demonstrated excellent performance. In all cases presented in this work, we adopt the same parameter setting and do not perform further parameter analysis.

\subsection{Case 1: 2D lid-driven cavity flow with an embedded stationary circular cylinder}
\label{Case1_cavity_cylinder}

\begin{figure}[htbp] 
 \centering  
  \subfigure[]{
  \label{Case1_domain}
  \includegraphics[scale=0.6]{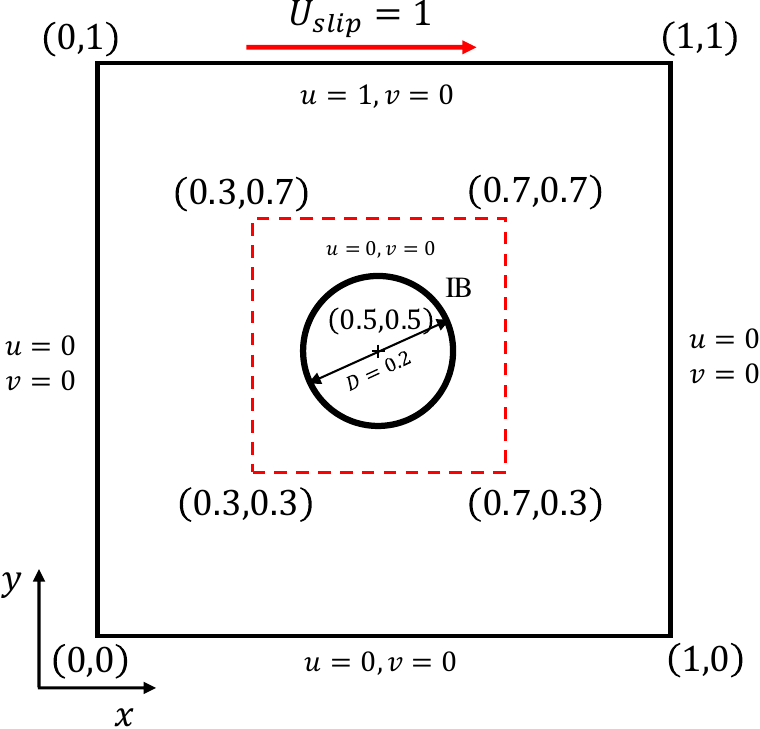}}
  \subfigure[]{
  \label{Case1_grid}
  \includegraphics[scale=0.26]{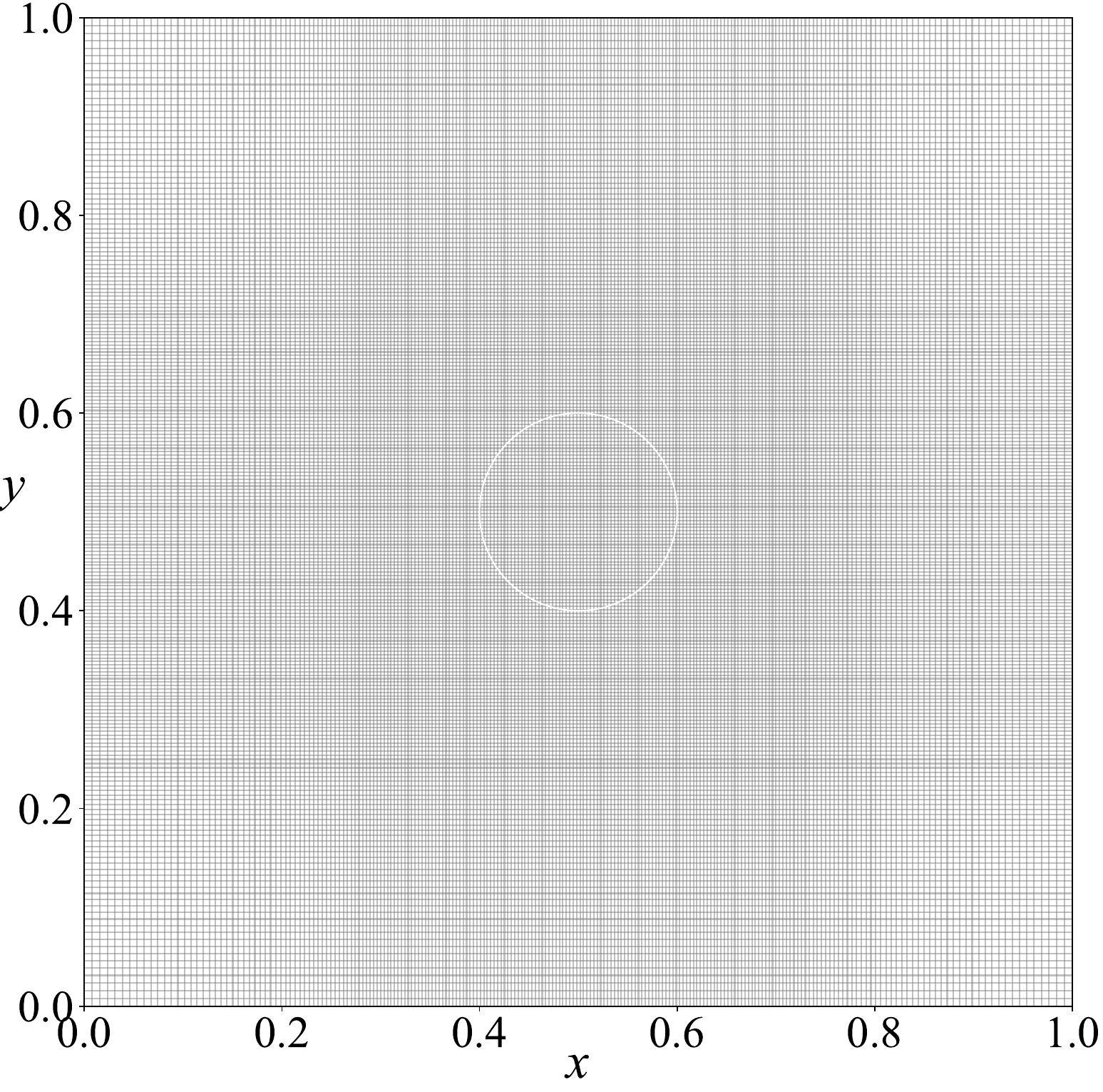}}
 \caption{Schematic diagram of 2D lid-driven cavity flow with an embedded stationary circular  cylinder. (a) Flow boundary conditions and geometric configuration. (b) Computational grid.}
 \label{Case1Domain_Grid}
\end{figure}

The flow boundary conditions, geometric configuration, and computational grid of the flow are illustrated in Fig.~\ref{Case1Domain_Grid}. The numerical simulation is performed on a square domain with identical horizontal and vertical side lengths $H_x=H_y=H=1$. A non-uniform Cartesian grid is implemented for computational domain discretization, featuring a locally refined uniform grid~($\Delta x=\Delta y=0.003478$) within a specified square region around the cylinder. Beyond this refined region, the grid undergoes progressive coarsening, resulting in $\Delta_{\max}/\Delta_{\min} \approx 2.3$. The computational grid consists of $50{,}625$ cells. The CFD time step size $\Delta t=0.002$, the kinematic viscosity~($\nu$) is 0.0002, and the diameter of cylinder $D=0.2$. The Reynolds number $Re=uH/\nu=5000$. We set $Num_{\mathrm{CG-type}}=3$ and $Num_{\mathrm{DLSM}}=2$, and the iteration terminates when the residual $L2$-norm falls below $\epsilon=10^{-6}$.

\begin{figure}[htbp] 
 \centering  
  \subfigure[]{
  \label{Case1_Rline_10steps_CG}
  \includegraphics[scale=0.21]{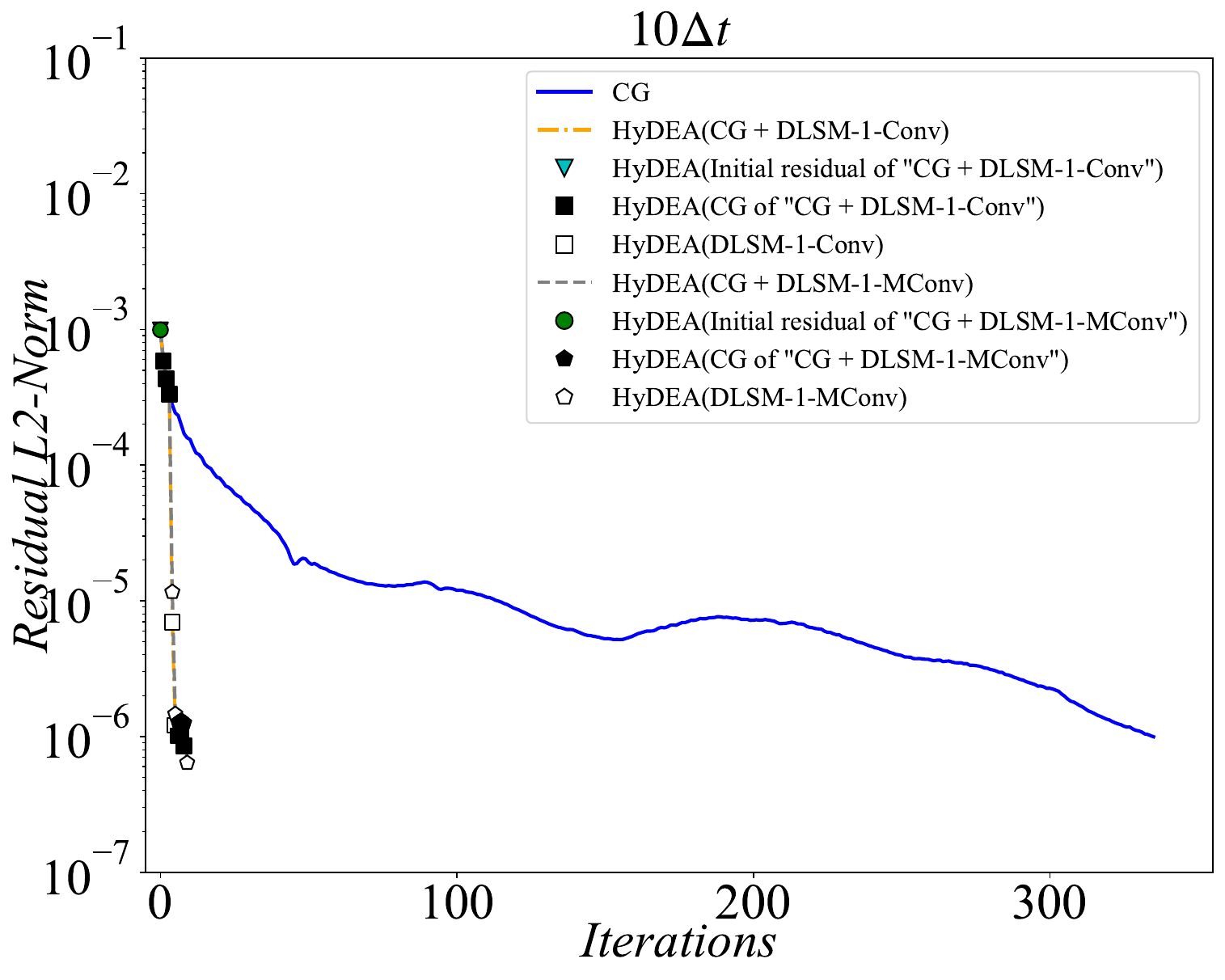}}
  \subfigure[]{
  \label{Case1_Rline_100steps_CG}
  \includegraphics[scale=0.21]{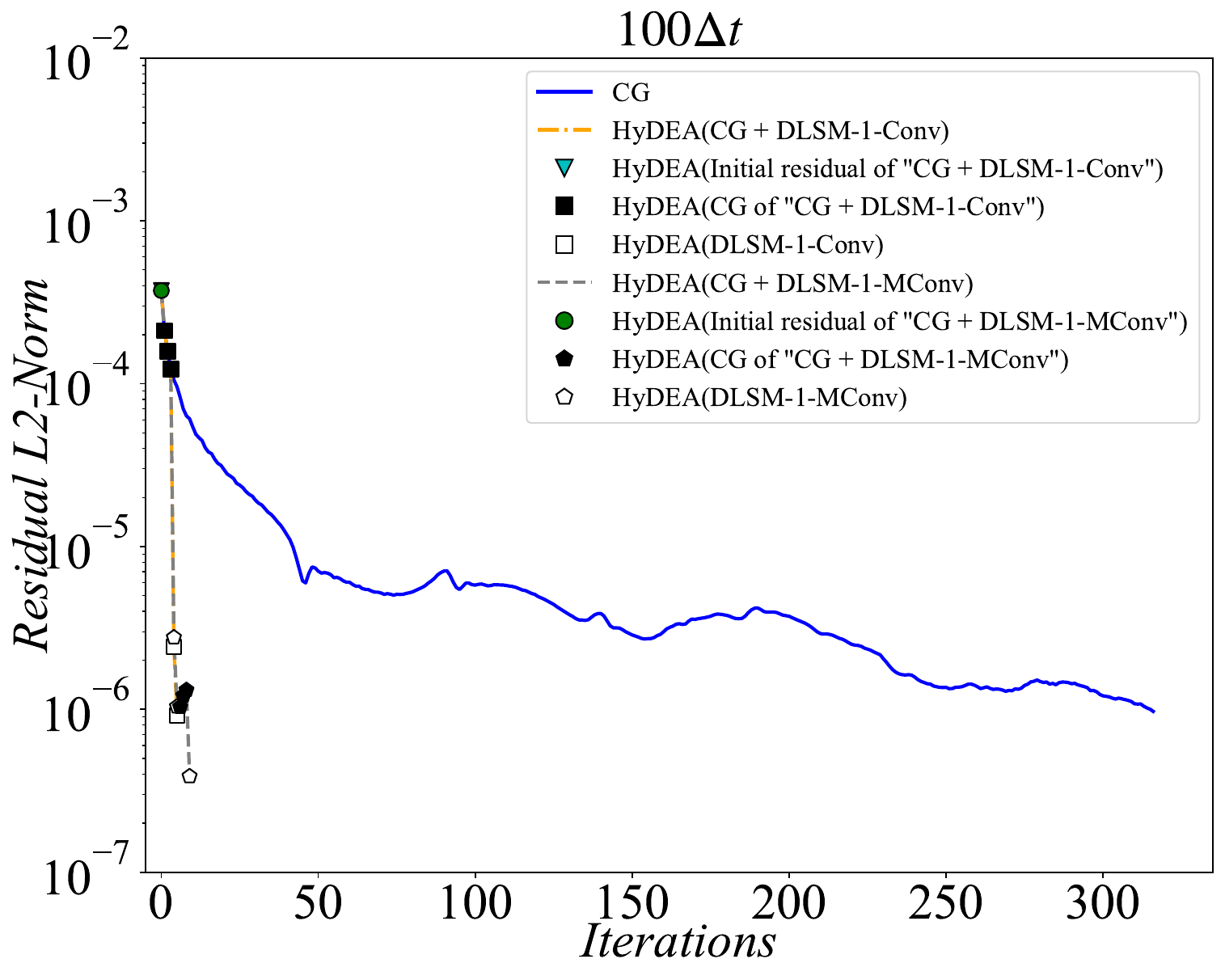}}
  \subfigure[]{
  \label{Case1_Rline_1000steps_CG}
  \includegraphics[scale=0.21]{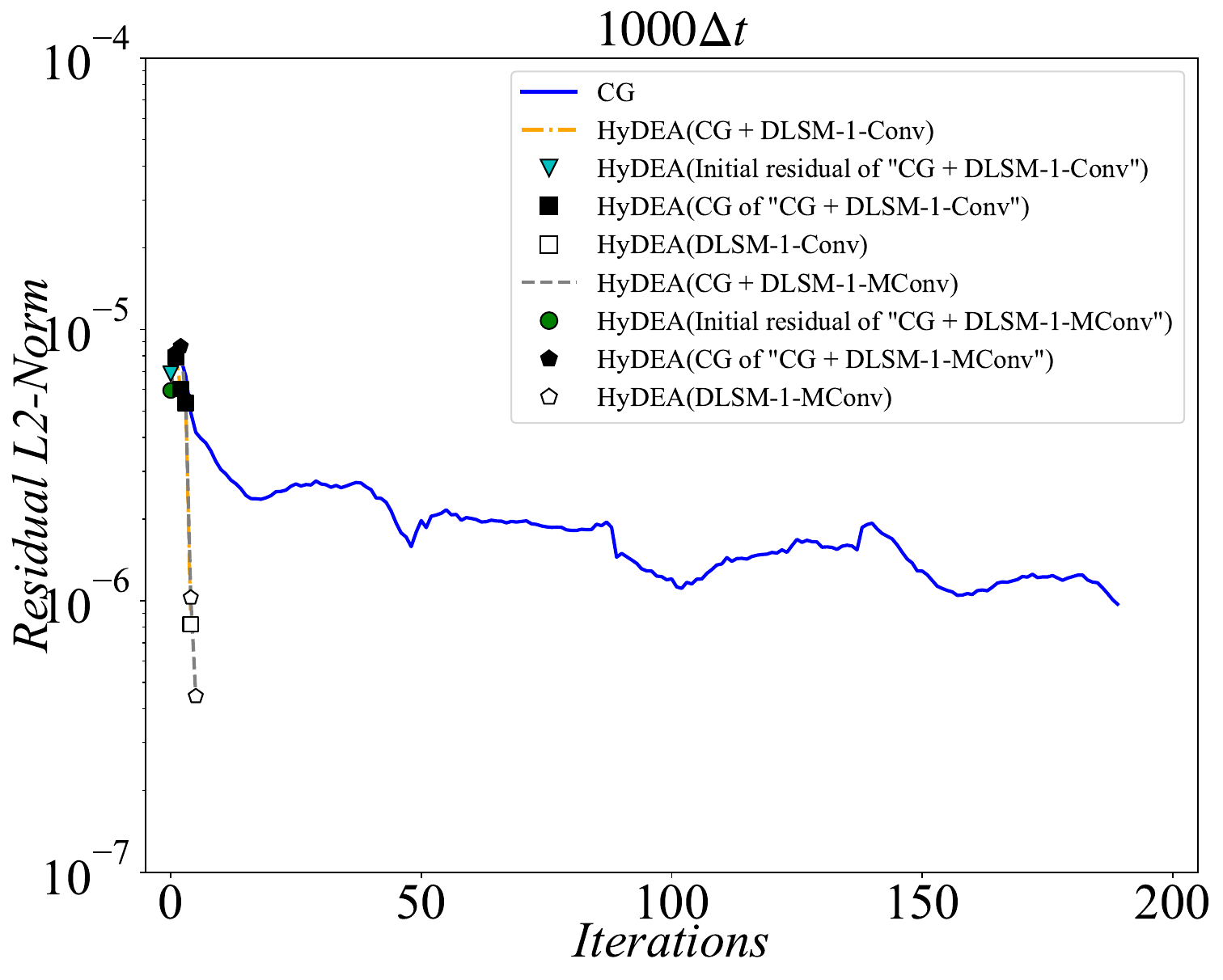}}  
  \subfigure[]{
  \label{Case1_Rline_10steps_ICPCG}
  \includegraphics[scale=0.21]{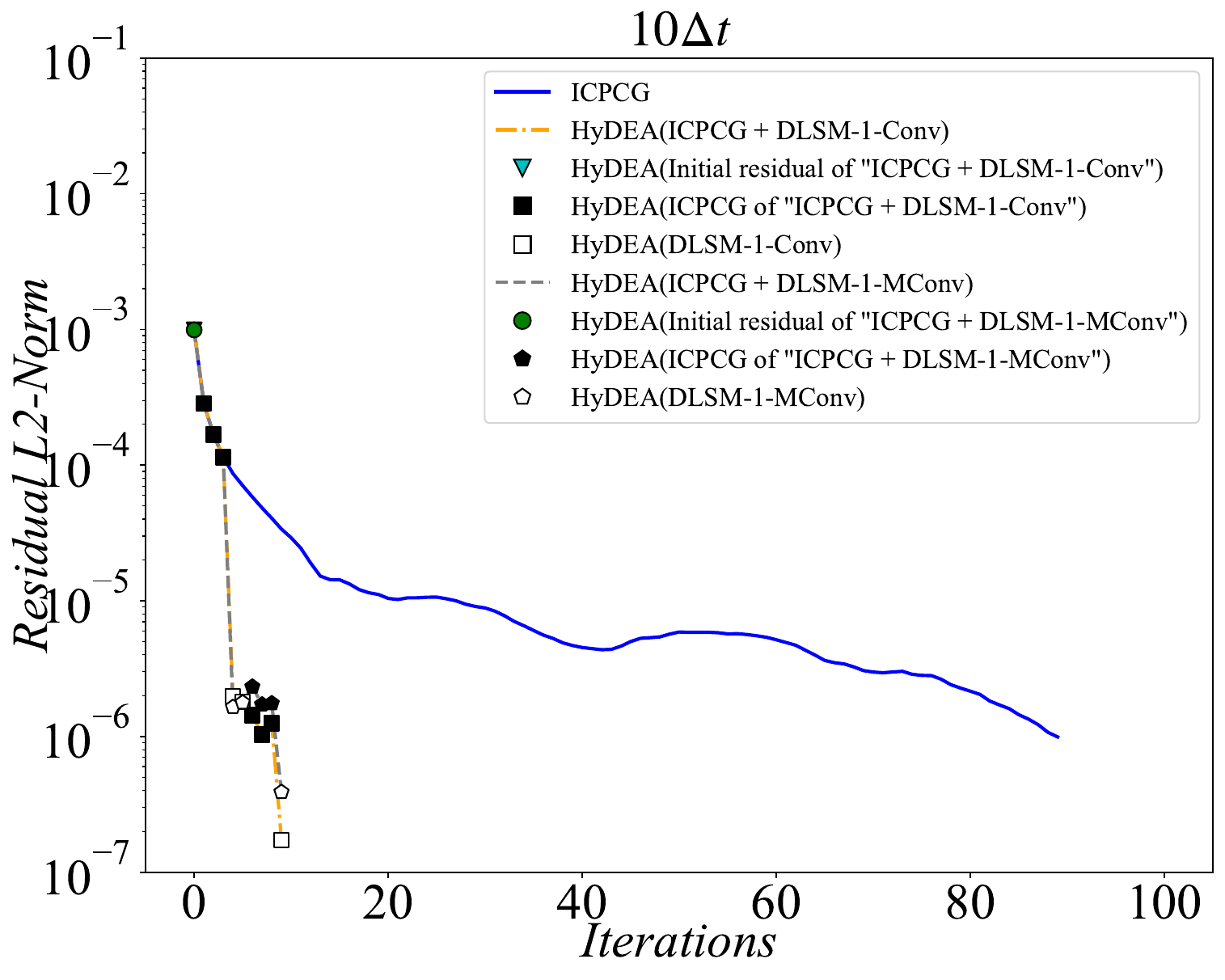}}
  \subfigure[]{
  \label{Case1_Rline_100steps_ICPCG}
  \includegraphics[scale=0.21]{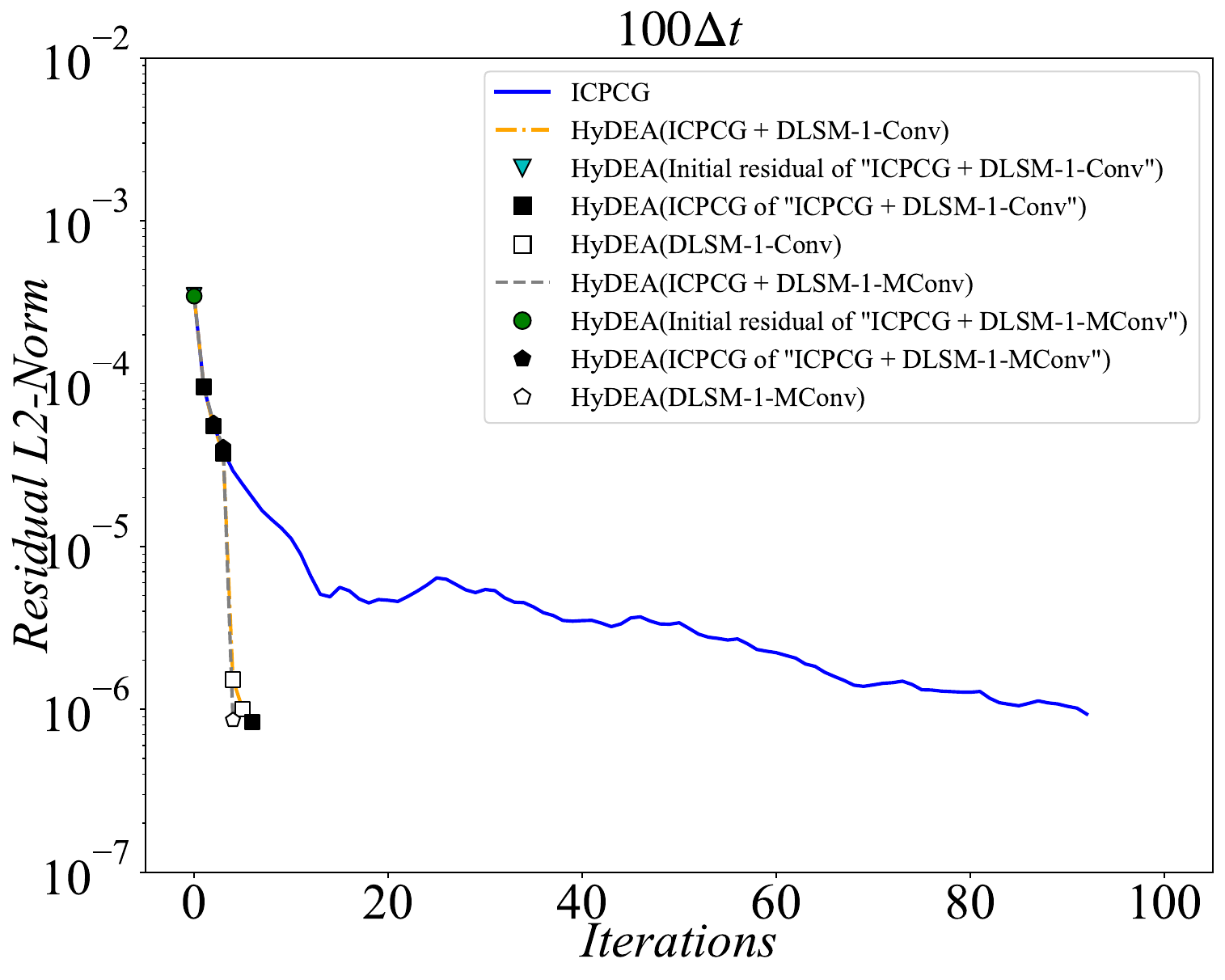}}
  \subfigure[]{
  \label{Case1_Rline_1000steps_ICPCG}
  \includegraphics[scale=0.21]{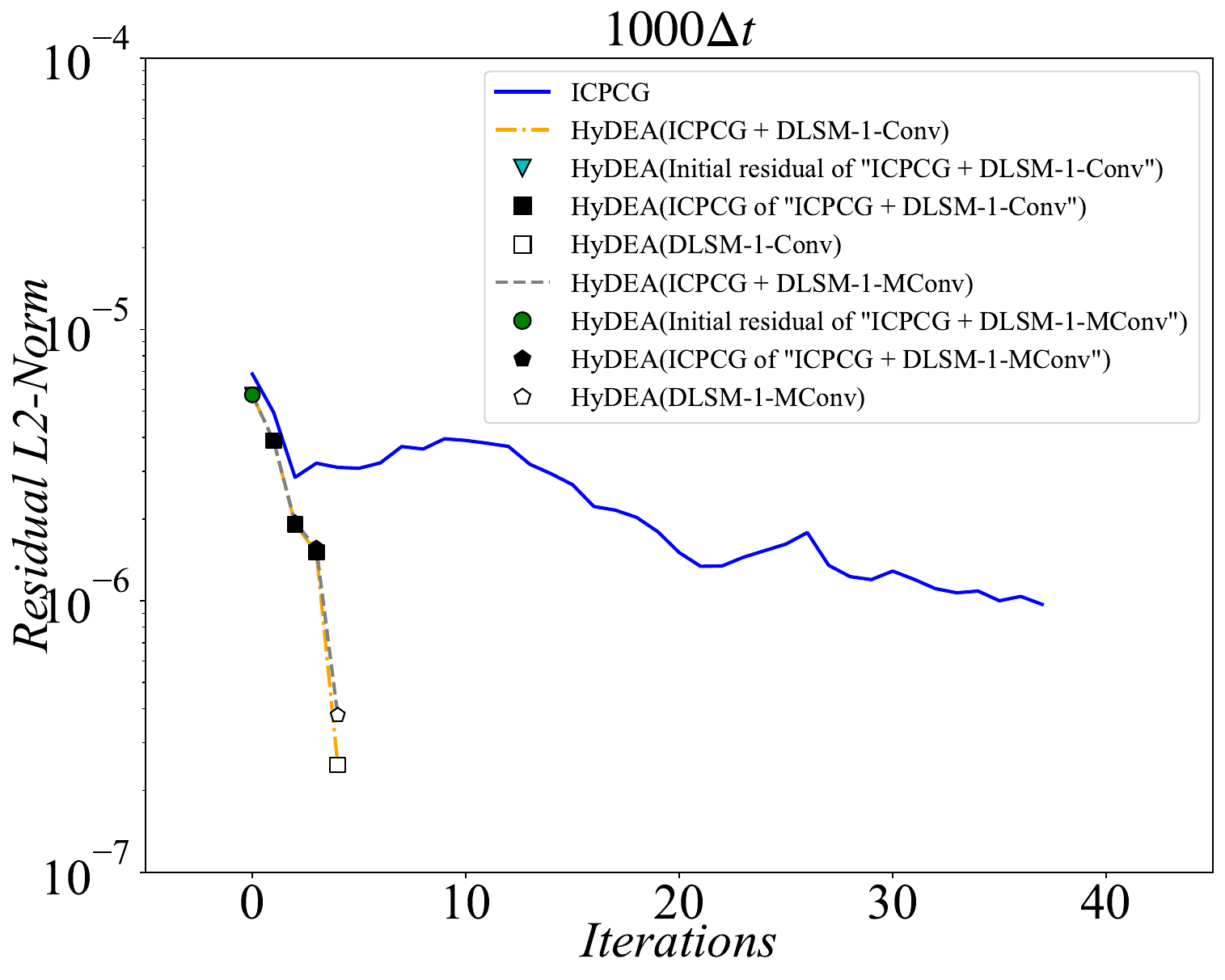}}
  \subfigure[]{
  \label{Case1_Rline_10steps_JPCG}
  \includegraphics[scale=0.21]{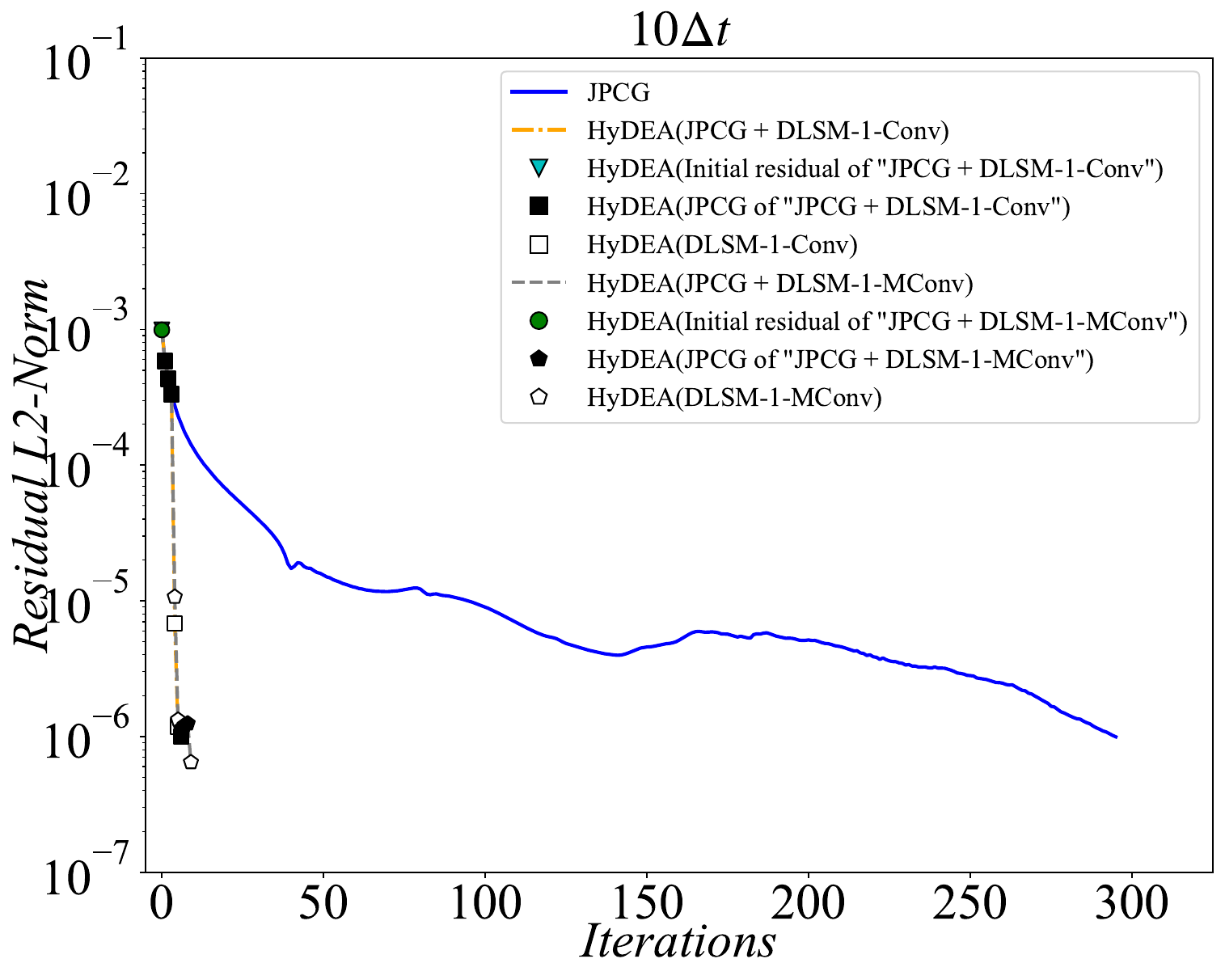}}
  \subfigure[]{
  \label{Case1_Rline_100steps_JPCG}
  \includegraphics[scale=0.21]{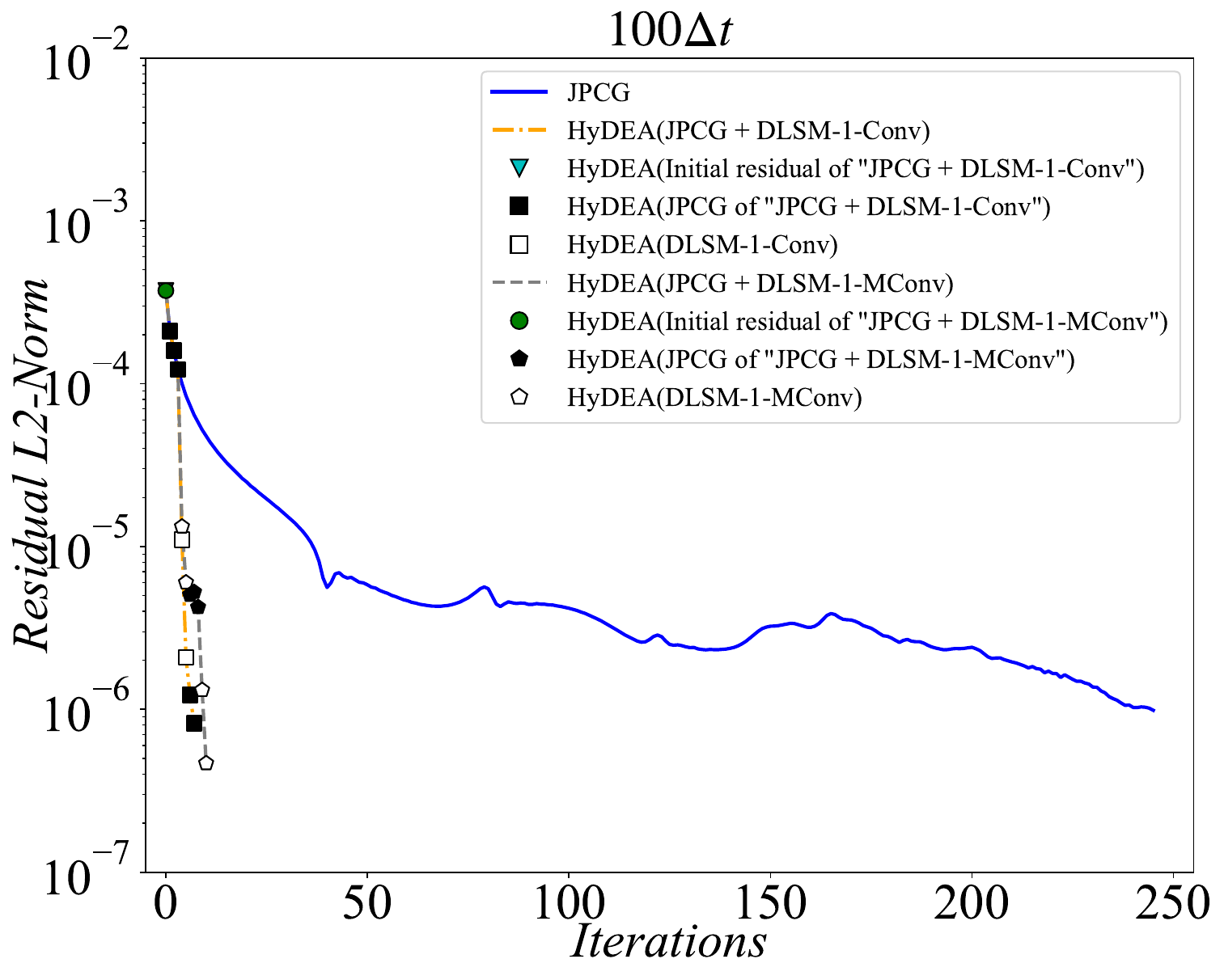}}
  \subfigure[]{
  \label{Case1_Rline_1000steps_JPCG}
  \includegraphics[scale=0.21]{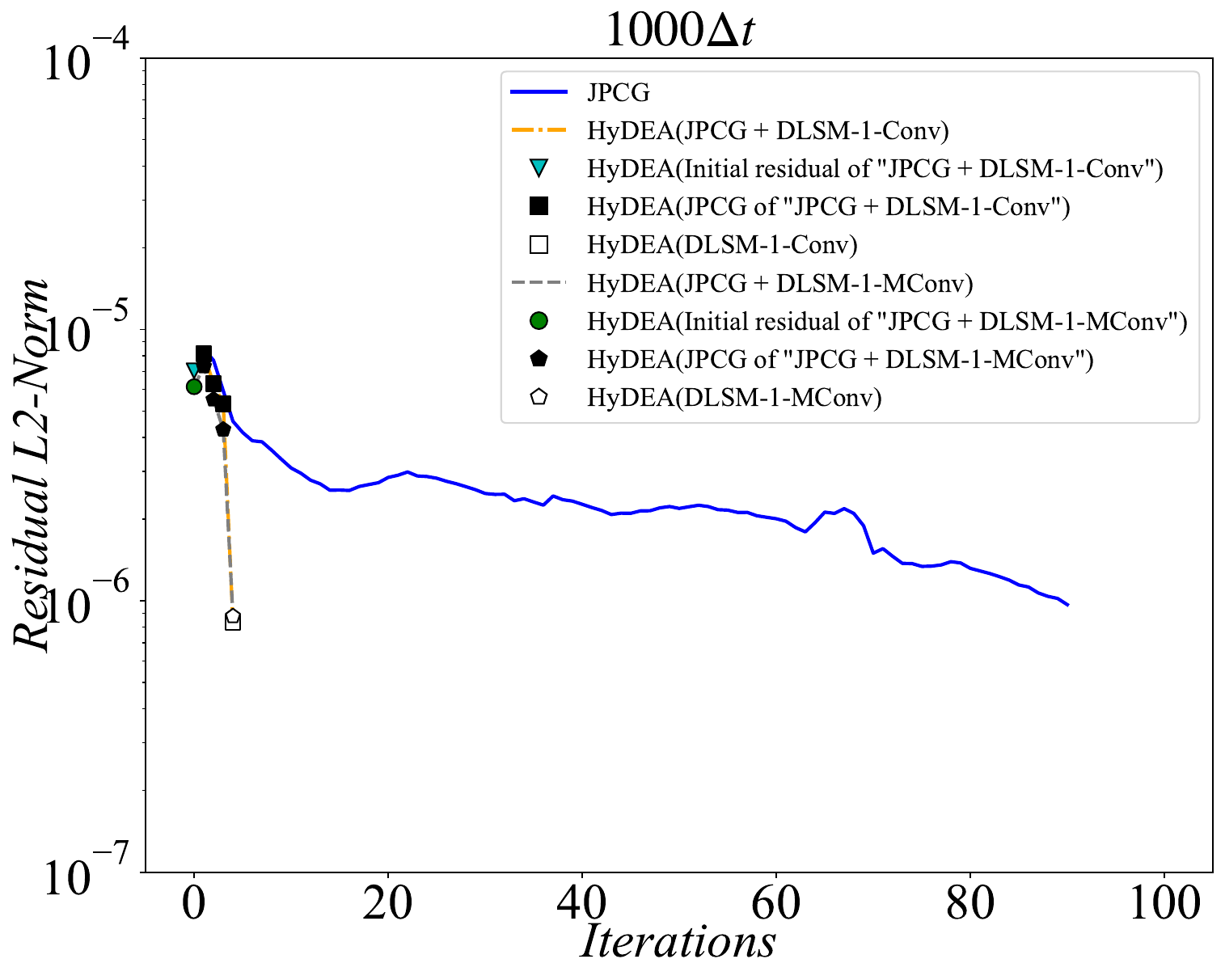}}
  \subfigure[]{
  \label{Case1_Rline_10steps_MGPCG}
  \includegraphics[scale=0.21]{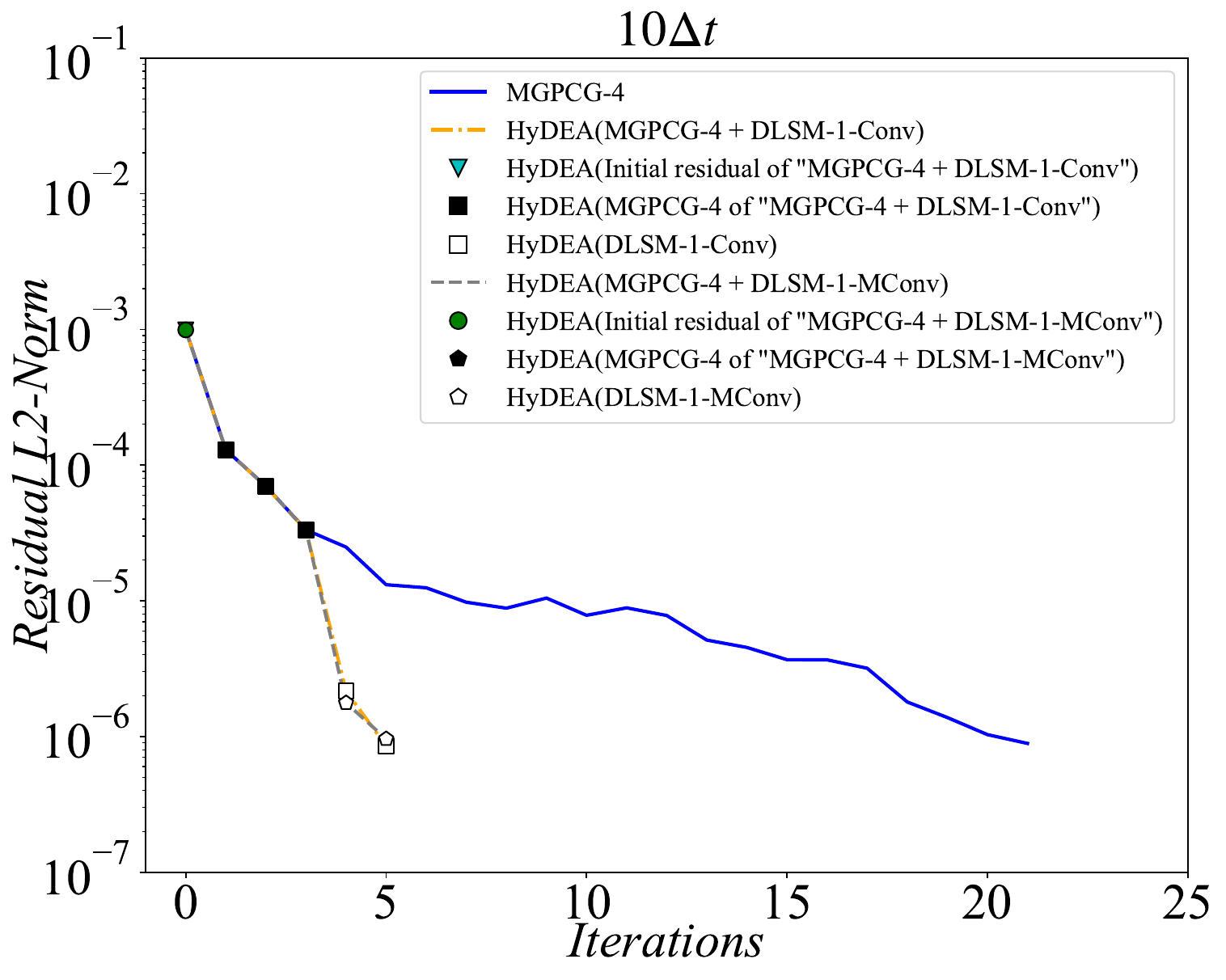}}
  \subfigure[]{
  \label{Case1_Rline_100steps_MGPCG}
  \includegraphics[scale=0.21]{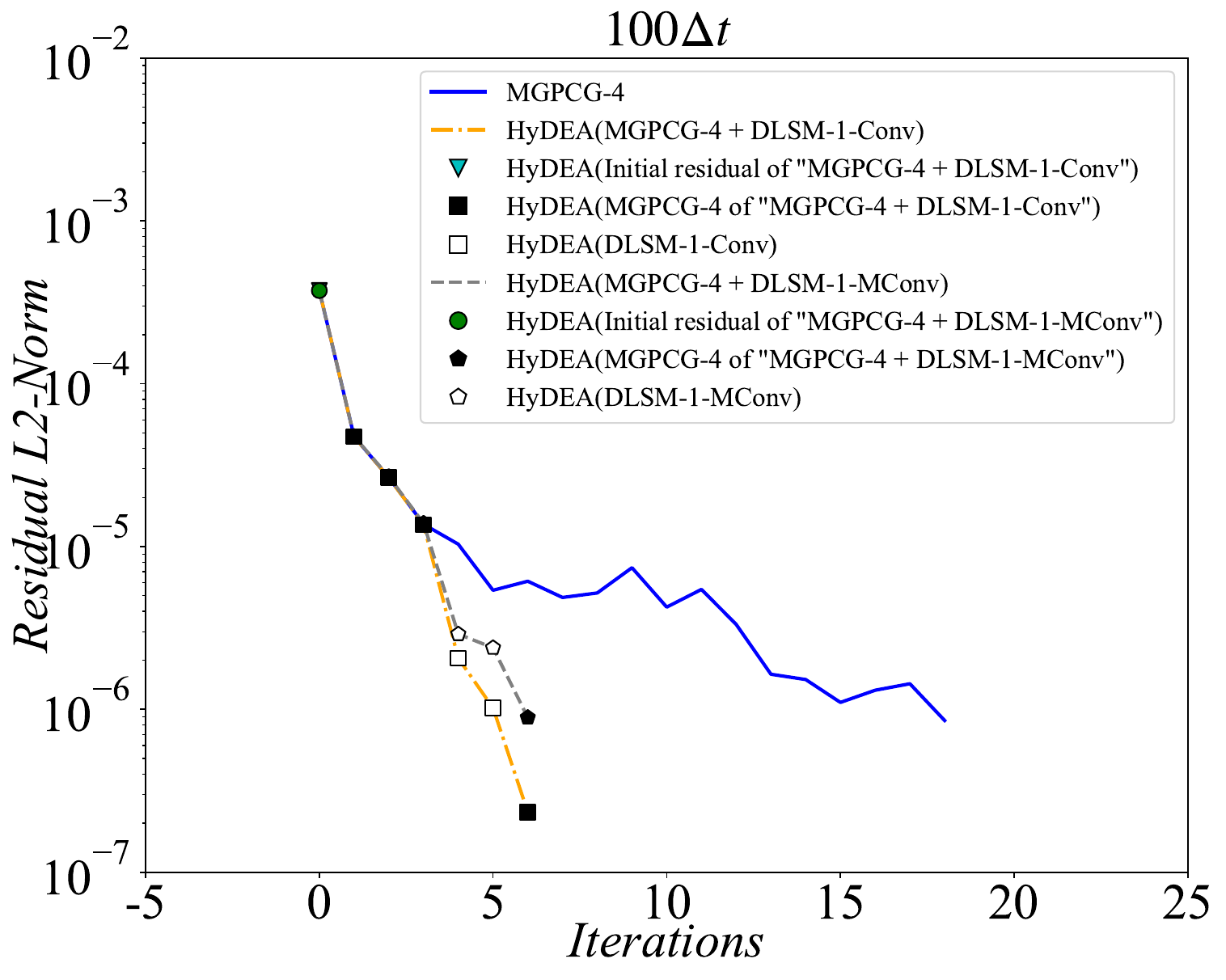}}
  \subfigure[]{
  \label{Case1_Rline_1000steps_MGPCG}
  \includegraphics[scale=0.21]{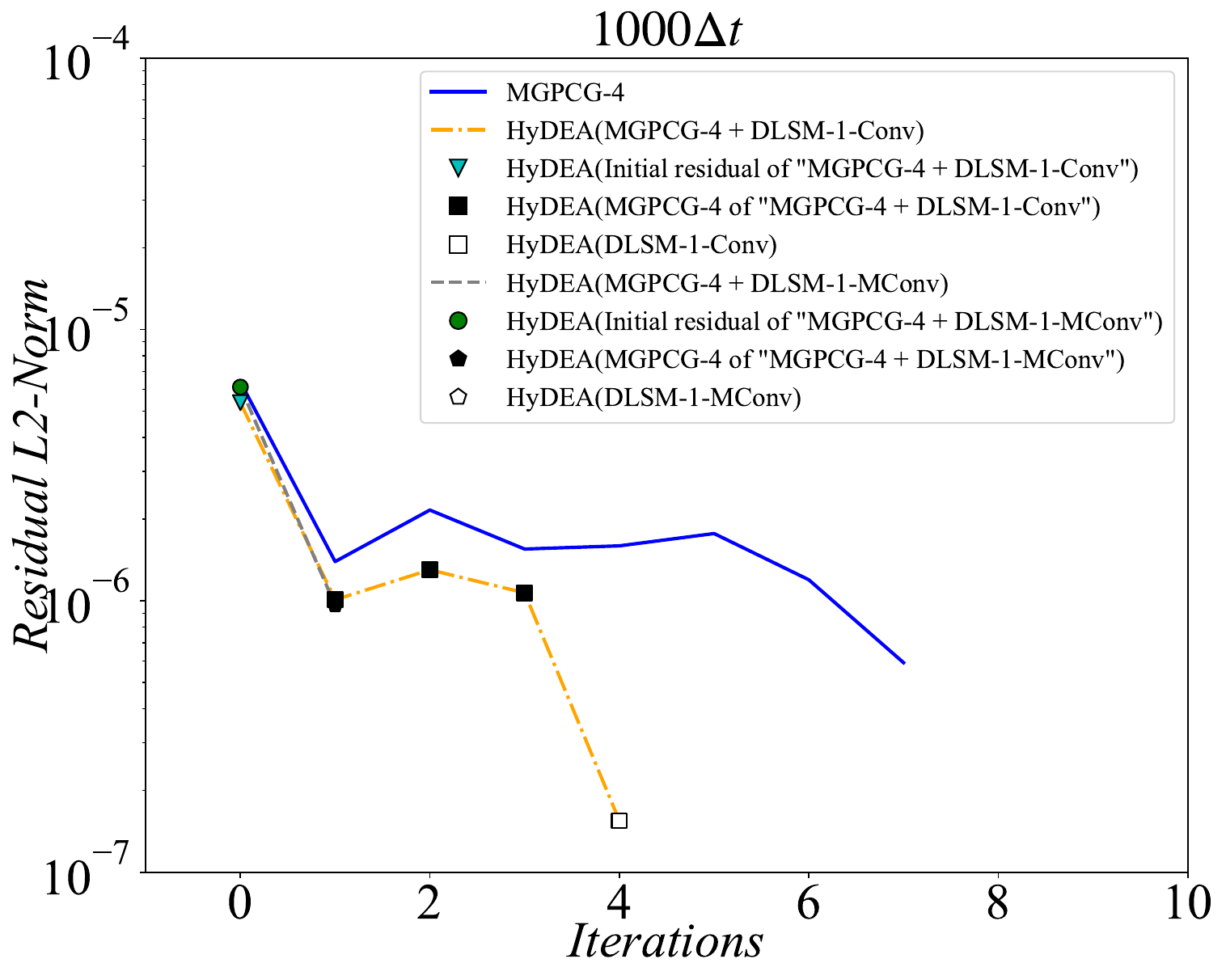}}
  \caption{Iterative residuals of solving the PPE for 2D lid-driven cavity flow with an embedded stationary circular cylinder at the $10th$, $100th$ and $1000th$ time steps. (a)-(c) HyDEA~(CG + DLSM-1-Conv/MConv). (d)-(f) HyDEA~(ICPCG + DLSM-1-Conv/MConv). (g)-(i) HyDEA~(JPCG + DLSM-1-Conv/MConv). (j)-(l) HyDEA~(MGPCG-4 + DLSM-1-Conv/MConv).}\label{Case1_Rline_3+2}
\end{figure}

The iterative residuals of solving the PPE using the CG-type methods, HyDEA~(CG-type + DLSM-1-Conv) and HyDEA~(CG-type + DLSM-1-MConv) are compared at three representative time steps~($10\Delta t$, $100\Delta t$ and $1000\Delta t$) in Fig.~\ref{Case1_Rline_3+2}. 
Both HyDEA~(CG-type + DLSM-1-Conv) and HyDEA~(CG-type + DLSM-1-MConv) demonstrate rapid residual reduction to the predefined tolerance, achieving significantly fewer iterations than the corresponding CG-type method. Taking HyDEA~(ICPCG + DLSM-1-MConv) at $10\Delta t$ as an example, HyDEA begins with comparable initial residual $L2$-norm~(denoted by solid circle) to the ICPCG method. Subsequently, HyDEA alternates between $Num_{\mathrm{CG-type}}=3$ ICPCG iterations~(solid pentagon) and $Num_{\mathrm{DLSM}}=2$ DLSM-1-MConv iterations~(empty pentagon) in a cyclic manner until the termination criterion is satisfied. HyDEA reaches the predefined tolerance by less than 2 hybrid rounds (9 iterations in total), an approximately ten-fold reduction in iteration count compared to around $90$ iterations of the ICPCG method alone. The reductions in iteration counts are even more pronounced for HyDEA~(CG + DLSM-1-MConv) and HyDEA~(JPCG + DLSM-1-MConv), but less impressive for HyDEA~(MGPCG-4 + DLSM-1-MConv), compared to their respective counterparts. The overall results of HyDEA~(CG-type + DLSM-1-Conv) and HyDEA~(CG-type + DLSM-1-MConv) exhibit comparable performance, indicating that when $\Delta_{\max}/\Delta_{\min}$ is relatively small, the geometric encoding of non-uniform grids is unnecessary.

Furthermore, Table~\ref{acceleration ratio compare case1} summarizes the computational time and wall-time acceleration ratio for solving the PPE over $20{,}000$ consecutive time steps using HyDEA and its corresponding standalone CG-type counterparts. The results indicate that HyDEA consistently achieves effective acceleration compared to the corresponding CG-type methods. Notably, in configurations utilizing the CG, ICPCG, and JPCG, the DLSM module accounts for a substantial proportion of the overall computational time. This is primarily attributed to the $\mathtt{Python}$ and $\mathtt{PyTorch}$ implementation of the DLSM; the inherent overhead of interpreted languages, coupled with the structural complexity of the neural network, introduces additional computational costs. However, HyDEA~(MGPCG-4 + DLSM-1-MConv) presents a different scenario. The MGPCG-4 baseline requires significantly fewer iterations to meet the termination residual threshold compared to other solvers like ICPCG (as compared between Fig.~\ref{Case1_Rline_3+2}(j-l) and (d-f)). Consequently, the margin for further reducing the iteration count via the DLSM module is relatively limited. Nevertheless, the time consumed by the DLSM module is noticeably lower than that of the MGPCG-4 method. This observation confirms that when integrated with an already highly efficient and computationally demanding preconditioned solver, HyDEA retains a clear dual advantage: it successfully accelerates iterative convergence while ensuring the neural network's computational overhead remains lower than that of such a highly intensive traditional solver.

\begin{table}[htbp]
\renewcommand{\arraystretch}{1.5}
\normalsize
\centering
\caption{The computational time and wall-time acceleration ratio for solving the PPE using HyDEA over $20{,}000\Delta t$ for 2D lid-driven cavity flow with an embedded stationary circular cylinder.}
\begin{tabular}{cccc}
\hline
  HyDEA  &   Computational time (s) & Acceleration ratio & \makecell{Computational time of \\ DLSM-1-MConv (s)}  \\
\hline
    (CG + DLSM-1-MConv) & 291 & $\times8.35$ & 231 \\
    (ICPCG + DLSM-1-MConv) & 308 & $\times3.36$ & 206 \\
    (JPCG + DLSM-1-MConv) & 291 & $\times6.79$ & 231 \\
    (MGPCG-4 + DLSM-1-MConv) & 724 & $\times1.73$ & 166 \\
\hline
\end{tabular}
\label{acceleration ratio compare case1}
\end{table}

\begin{figure}[htbp] 
 \centering  
  \subfigure[]{
  \label{U_IC_1500step_Case1}
  \includegraphics[scale=0.135]{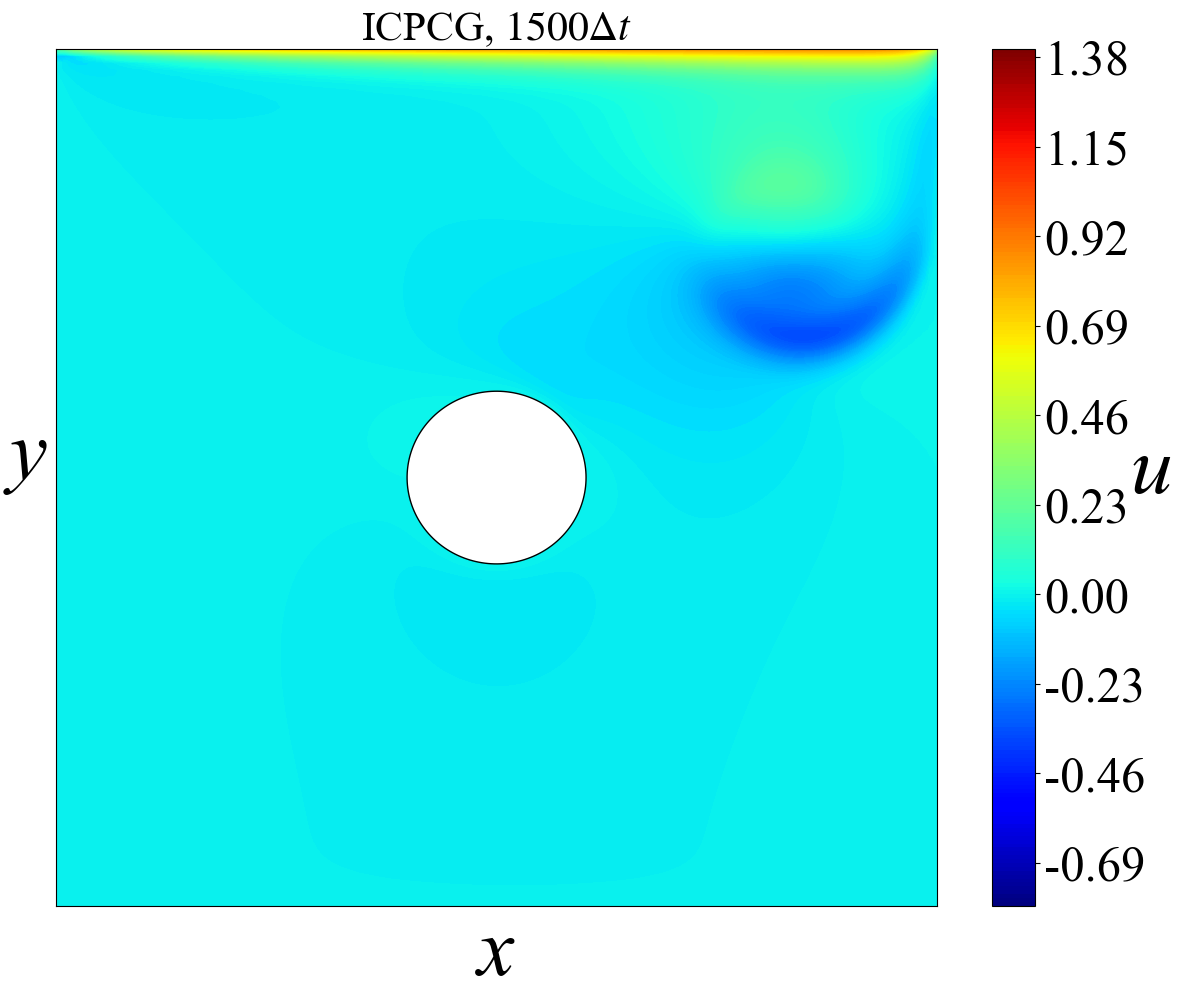}}
  \subfigure[]{
  \label{U_PMHIC_1500step_Case1}
  \includegraphics[scale=0.135]{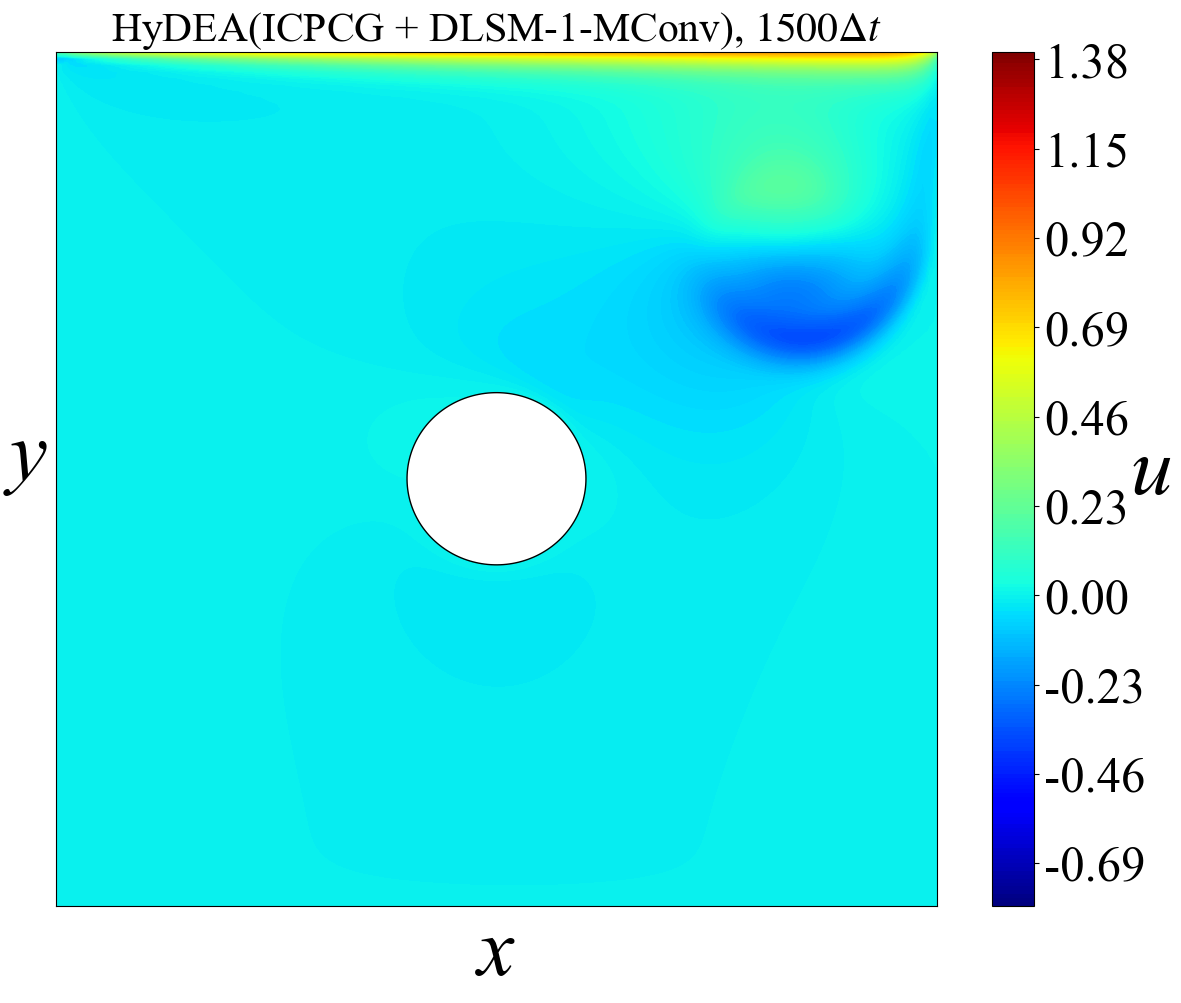}}
  \subfigure[]{
  \label{V_IC_1500step_Case1}
  \includegraphics[scale=0.135]{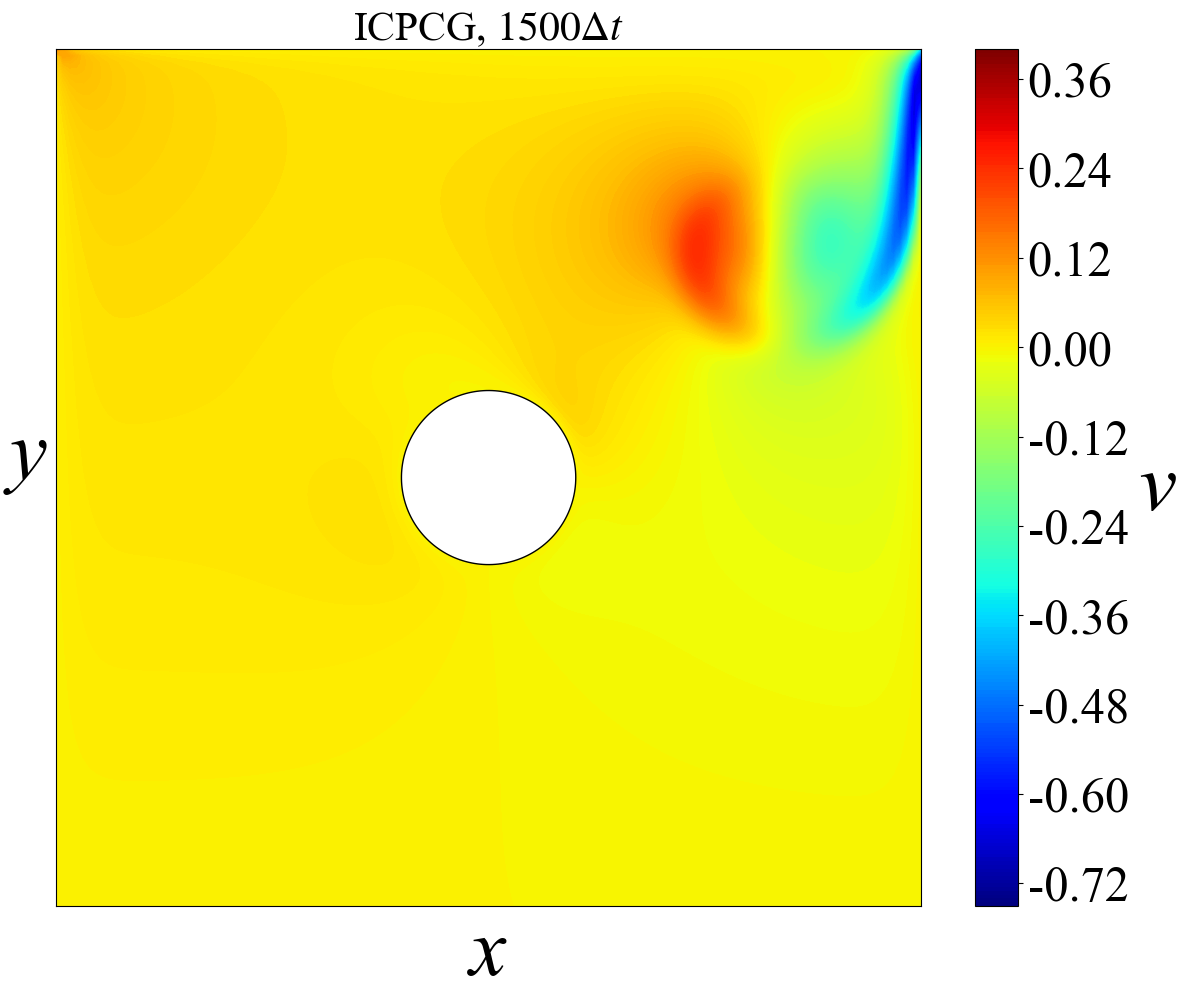}}
  \subfigure[]{
  \label{V_PMHIC_1500step_Case1}
  \includegraphics[scale=0.135]{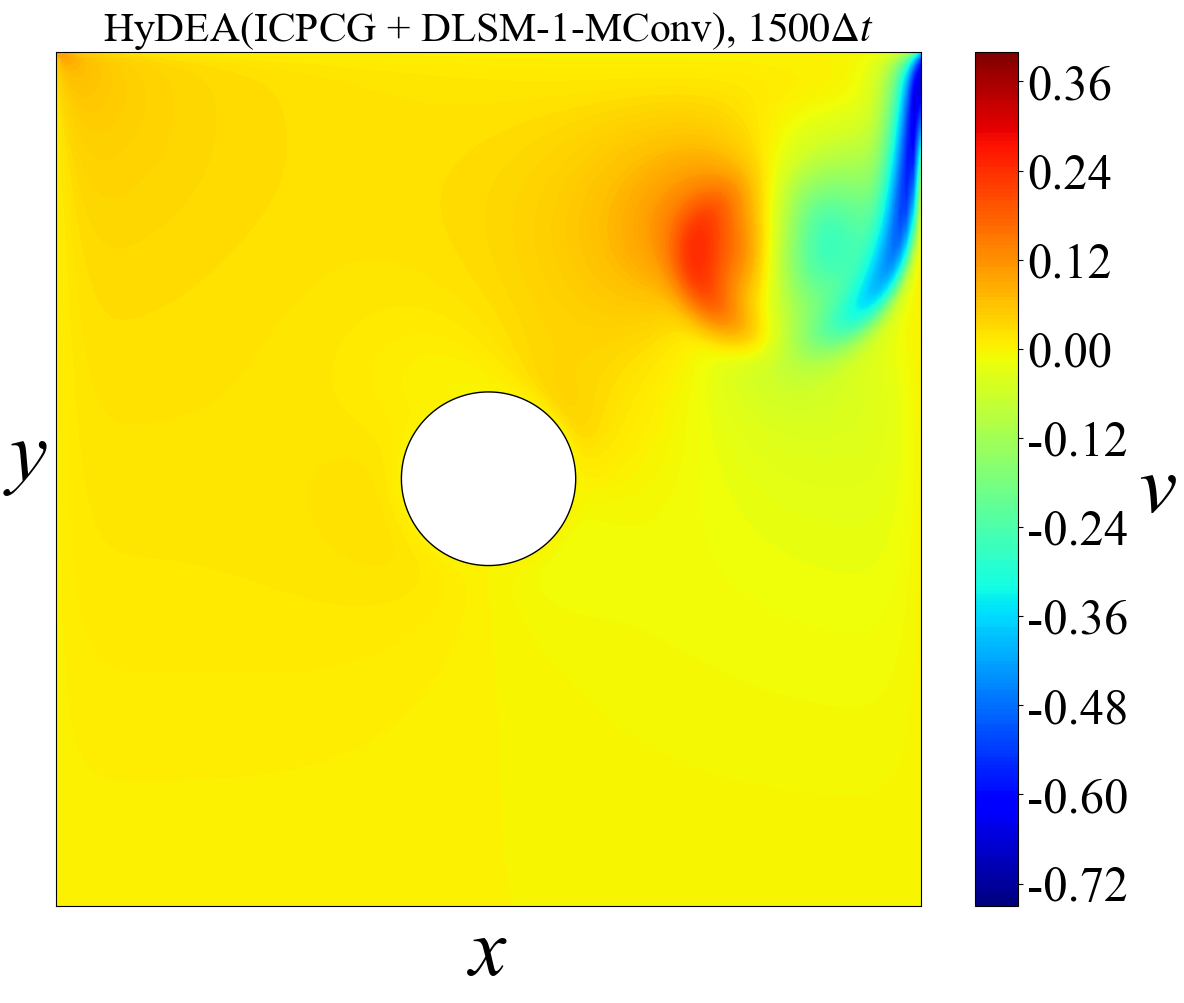}}
  \subfigure[]{
  \label{U_IC_6000step_Case1}
  \includegraphics[scale=0.135]{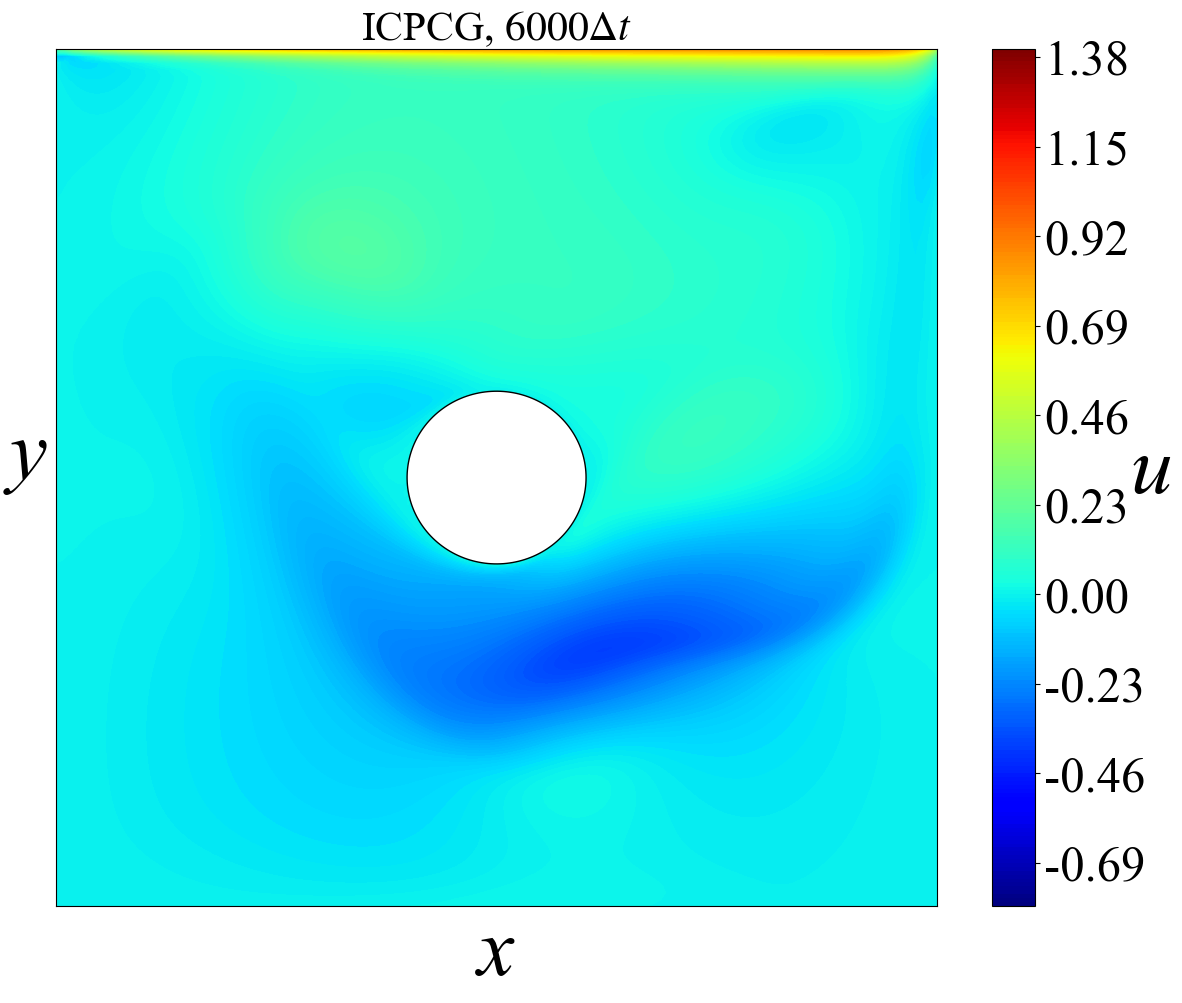}}
  \subfigure[]{
  \label{U_PMHIC_6000step_Case1}
  \includegraphics[scale=0.135]{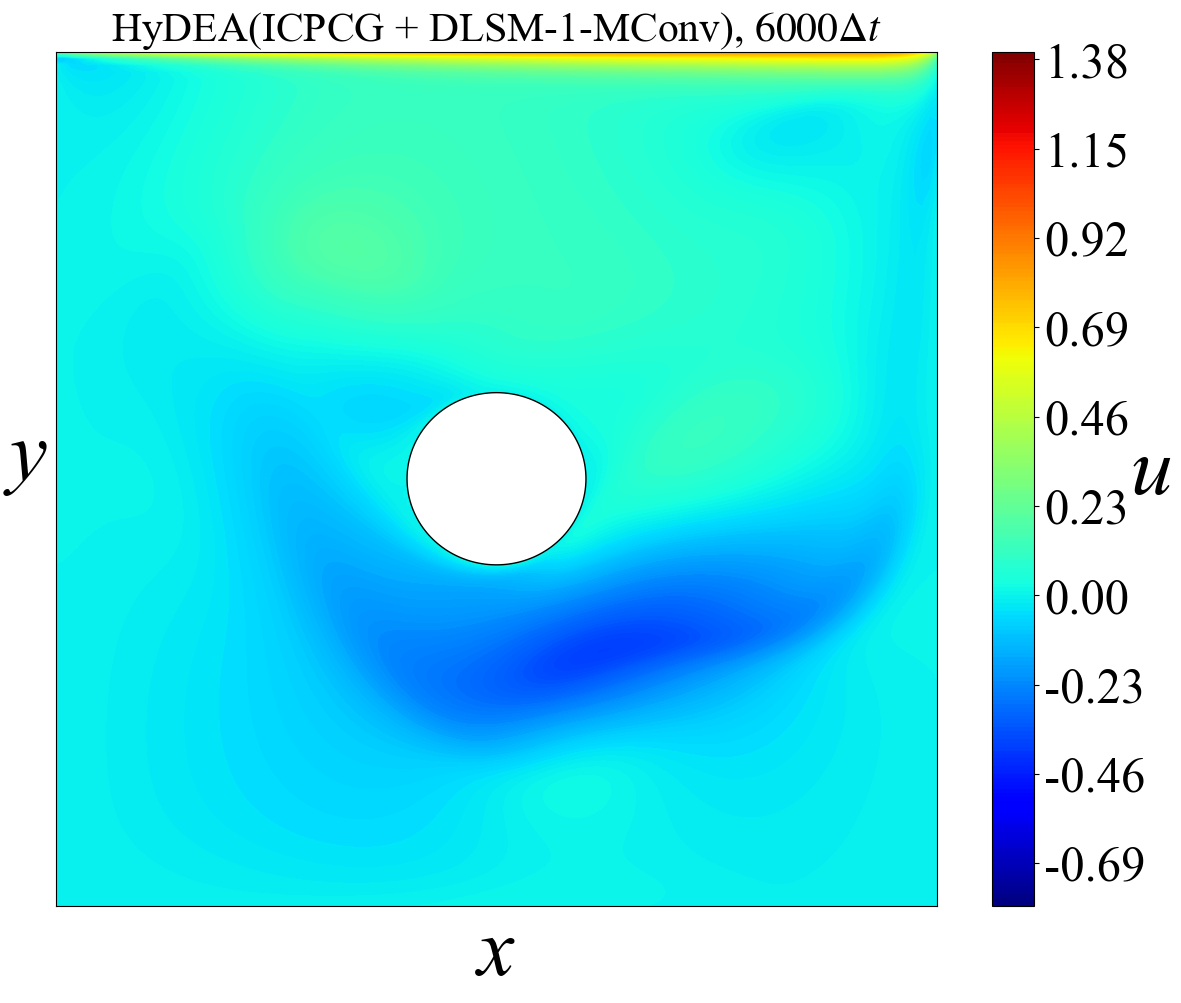}}
  \subfigure[]{
  \label{V_IC_6000step_Case1}
  \includegraphics[scale=0.135]{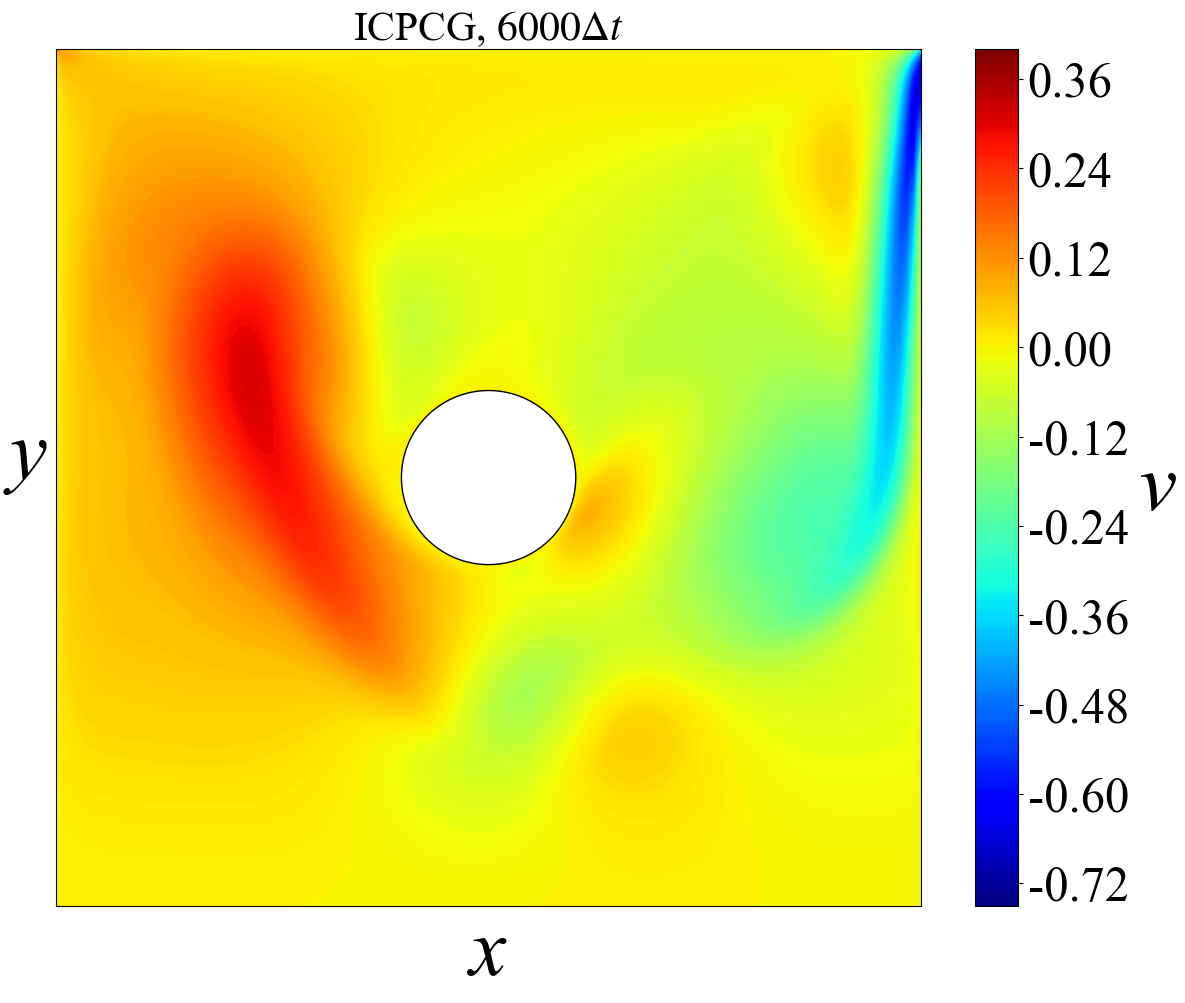}}
  \subfigure[]{
  \label{V_PMHIC_6000step_Case1}
  \includegraphics[scale=0.135]{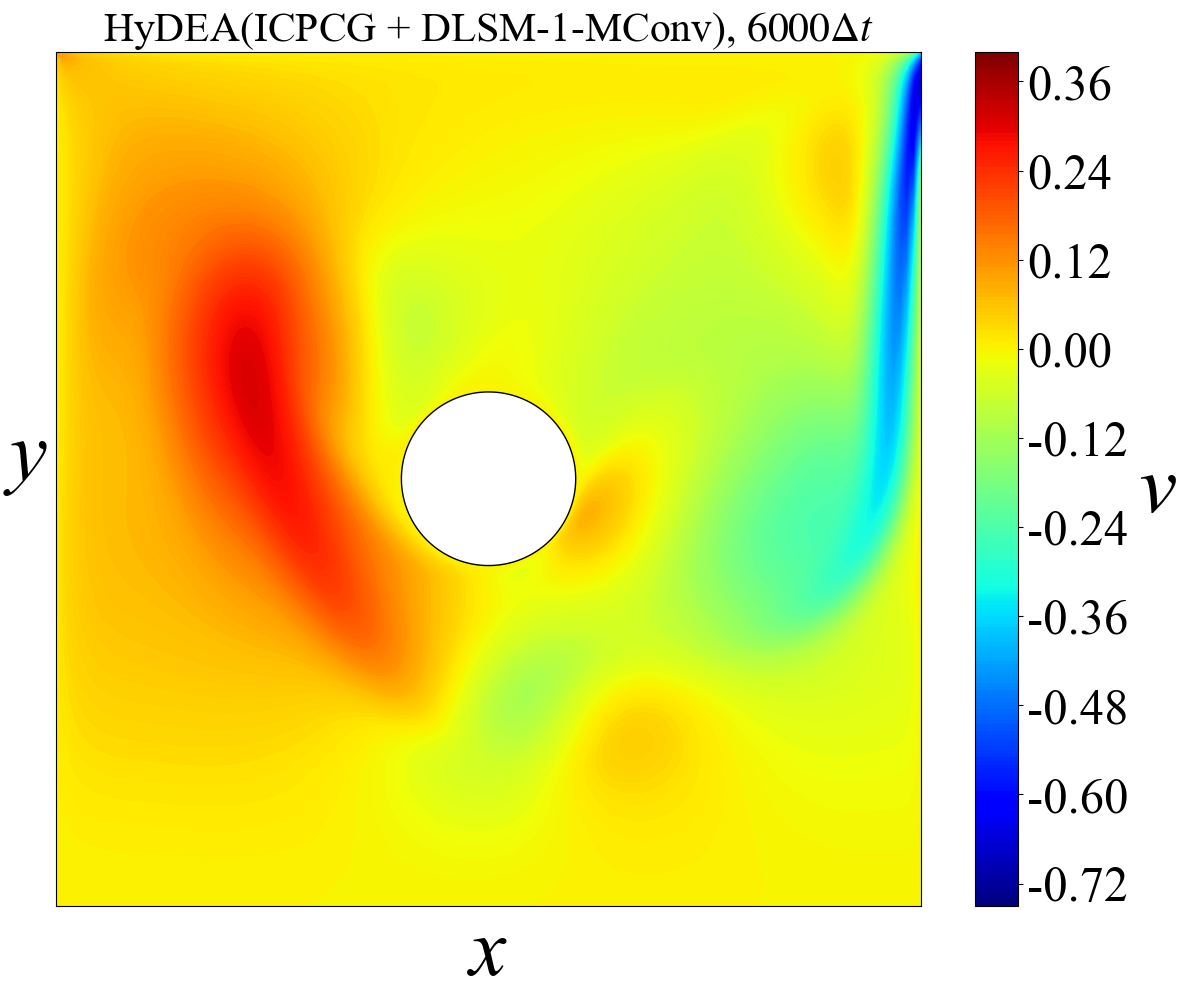}}
 \caption{Velocity fields for 2D lid-driven cavity flow with an embedded stationary circular cylinder by ICPCG and HyDEA~(ICPCG + DLSM-1-MConv). (a)-(d) $u$ and $v$ at the $1500th$ time step. (e)-(h) $u$ and $v$ at the $6000th$ time step.}
 \label{Case1_flowfield}
\end{figure}

Furthermore, taking ICPCG and HyDEA~(ICPCG + DLSM-1-MConv) as representative cases, the velocity fields at time steps $1500th$ and $6000th$ are depicted in Fig.~\ref{Case1_flowfield}. The results demonstrate that the temporal evolution of the flow field is accurately calculated.

\subsection{Case 2: 2D flows past an obstacle}
\label{flow_around_obstacle}

\subsubsection{2D flow past a circular cylinder at $Re=100$}
\label{Re100Cylinder}
\begin{figure}[htbp] 
 \centering  
  \subfigure[]{
  \label{Case2_domain}
  \includegraphics[scale=0.33]{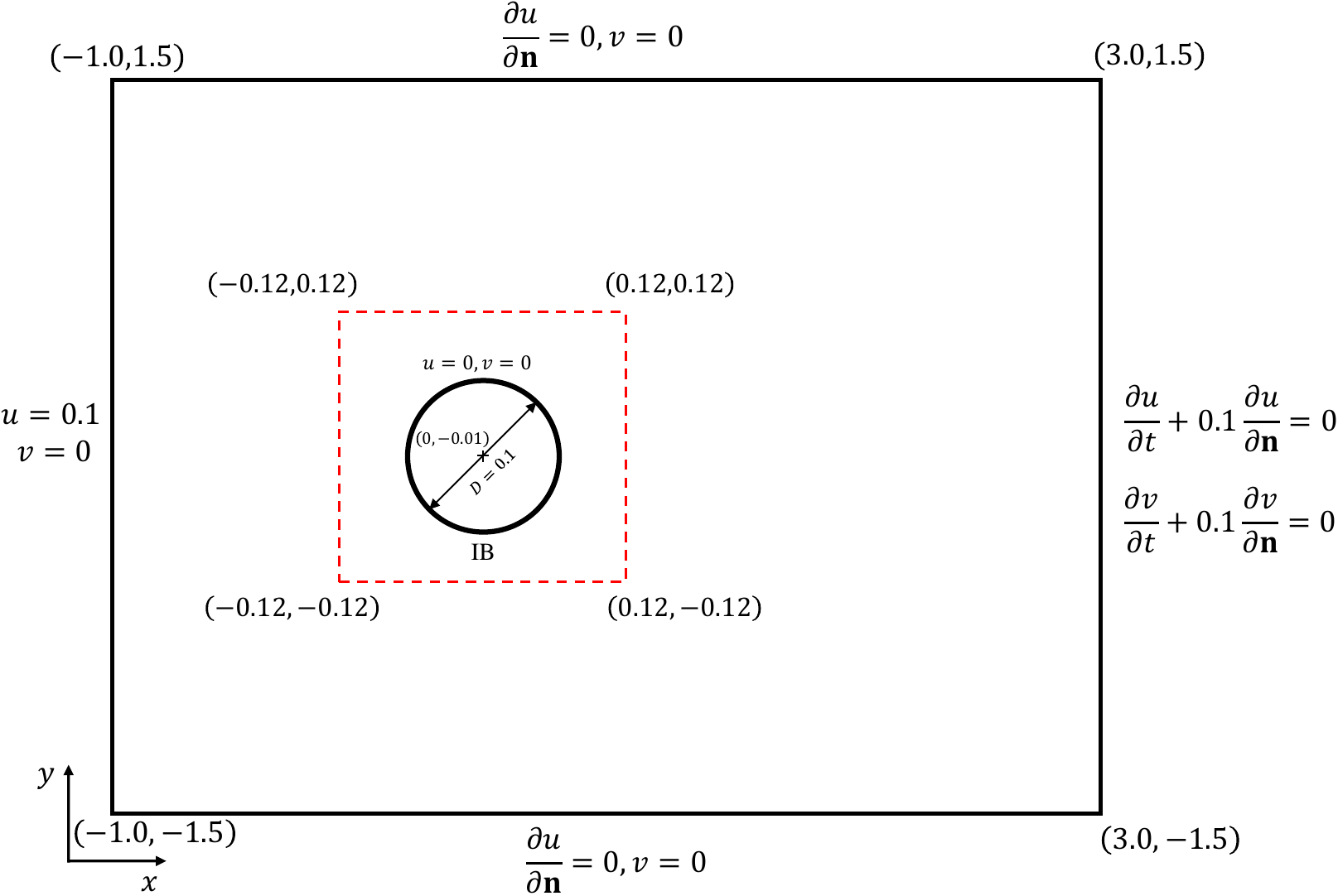}}
  \subfigure[]{
  \label{Case2_grid}
  \includegraphics[scale=0.28]{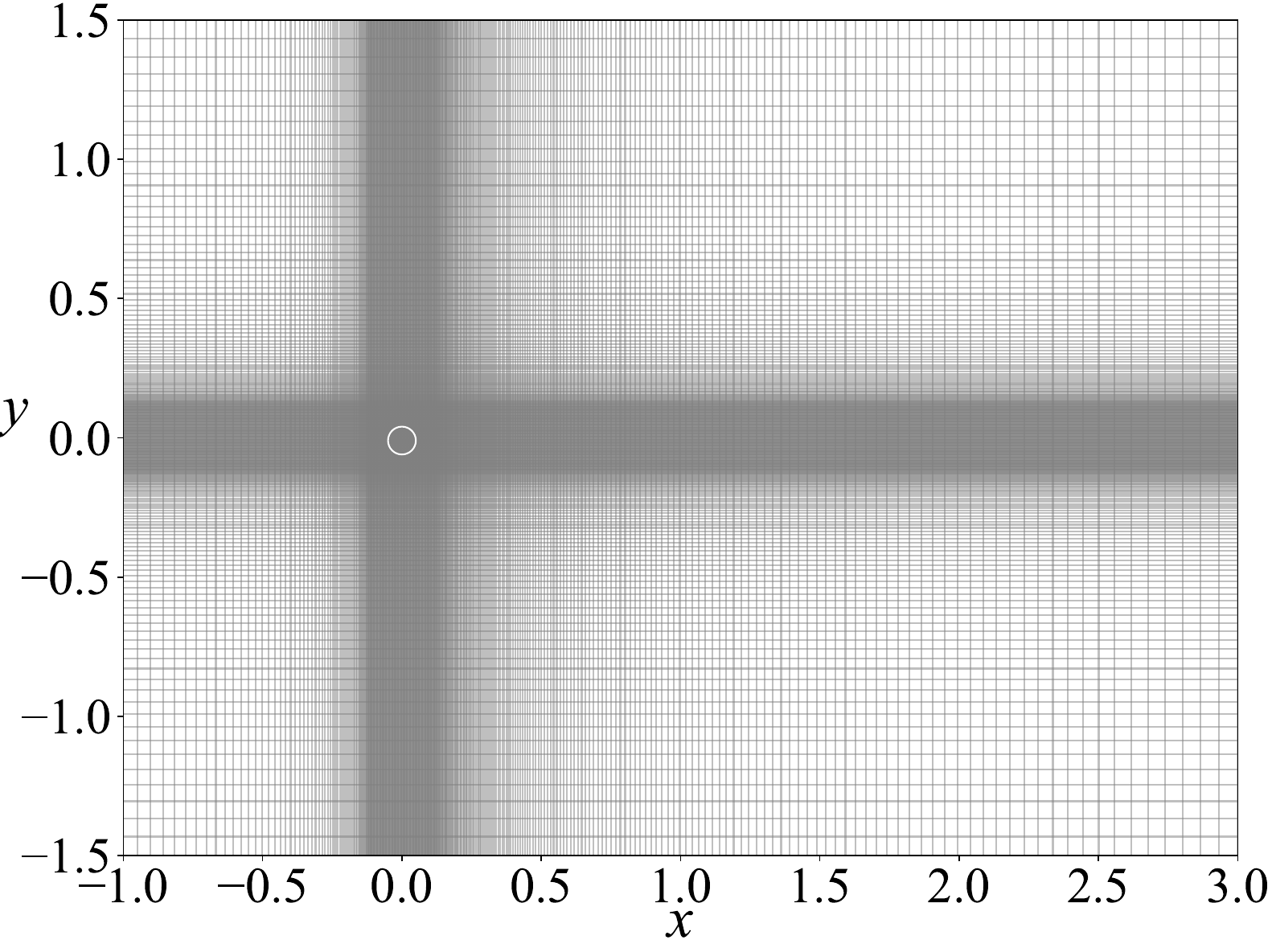}}
 \caption{Schematic diagram of 2D flow past a circular cylinder at $Re=100$. (a) Flow boundary conditions and geometric configuration. (b) Computational grid.}
 \label{Case2Domain_cylinder_Grid}
\end{figure}

The flow boundary conditions, geometric configuration, and computational grid of the flow are illustrated in Fig.~\ref{Case2Domain_cylinder_Grid}. The computational domain is discretized using a non-uniform Cartesian grid, featuring a locally refined uniform grid~($\Delta x=\Delta y=0.0016667$) within a specified square region around the cylinder. Beyond this refined region, the grid undergoes progressive coarsening, resulting in $\Delta_{\max}/\Delta_{\min} \approx 41$. The computational grid consists of $107{,}016$ cells. The inlet velocity is prescribed using a Dirichlet boundary condition, while the upper and lower boundaries are treated with free-slip boundary conditions. The outlet velocity is prescribed using a convective boundary condition. $\Delta t=0.01$, $\nu=0.0001$ and the diameter of cylinder $D=0.1$. We set $Num_{\mathrm{CG-type}}=3$ and $Num_{\mathrm{DLSM}}=2$, and a detailed analysis of these two parameters is provided in~\ref{appendixA}. The iteration terminates when the residual $L2$-norm falls below $\epsilon=10^{-6}$.

\begin{figure}[htbp] 
 \centering  
 \subfigure[]{
  \label{Case2_cylinder_Rline_1steps_ICC}
  \includegraphics[scale=0.21]{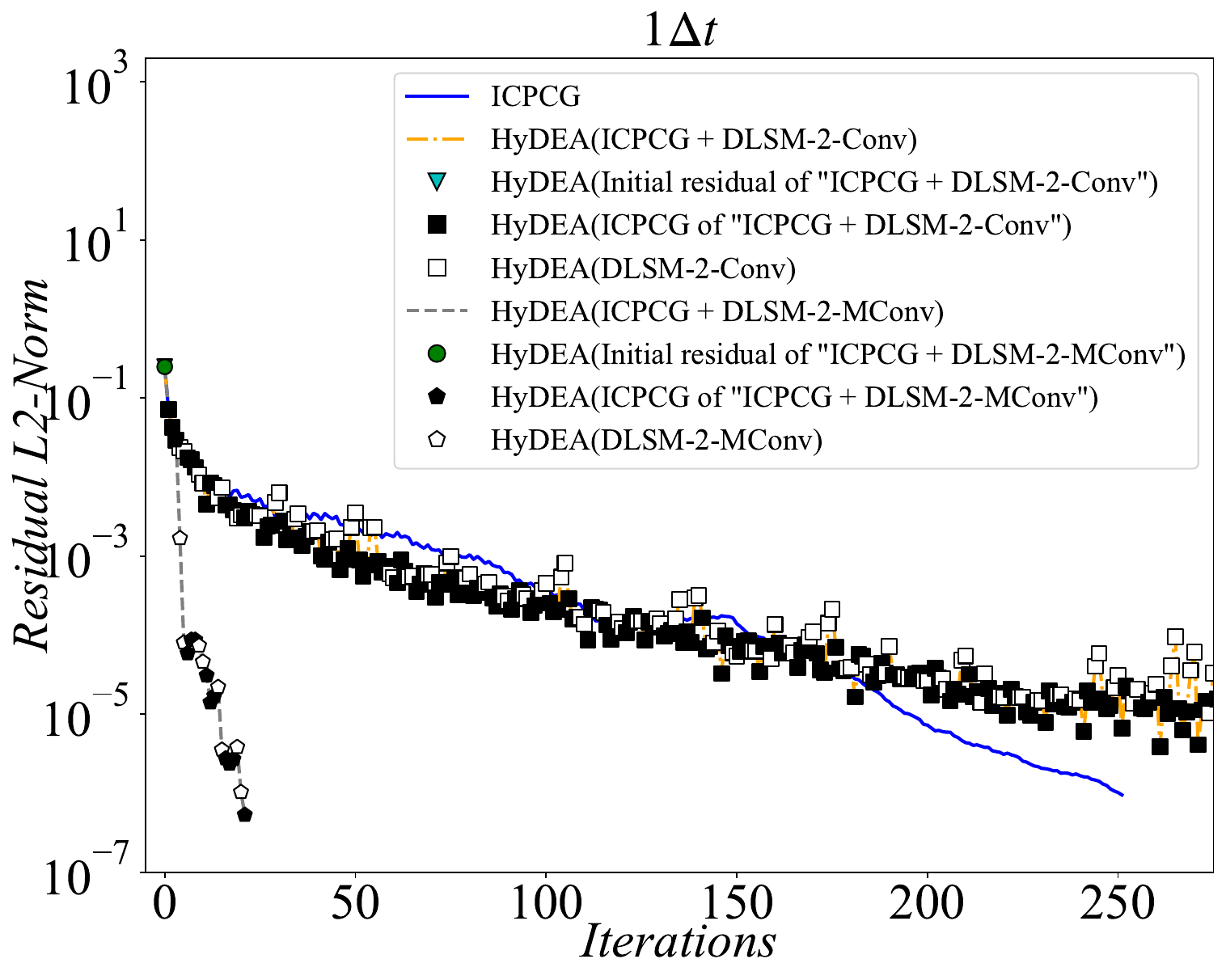}}
  \subfigure[]{
  \label{Case2_cylinder_Rline_100steps_ICC}
  \includegraphics[scale=0.21]{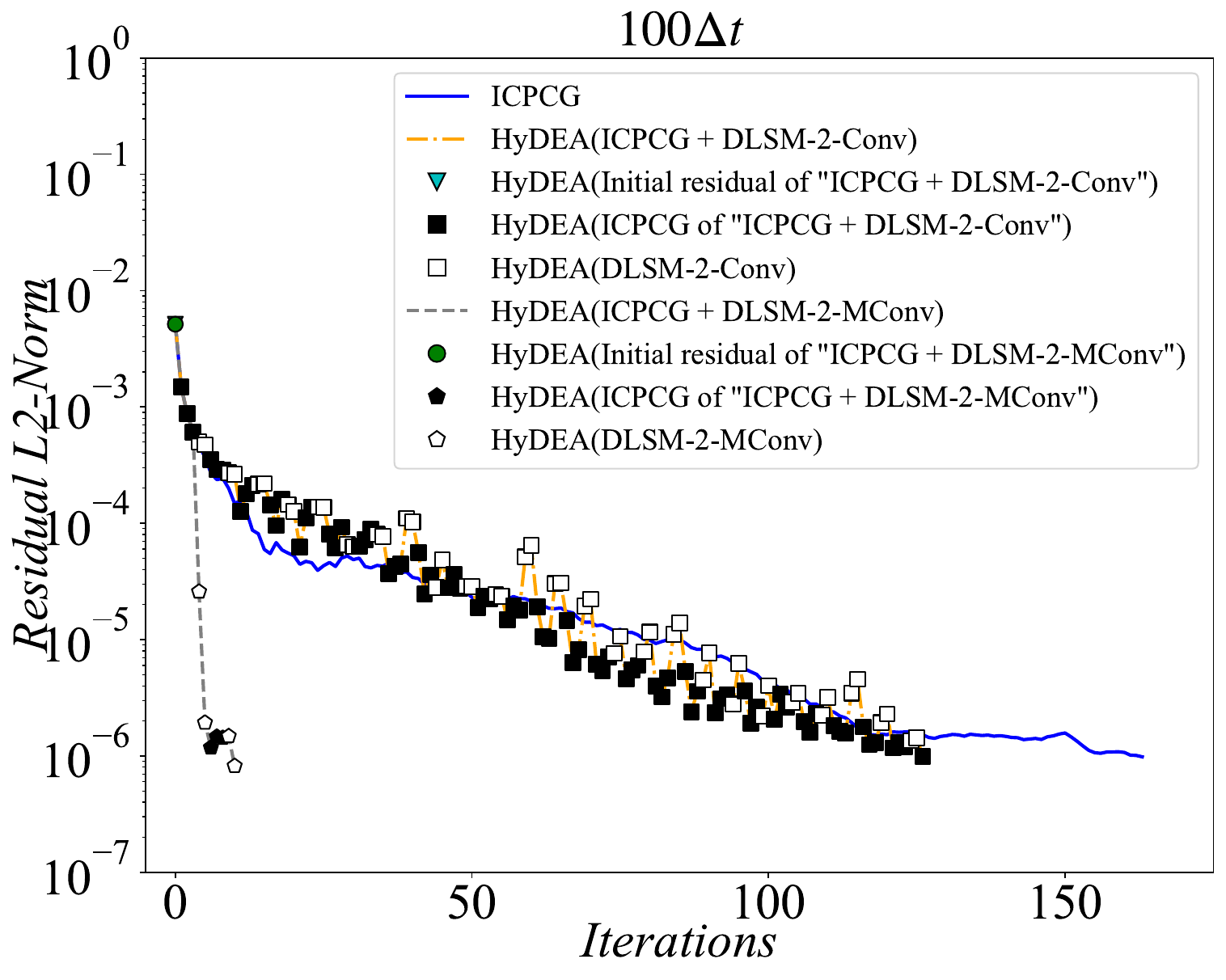}}
  \subfigure[]{
  \label{Case2_cylinder_Rline_10000steps_ICC}
  \includegraphics[scale=0.21]{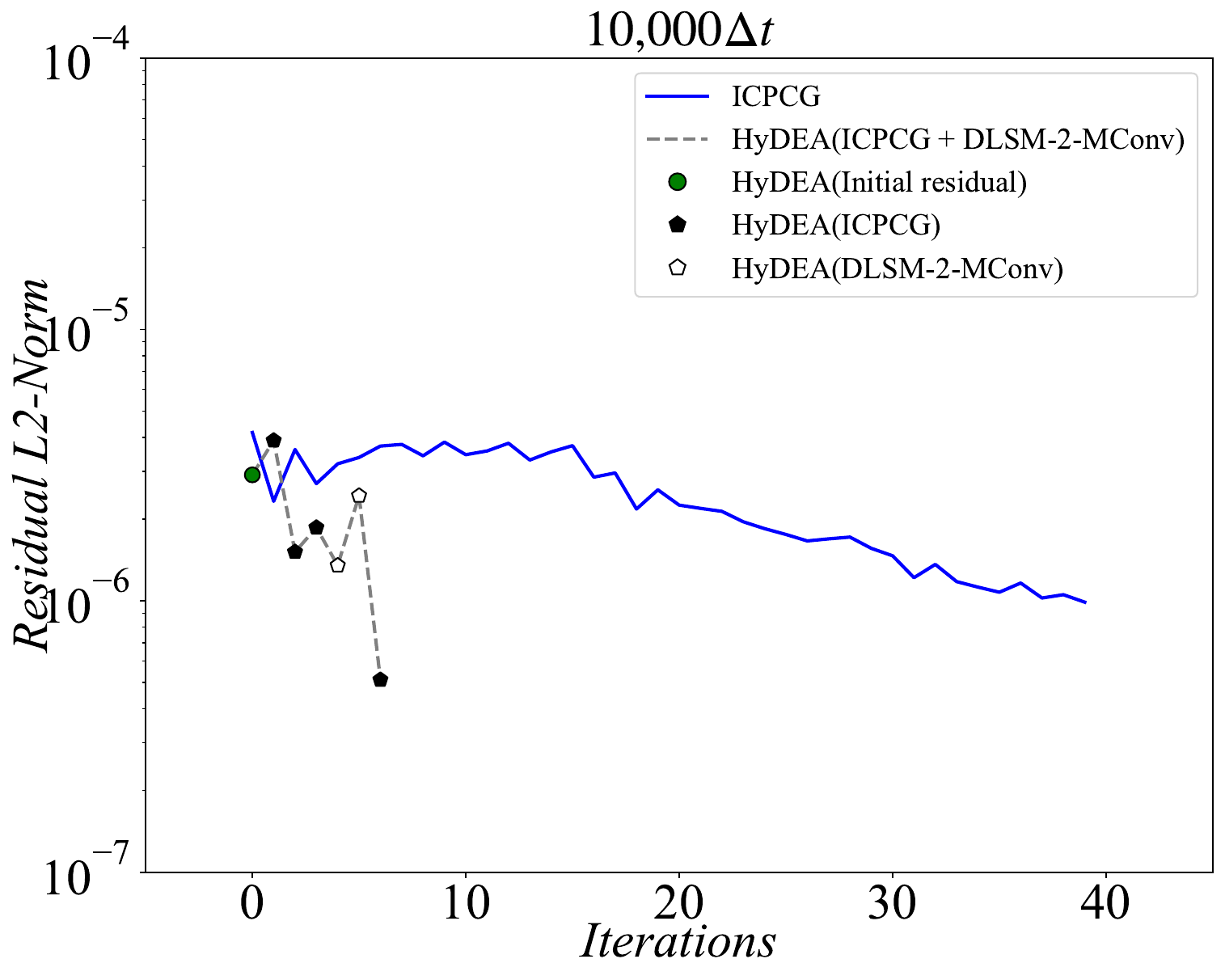}} 
  \subfigure[]{
  \label{Case2_cylinder_Rline_1steps_MG4}
  \includegraphics[scale=0.21]{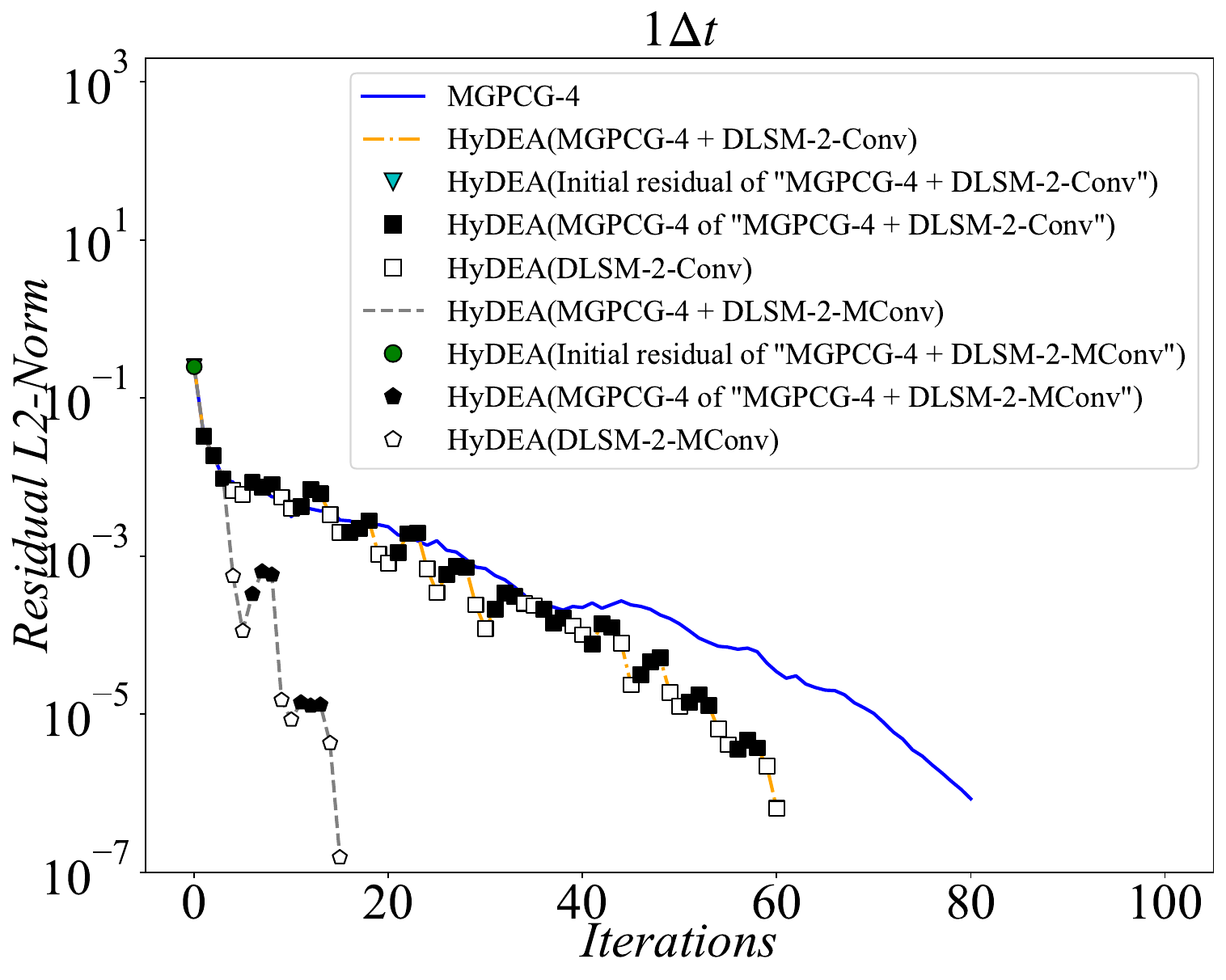}}
  \subfigure[]{
  \label{Case2_cylinder_Rline_100steps_MG4}
  \includegraphics[scale=0.21]{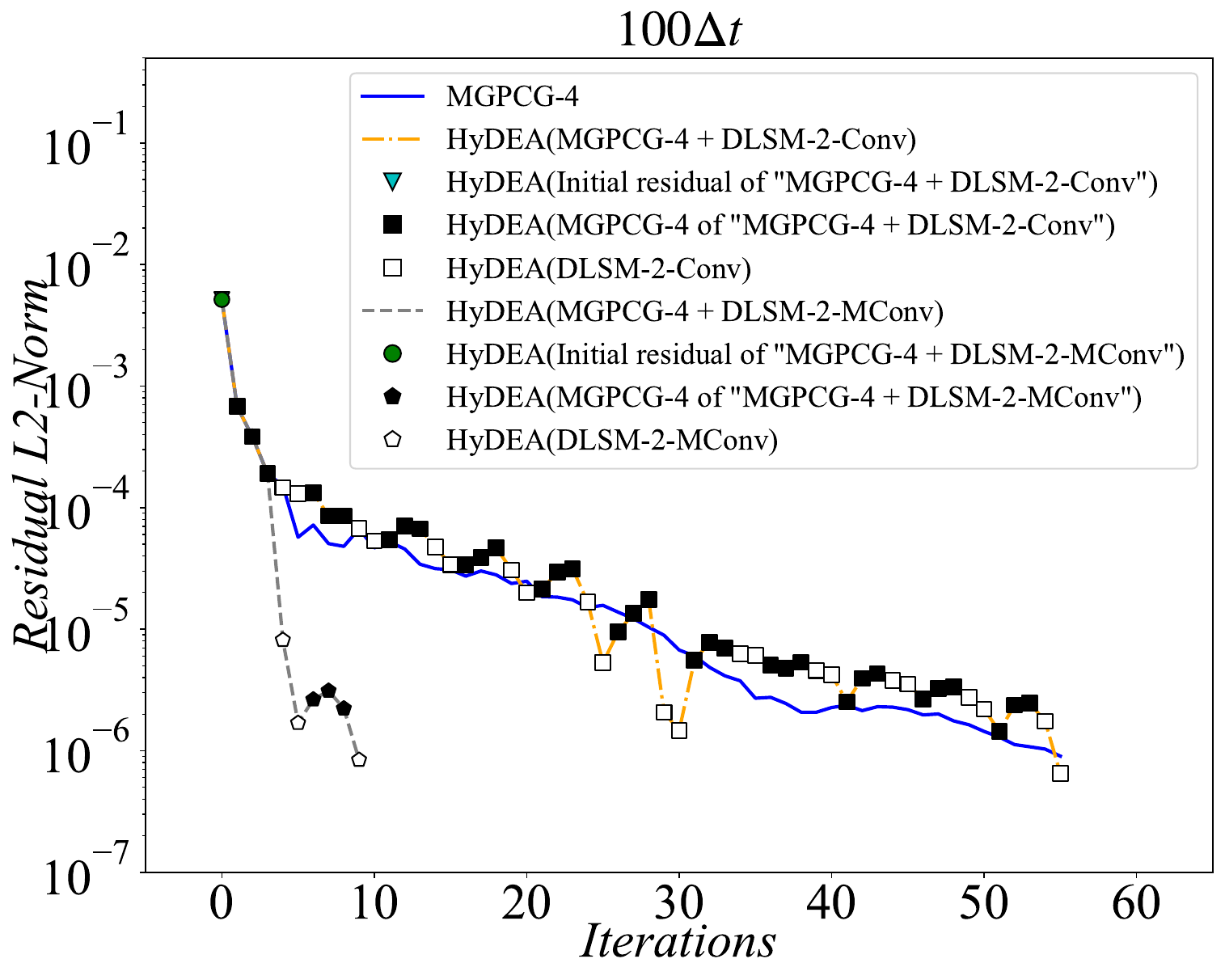}}
  \subfigure[]{
  \label{Case2_cylinder_Rline_10000steps_MG4}
  \includegraphics[scale=0.21]{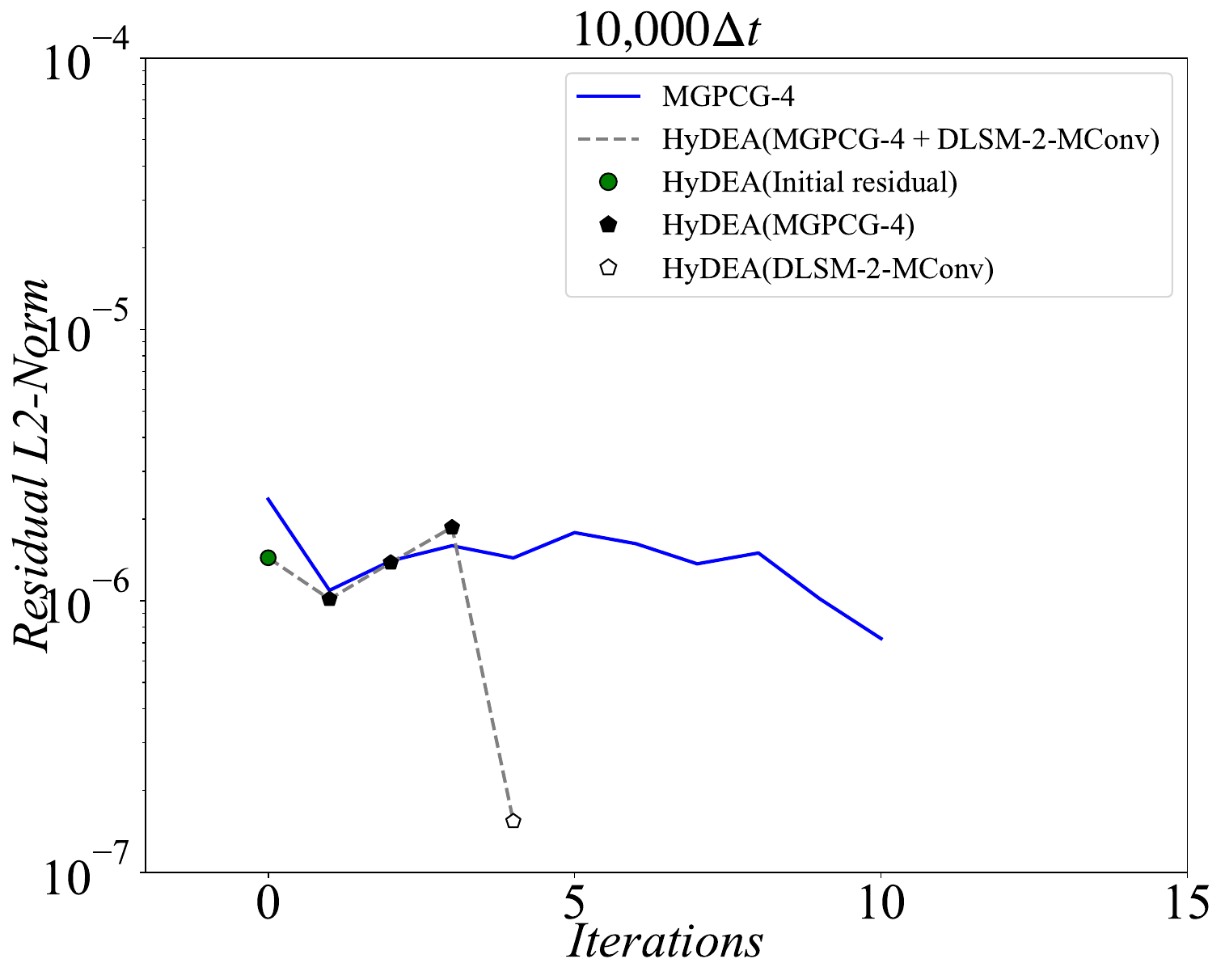}} 
  \caption{Iterative residuals of solving the PPE at $1st$, $100th$ and $10{,}000th$ time steps for 2D flow past a circular cylinder at $Re=100$. (a)-(c) HyDEA~(ICPCG + DLSM-2-Conv/MConv). (d)-(f) HyDEA~(MGPCG-4 + DLSM-2-Conv/MConv).}\label{Case2_cylinder_Rline_3+2}
\end{figure}

To avoid extensive comparisons, we report only the results of HyDEA~(ICPCG + DLSM-2-Conv/MConv) and HyDEA~(MGPCG-4 + DLSM-2-Conv/MConv), together with the corresponding ICPCG and MGPCG-4 methods. The iterative residuals of solving the PPE at the $1st$, $100th$ and $10{,}000th$ time steps are compared and presented in Fig.~\ref{Case2_cylinder_Rline_3+2}. 

As evidenced in Fig.~\ref{Case2_cylinder_Rline_3+2}(a-b, d-e), HyDEA~(CG-type + DLSM-2-Conv) fails to reduce the number of iterations relative to its standalone CG-type counterparts. Conversely, HyDEA~(CG-type + DLSM-2-MConv) exhibits markedly superior performance, achieving faster convergence than both HyDEA~(CG-type + DLSM-2-Conv) and the baseline CG-type methods. Specifically, taking HyDEA~(ICPCG + DLSM-2-MConv) as an example, it takes less than $5$ hybrid rounds (totaling $21$ iterations) at $t=1\Delta t$, $2$ rounds ($10$ iterations) at $t=100\Delta t$, and less than $2$ rounds~($6$ iterations) at $t=10{,}000\Delta t$. These findings underscore the critical necessity of introducing the Mesh-Conv operator in HyDEA for handling non-uniform Cartesian grid discretization conditions.

Furthermore, Table~\ref{acceleration ratio compare FAC} summarizes the computational time and wall-time acceleration ratio for solving the PPE over $20{,}000$ consecutive time steps using HyDEA and its corresponding standalone CG-type counterparts. The results indicate that HyDEA consistently provides effective acceleration compared to the corresponding CG-type methods.

\begin{table}[htbp]
\renewcommand{\arraystretch}{1.5}
\normalsize
\centering
\caption{The computational time and wall-time acceleration ratio for solving the PPE using HyDEA over $20{,}000\Delta t$ for flow past a circular cylinder at $Re=100$.}
\begin{tabular}{cccc}
\hline
  HyDEA  &   Computational time (s)  & Acceleration ratio &  \makecell{Computational time of \\ DLSM-2-MConv (s)} \\
\hline
   (ICPCG + DLSM-2-MConv) & 730 & $\times3.35$ &  447\\
   (MGPCG-4 + DLSM-2-MConv) & 1557 & $\times3.89$ & 239 \\

\hline
\end{tabular}
\label{acceleration ratio compare FAC}
\end{table}

\begin{figure}[htbp] 
 \centering  
  \subfigure[]{
  \label{Vorticity_ICPCG_10000step_cylinder_Case2}
  \includegraphics[scale=0.22]{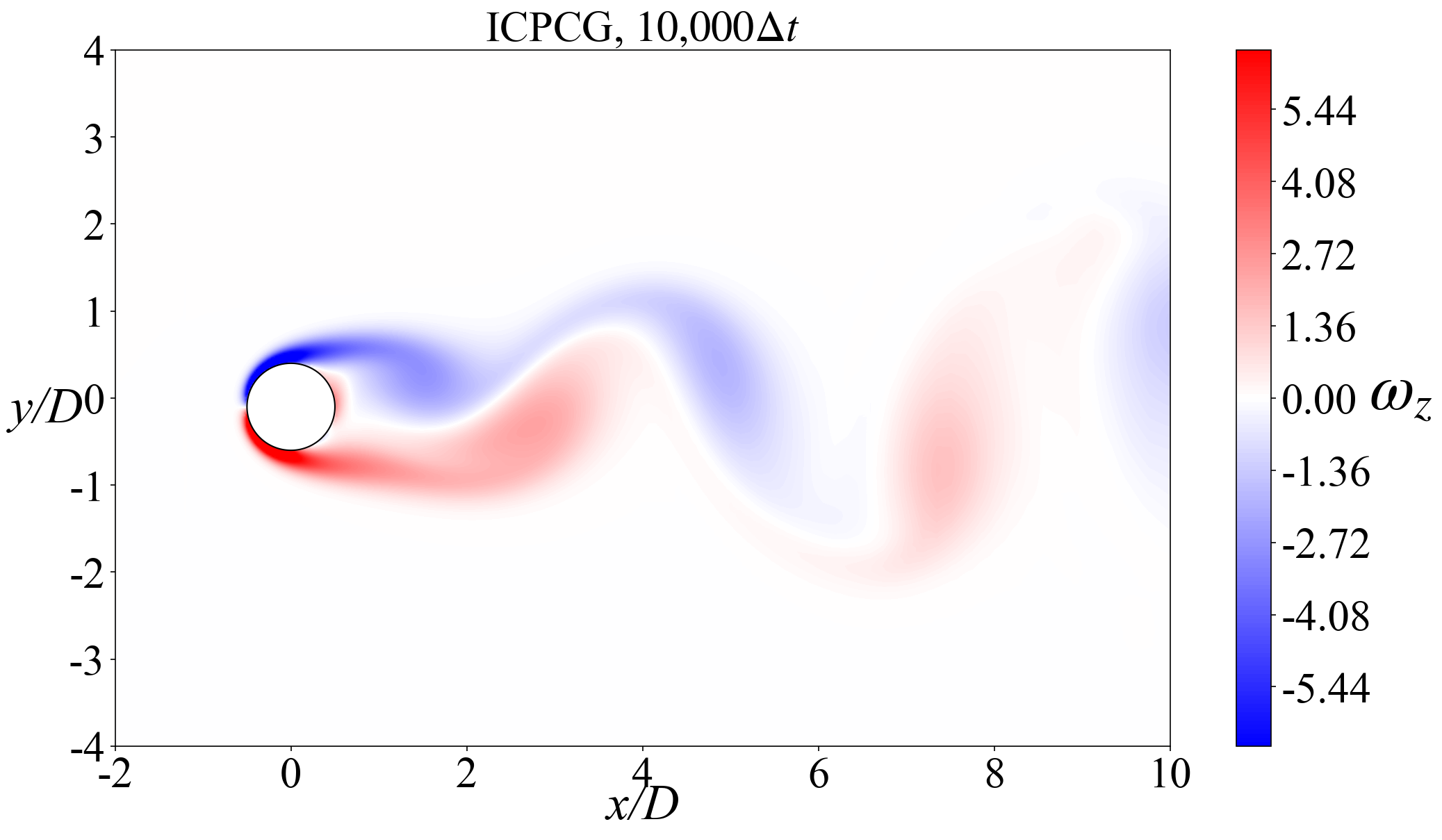}}
  \subfigure[]{
  \label{Vorticity_HyDEA_10000step_cylinder_Case2}
  \includegraphics[scale=0.22]{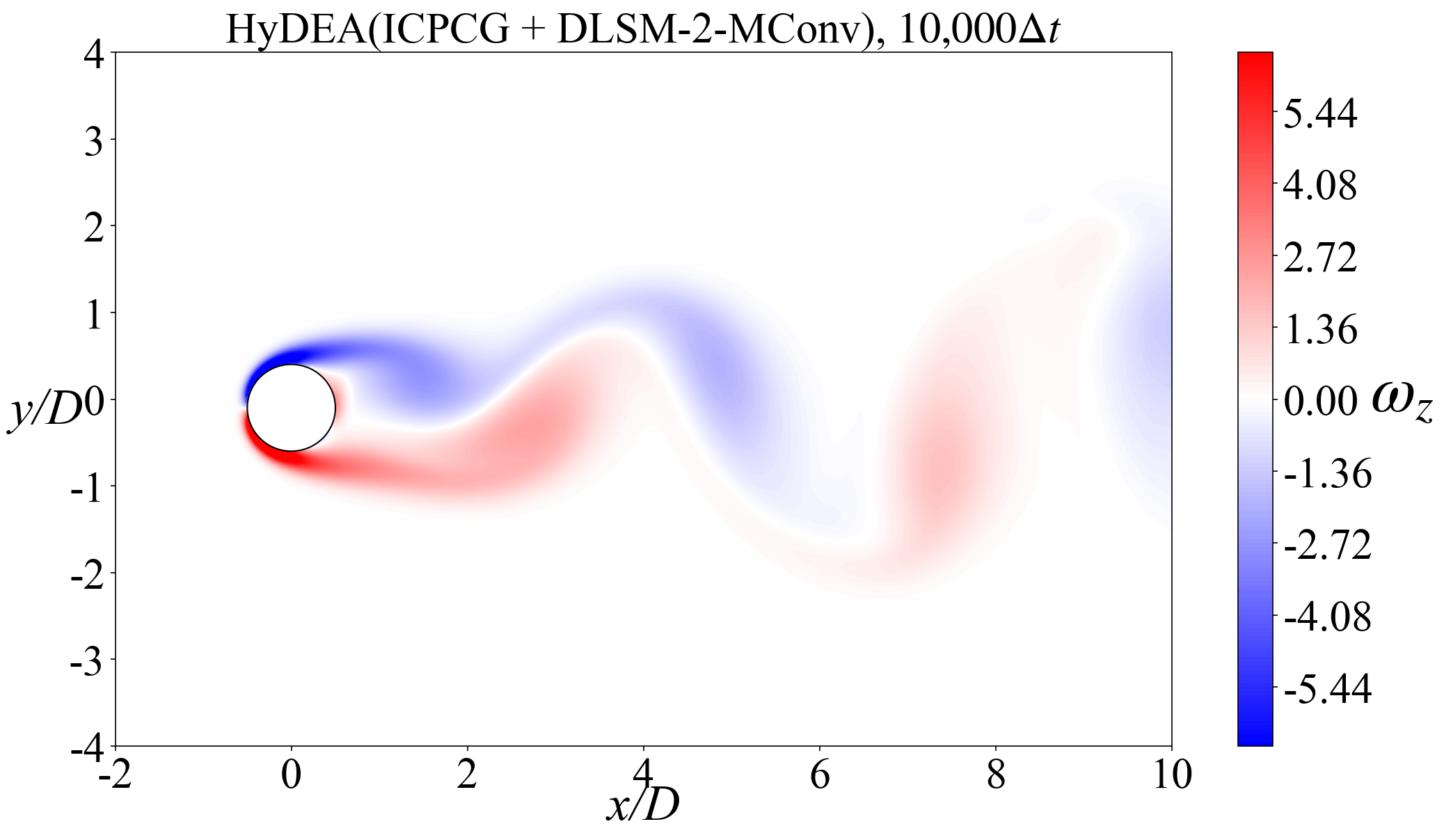}}
  \subfigure[]{
  \label{Vorticity_ICPCG_12000step_cylinder_Case2}
  \includegraphics[scale=0.22]{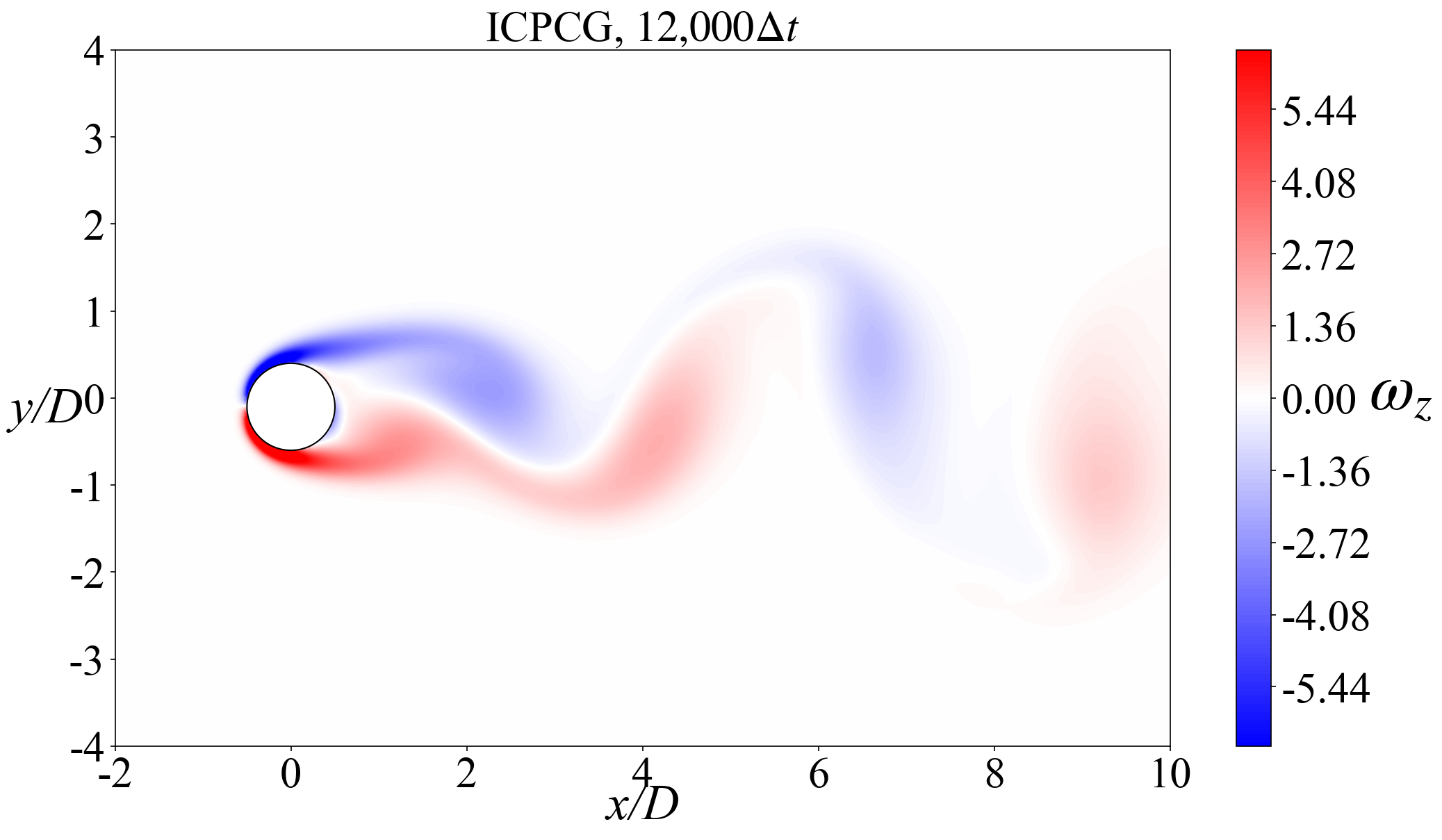}}
  \subfigure[]{
  \label{Vorticity_HyDEA_12000step_cylinder_Case2}
  \includegraphics[scale=0.22]{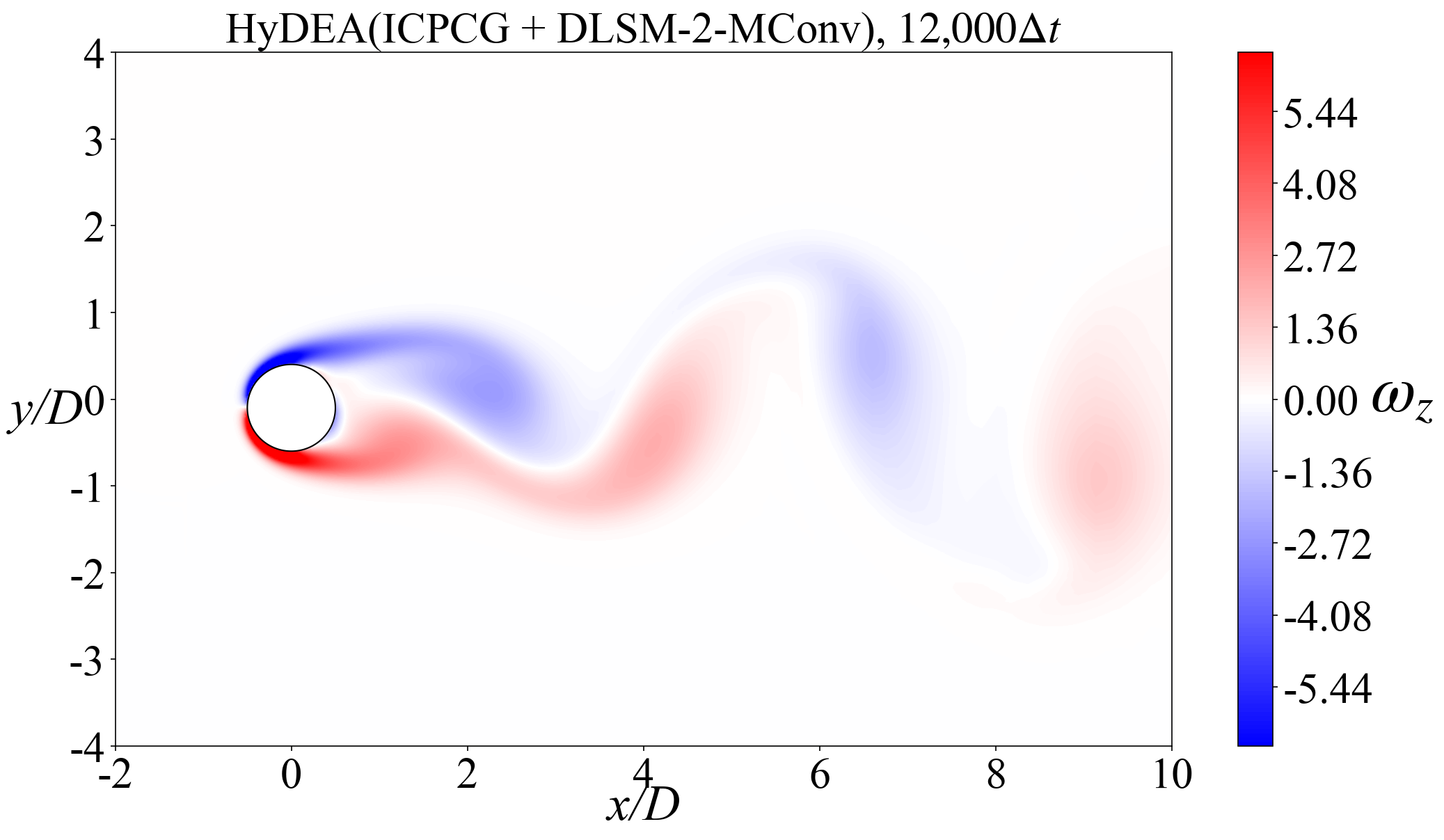}}
  \subfigure[]{
  \label{Vorticity_ICPCG_14000step_cylinder_Case2}
  \includegraphics[scale=0.22]{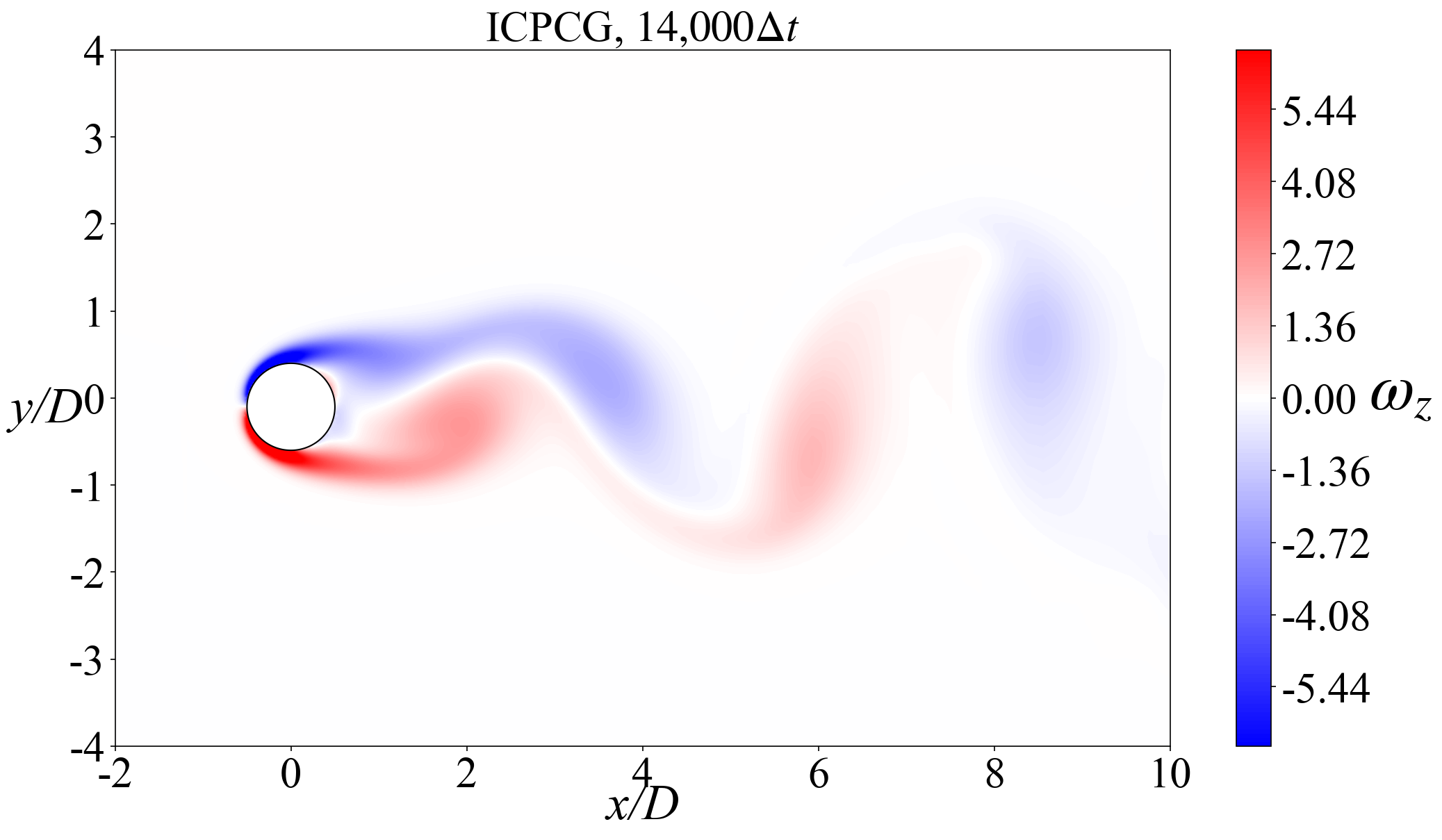}}
  \subfigure[]{
  \label{Vorticity_HyDEA_14000step_cylinder_Case2}
  \includegraphics[scale=0.22]{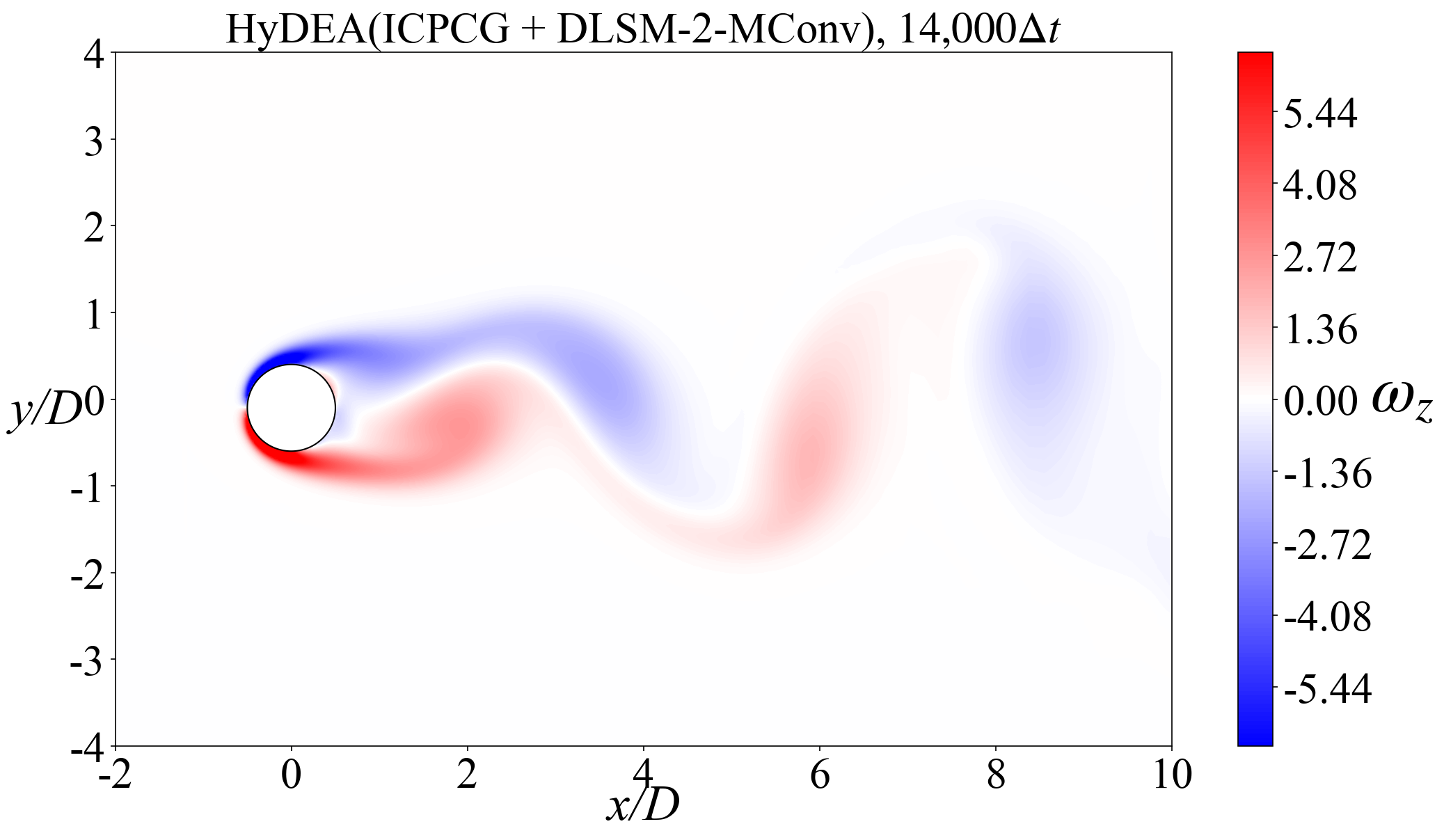}}
 \caption{Vorticity fields for 2D flow past a circular cylinder at $Re=100$ by ICPCG and HyDEA~(ICPCG + DLSM-2-MConv). (a)-(b) $10{,}000th$ time step. (c)-(d) $12{,}000th$ time step. (e)-(f) $14{,}000th$ time step.}
 \label{Case2_flowfield_cylinder}
\end{figure}

Taking the ICPCG and HyDEA~(ICPCG + DLSM-2-MConv) as representative cases, the vorticity fields at the $10{,}000th$, $12{,}000th$ and $14{,}000th$ time steps are presented in Fig.~\ref{Case2_flowfield_cylinder}, which clearly illustrate the accurate temporal evolution of the flow field.

Table~\ref{Global coefficients of FAC} compares the mean drag coefficient~($\overline{C_{D}}$), maximum lift coefficient~($C_{L,\max}$), and Strouhal number~($\mathit{St}=fD/U$) of flow past a circular cylinder at $Re=100$ with those from previous studies~\cite{persillon1998physical_cylinder, posdziech2007systematic_cylinder, zhang2022unsteady_cylinder}, where $f$, $D$, and $U$ represent the vortex shedding frequency, cylinder diameter, and free-stream velocity, respectively. The present results, obtained using HyDEA~(ICPCG + DLSM-2-MConv) to solve the PPE, demonstrate excellent quantitative agreement with the literature.

\begin{table}[htbp]
\renewcommand{\arraystretch}{1.5}
\normalsize
\centering
\caption{Some global coefficients of flow past a circular cylinder at $Re=100$.}
\begin{tabular}{p{3cm}p{3cm}p{3cm}p{3cm}}
\hline
   Data from  &     $\overline{C_{D}}$ & $C_{L,\max}$  & $\mathit{St}$   \\
\hline
   Persillon et al.~\cite{persillon1998physical_cylinder}  & 1.33 & 0.3632 & 0.175  \\

   Posdziech et al.~\cite{posdziech2007systematic_cylinder}  & 1.3504 & 0.3309 & 0.1667  \\

   Zhang et al.~\cite{zhang2022unsteady_cylinder}  & 1.3657 & 0.3413 & 0.1750  \\

   Present work  & 1.3416 & 0.3360 & 0.1631  \\
\hline
\end{tabular}
\label{Global coefficients of FAC}
\end{table}

\subsubsection{2D flow past an elliptical cylinder}
\label{elliptical_cylinder}

\begin{figure}[htbp]
\centering
  \includegraphics[scale=0.35]{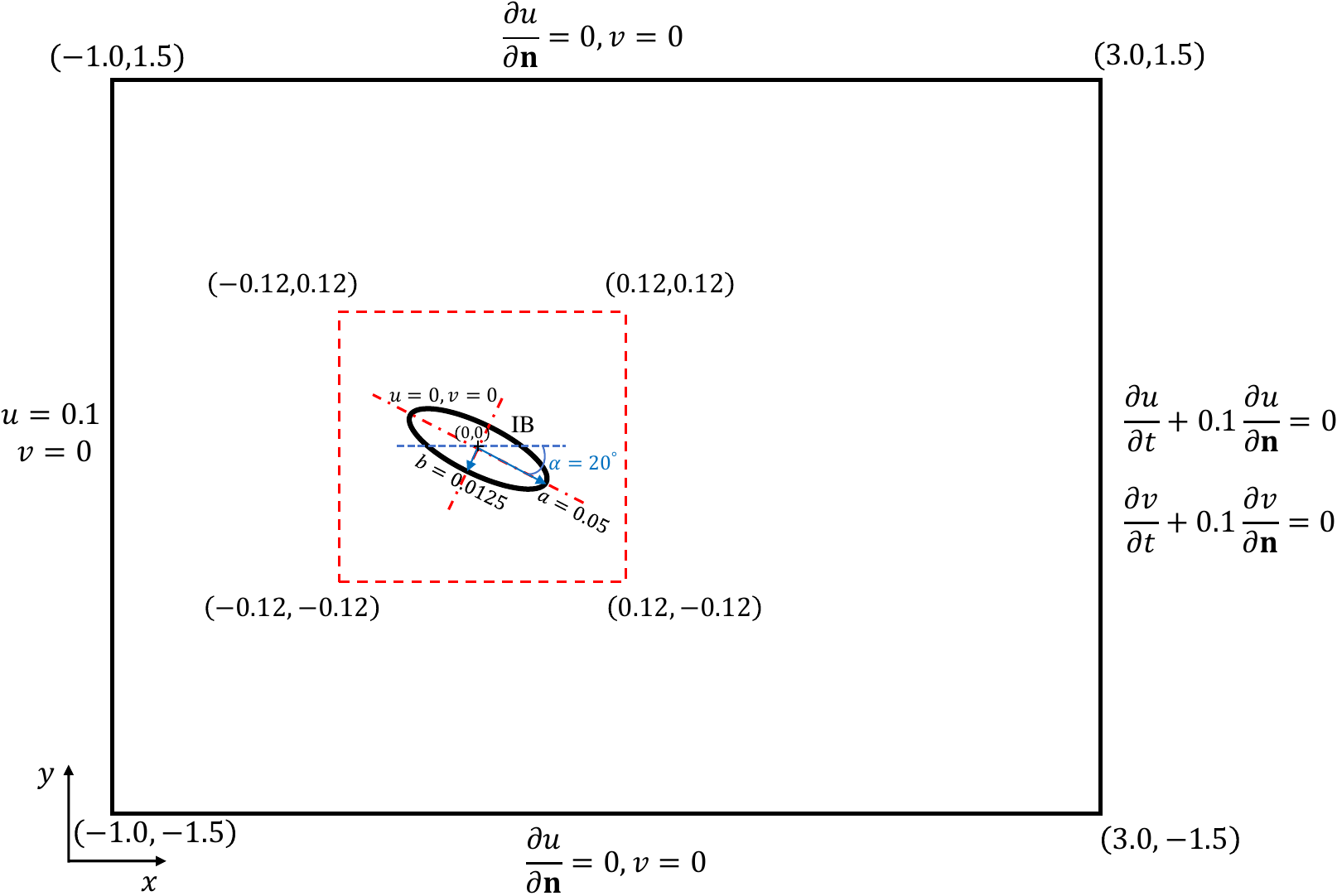}
  \caption{Schematic diagram of 2D flow past an elliptical cylinder.}\label{Case2_ellipse_configurations}
\end{figure}

This section evaluates the generalizability of HyDEA by simulating flow past an elliptical cylinder. The numerical experiment follows the configurations in Section~\ref{Re100Cylinder}, except that the circular cylinder is replaced by an elliptical cylinder as depicted in Fig.~\ref{Case2_ellipse_configurations}. The elliptical cylinder has an aspect ratio of $a/b=4$~(where $a$ and $b$ denote the lengths of semi-major and semi-minor axes, respectively) and is inclined at an angle of attack of $\alpha=20^{\circ}$. \textit{Notably, the network architecture and weights are kept exactly identical to those used in Section~\ref{Re100Cylinder}}. We set $Num_{\mathrm{CG-type}} = 3$ and $Num_{\mathrm{DLSM}} = 2$, and the iterative process is terminated when the residual $L2$-norm falls below $\epsilon=10^{-6}$.

\begin{figure}[htbp] 
 \centering  
  \subfigure[]{
  \label{Case2_ellipse_Rline_10steps_4bi1_20}
  \includegraphics[scale=0.21]{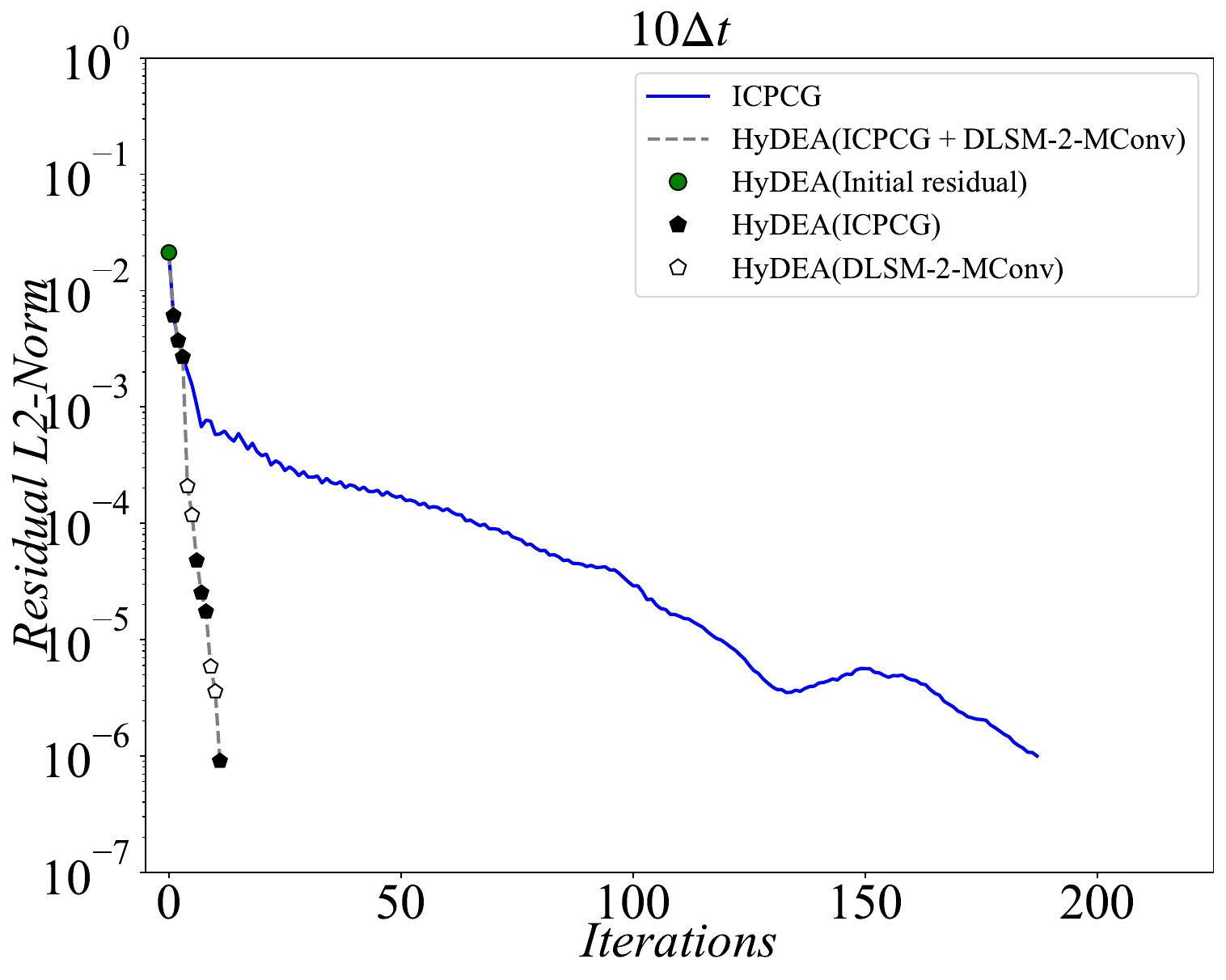}}
  \subfigure[]{
  \label{Case2_ellipse_Rline_100steps_4bi1_20}
  \includegraphics[scale=0.21]{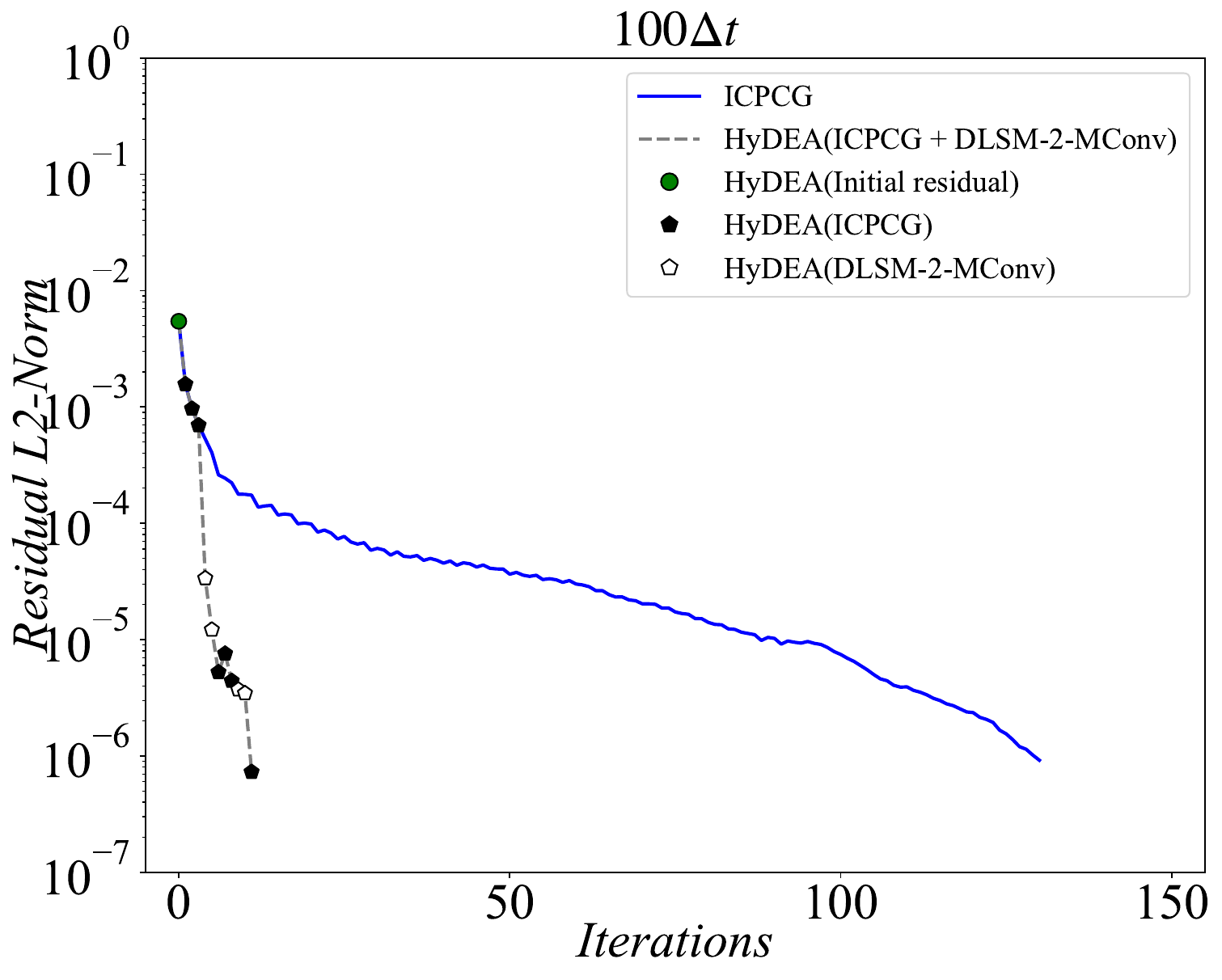}}
  \subfigure[]{
  \label{Case2_ellipse_Rline_10000steps_4bi1_20}
  \includegraphics[scale=0.21]{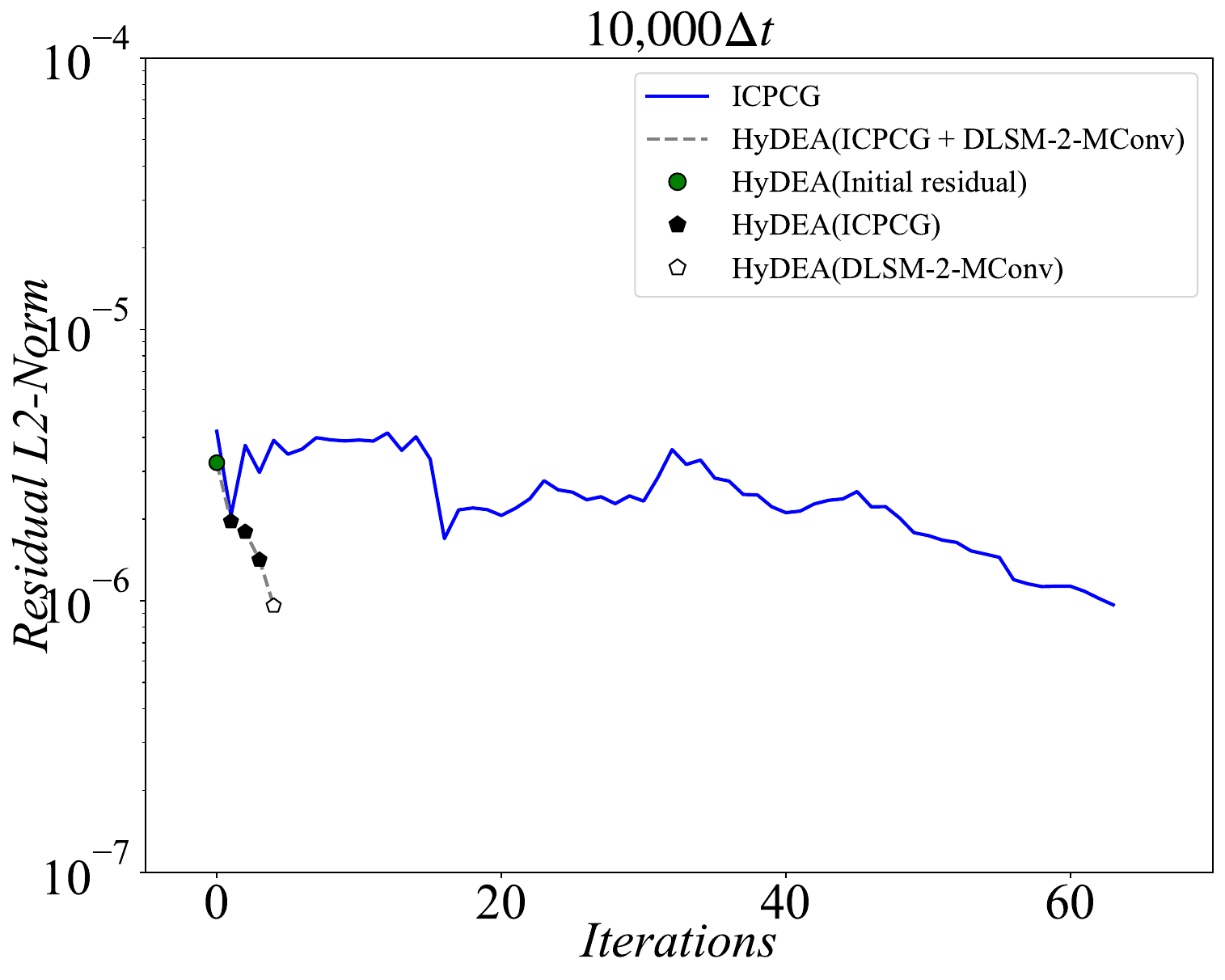}} 
  \caption{Iterative residuals of solving the PPE for 2D flow past an elliptical cylinder. (a) $10th$ time step. (b) $100th$ time step. (c) $10{,}000th$ time step.}\label{Case2_ellipse_Rline_3+2}
\end{figure}

To avoid extensive comparisons, we report only the results of HyDEA~(ICPCG + DLSM-2-MConv), together with the corresponding ICPCG method. Fig.~\ref{Case2_ellipse_Rline_3+2} presents the iterative residuals of solving the PPE at the $10th$, $100th$, and $10{,}000th$ time steps. The results demonstrate that HyDEA~(ICPCG + DLSM-2-MConv) requires significantly fewer iterations to reach the predefined tolerance compared to the standalone ICPCG method. Specifically, HyDEA~(ICPCG + DLSM-2-MConv) takes less than $3$ hybrid rounds (totaling 11 iterations) at both $t=10\Delta t$ and $t=100\Delta t$, and requires less than $1$ round (4 iterations) at $t=10{,}000\Delta t$. This robust convergence behavior further confirms the generalizability of HyDEA for non-uniform Cartesian grid configurations. 

\begin{figure}[htbp]
\centering
  \includegraphics[scale=0.2]{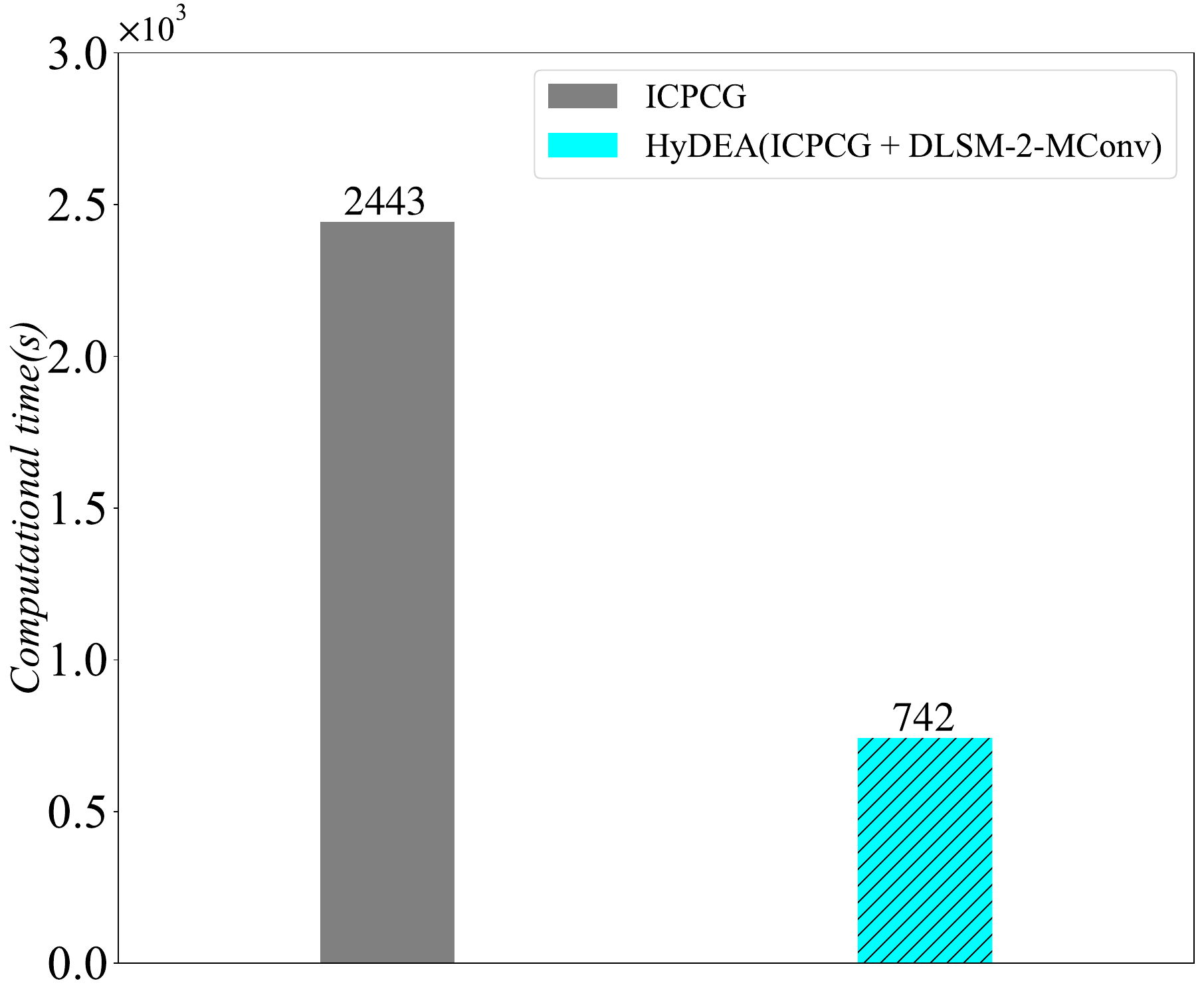}
  \caption{Computational time of the PPE solution over $20{,}000\Delta t$ for 2D flow past an elliptical cylinder.}\label{Case2_ellipse_timecomapre}
\end{figure}

Fig.~\ref{Case2_ellipse_timecomapre} compares the computational time required for the PPE solution over $20{,}000$ consecutive time steps between HyDEA~(ICPCG + DLSM-2-MConv) and the standalone ICPCG method. As clearly observed, HyDEA delivers a substantial reduction in computational time, effectively accelerating the overall solving procedure.

\begin{figure}[htbp] 
 \centering  
  \subfigure[]{
  \label{Vorticity_ICPCG_10000step_ellipse_4bi1_20_Case2}
  \includegraphics[scale=0.22]{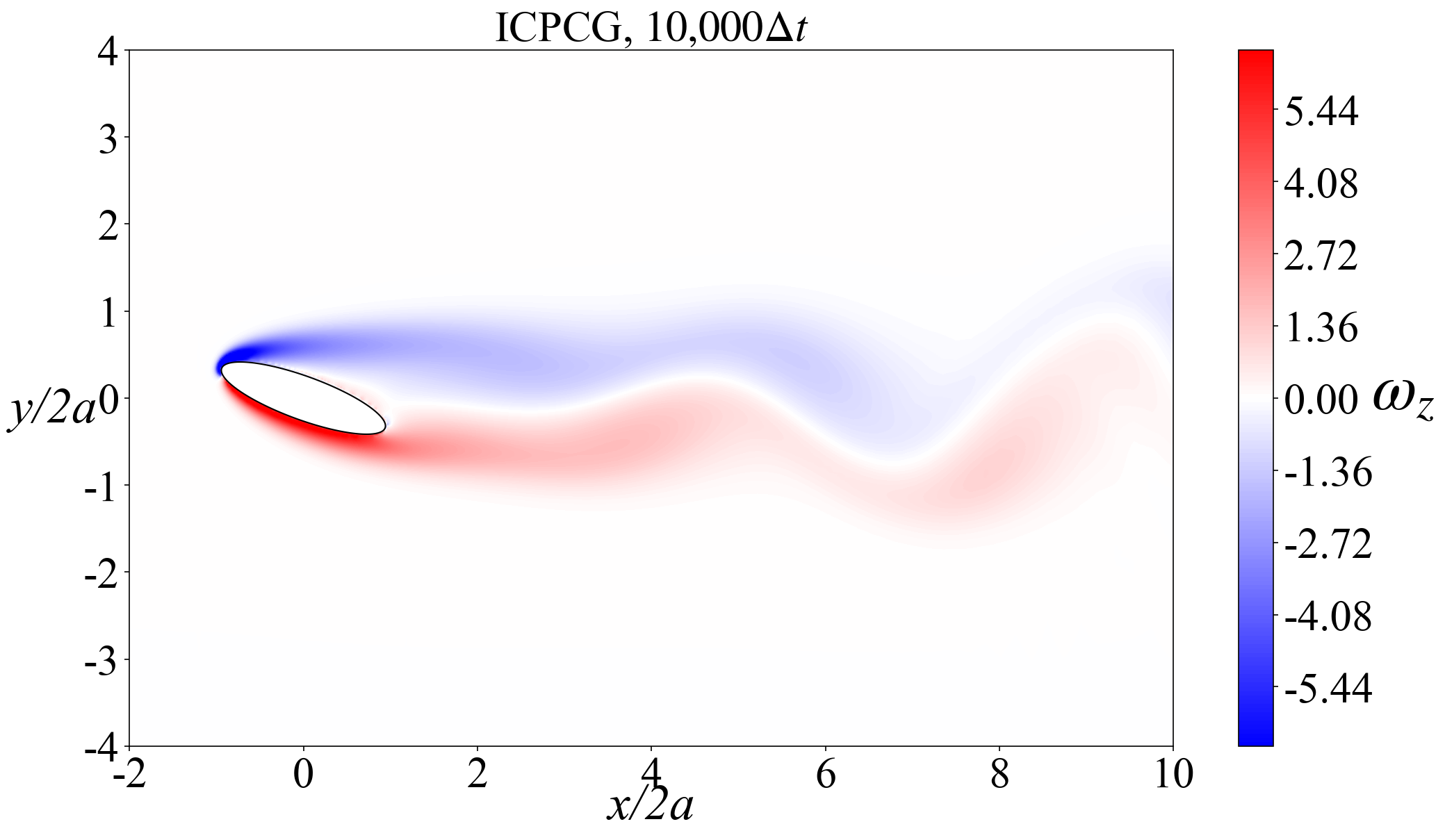}}
  \subfigure[]{
  \label{Vorticity_HyDEA_10000step_ellipse_4bi1_20_Case2}
  \includegraphics[scale=0.22]{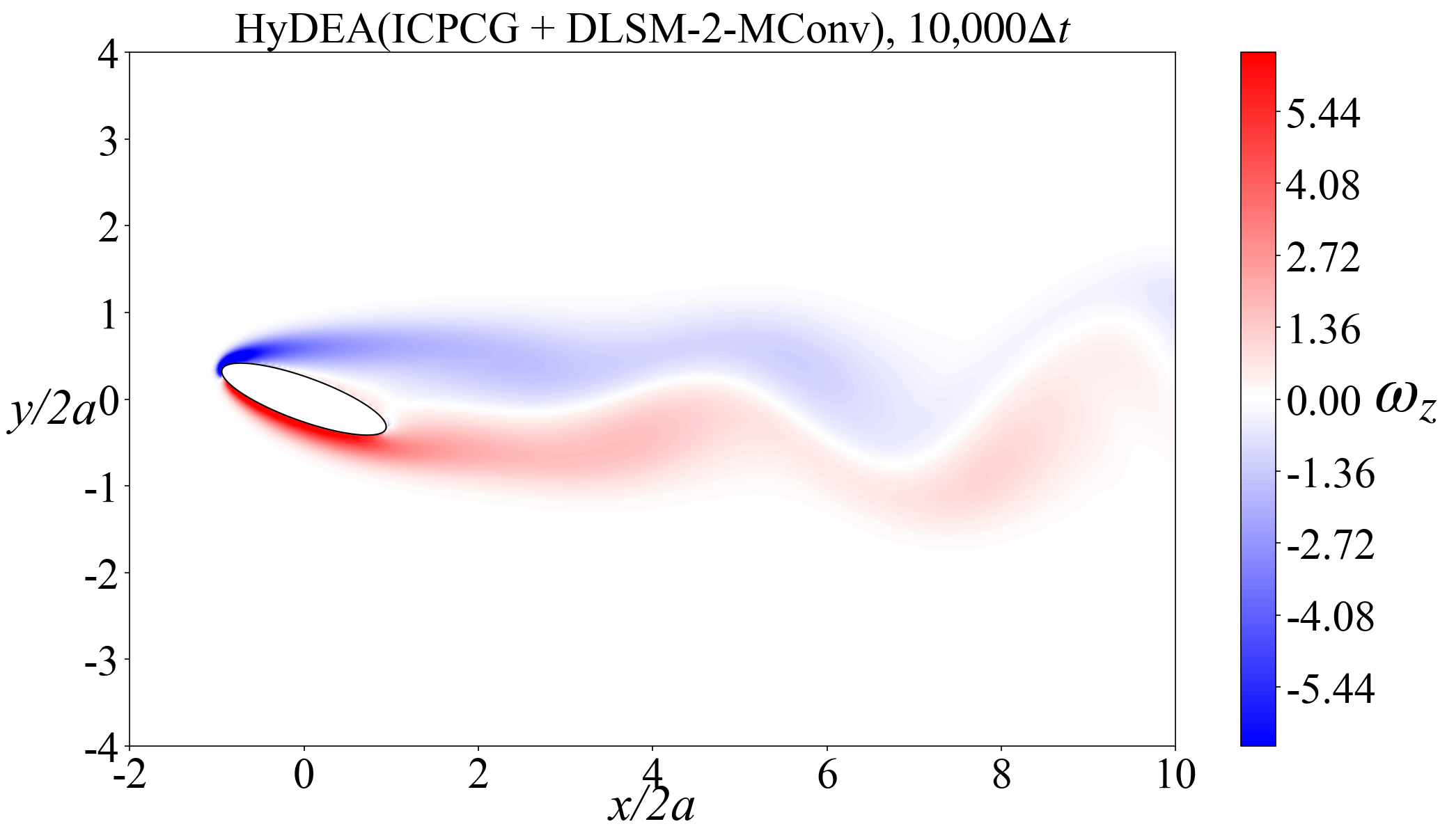}}
  \subfigure[]{
  \label{Vorticity_ICPCG_12000step_ellipse_4bi1_20_Case2}
  \includegraphics[scale=0.22]{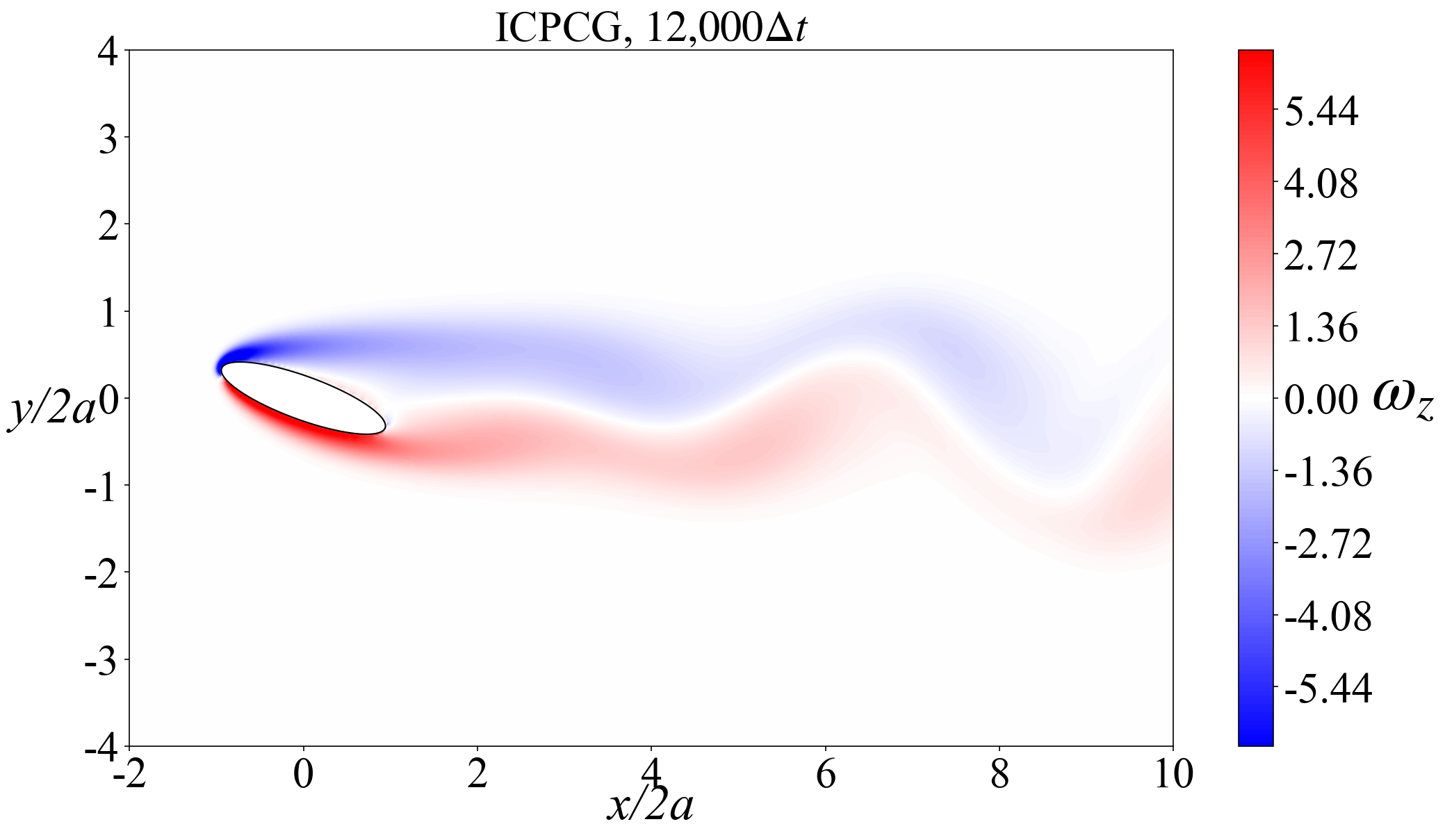}}
  \subfigure[]{
  \label{Vorticity_HyDEA_12000step_ellipse_4bi1_20_Case2}
  \includegraphics[scale=0.22]{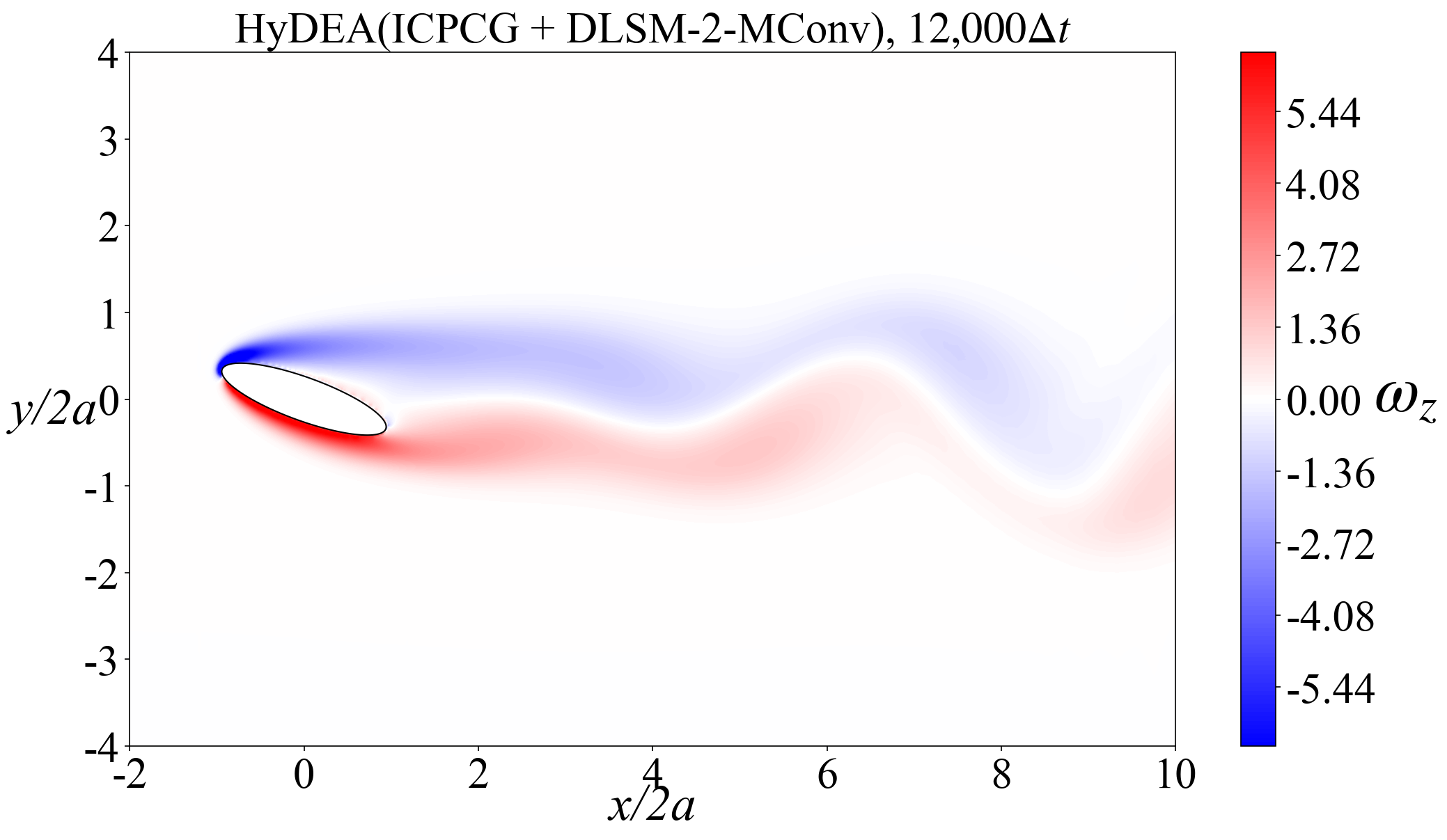}}
  \subfigure[]{
  \label{Vorticity_ICPCG_14000step_ellipse_4bi1_20_Case2}
  \includegraphics[scale=0.22]{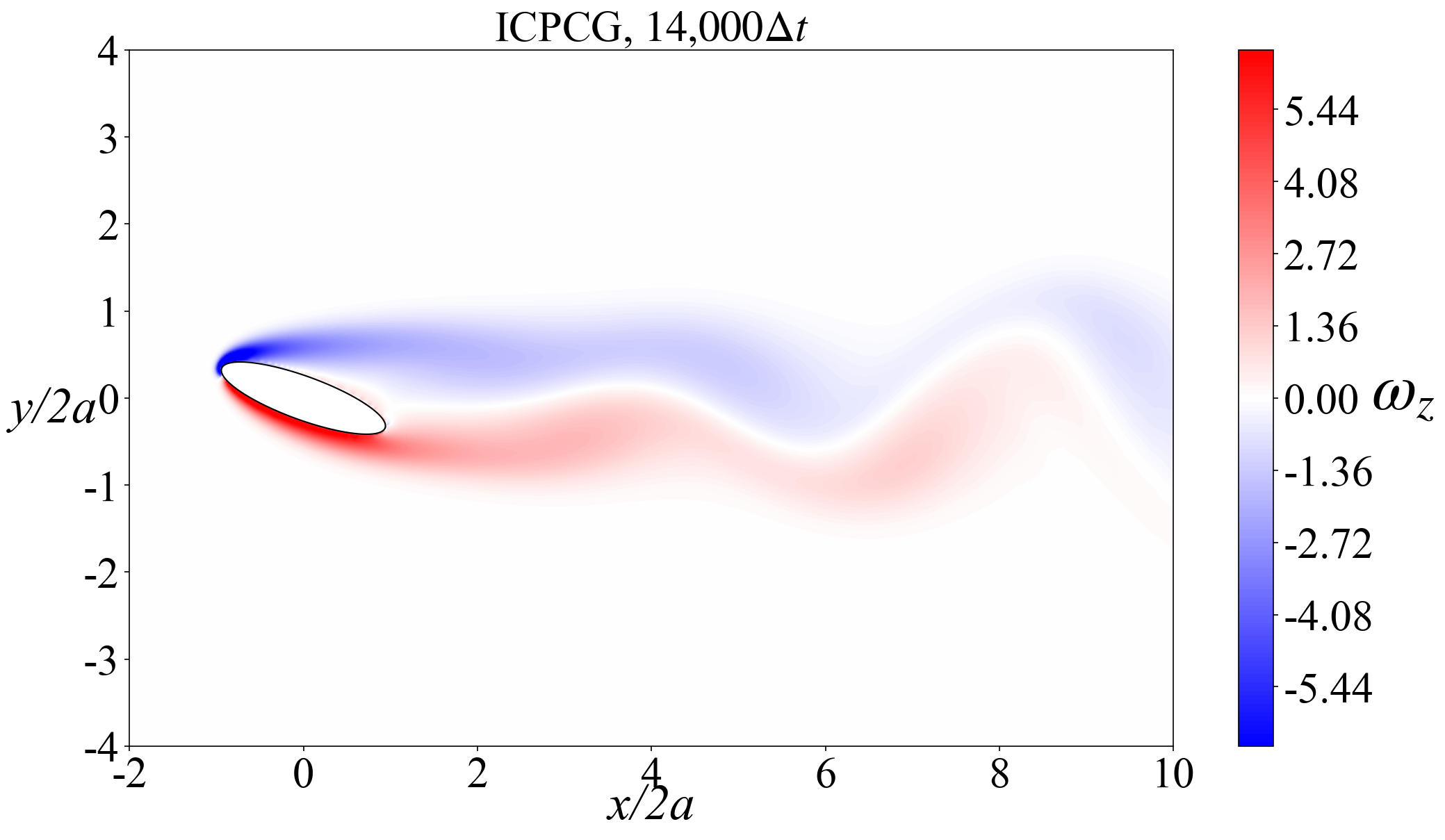}}
  \subfigure[]{
  \label{Vorticity_HyDEA_14000step_ellipse_4bi1_20_Case2}
  \includegraphics[scale=0.22]{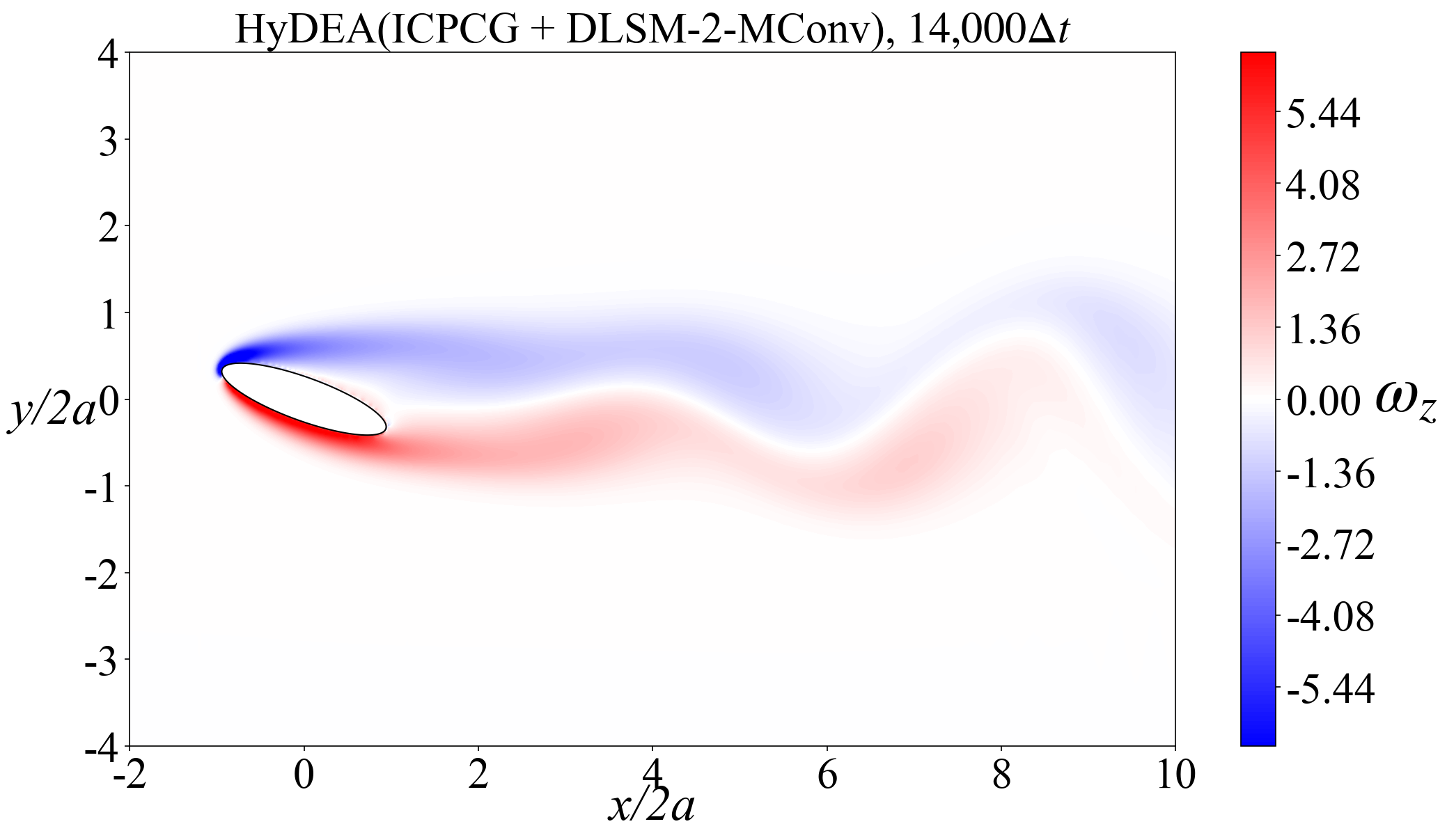}}
 \caption{Vorticity fields for 2D flow past an elliptical cylinder by ICPCG and HyDEA~(ICPCG + DLSM-2-MConv). (a) $10{,}000th$ time step. (b) $12{,}000th$ time step. (c) $14{,}000th$ time step.}
 \label{Case2_flowfield_ellipse}
\end{figure}

Furthermore, the vorticity fields at the $10{,}000th$, $12{,}000th$ and $14{,}000th$ time steps are presented in Fig.~\ref{Case2_flowfield_ellipse}, which clearly demonstrate that the temporal evolution of the flow field is accurately calculated.

\subsubsection{2D flow past the DARPA SUBOFF profile}
\label{flow past suboff}

\begin{figure}[htbp]
\centering
  \includegraphics[scale=0.35]{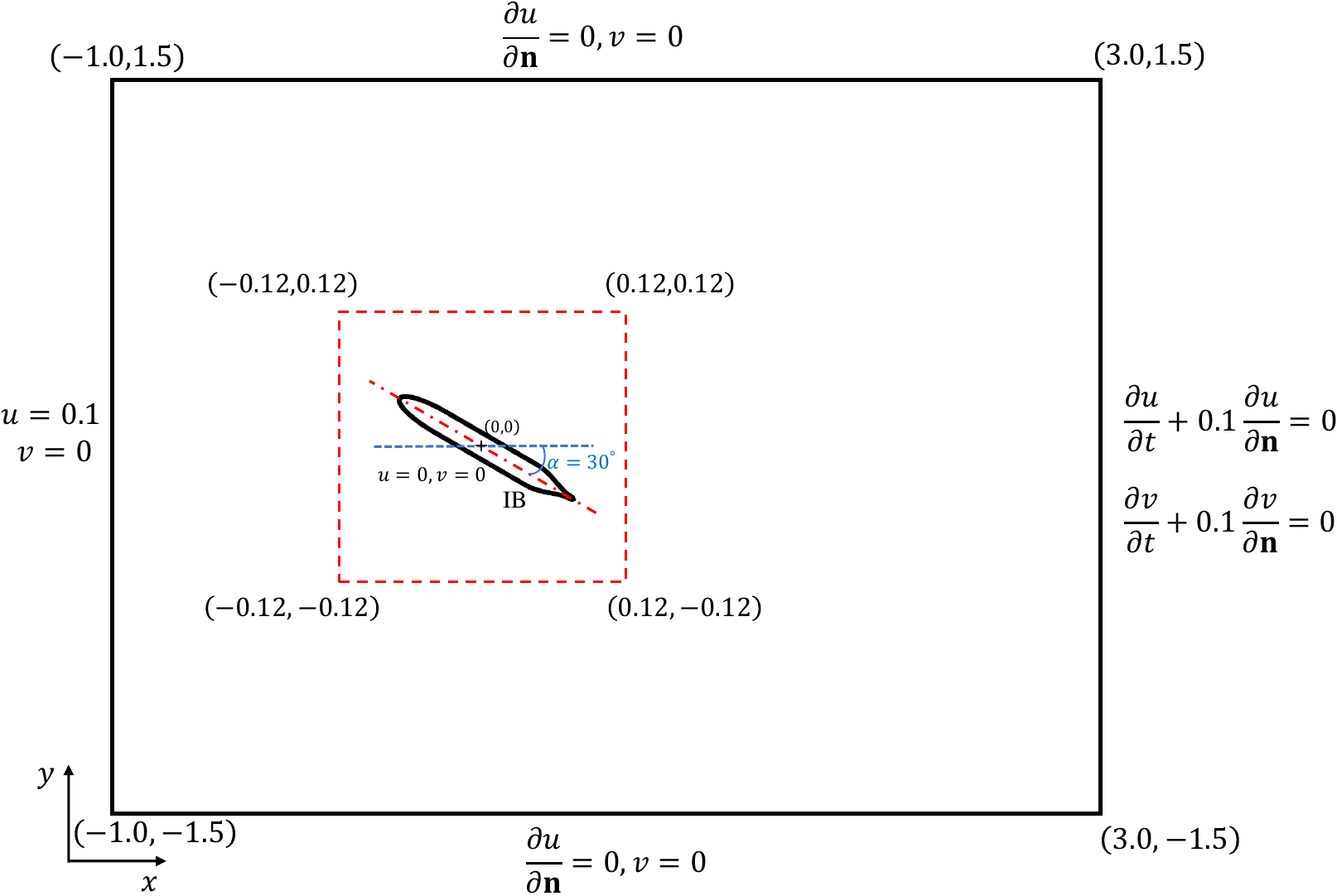}
  \caption{Schematic diagram of 2D flow past the DARPA SUBOFF profile.}\label{Case2_suboff_configurations}
\end{figure}

This section evaluates the generalizability of HyDEA by simulating flow past the DARPA SUBOFF profile~\cite{groves1989geometric}. The DARPA SUBOFF profile is proportionally scaled down from its original size, yielding an overall length of $L = 0.2$ while maintaining all other characteristic ratios. The simulation adopts the same numerical configurations as in Section~\ref{Re100Cylinder}, with the modification being the introduction of the DARPA SUBOFF profile inclined at an angle of attack of $\alpha=30^{\circ}$ as depicted in Fig.~\ref{Case2_suboff_configurations} and a time step size $\Delta t=0.005$. \textit{It is crucial to emphasize that the network architecture and weights are kept identical to those used in Section~\ref{Re100Cylinder}}. We set $Num_{\mathrm{CG-type}} = 3$ and $Num_{\mathrm{DLSM}} = 2$, and the iterative process is terminated when the residual $L2$-norm falls below $\epsilon=10^{-6}$.

\begin{figure}[htbp] 
 \centering  
  \subfigure[]{
  \label{Case2_suboff_Rline_10steps_20}
  \includegraphics[scale=0.21]{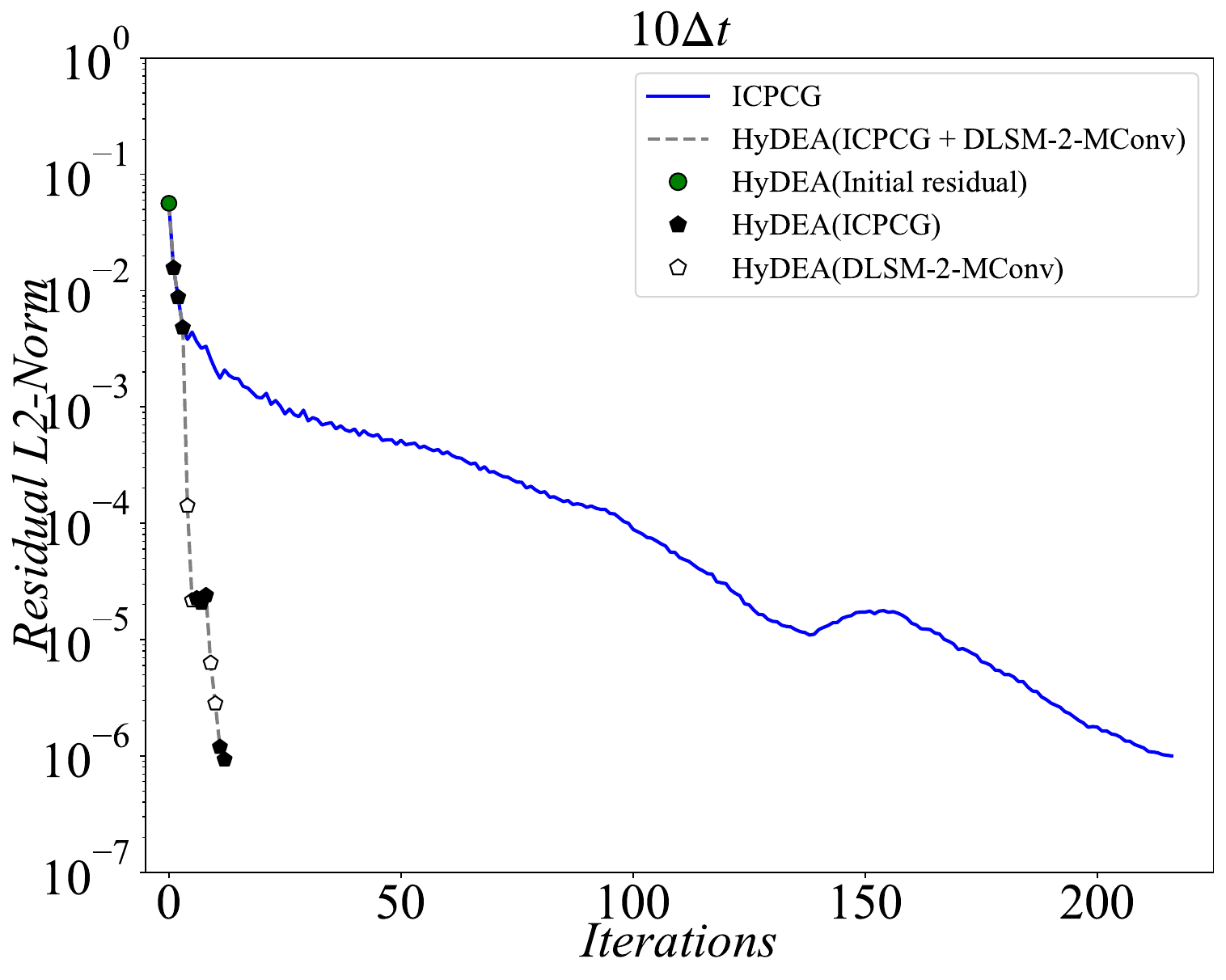}}
  \subfigure[]{
  \label{Case2_suboff_Rline_100steps_20}
  \includegraphics[scale=0.21]{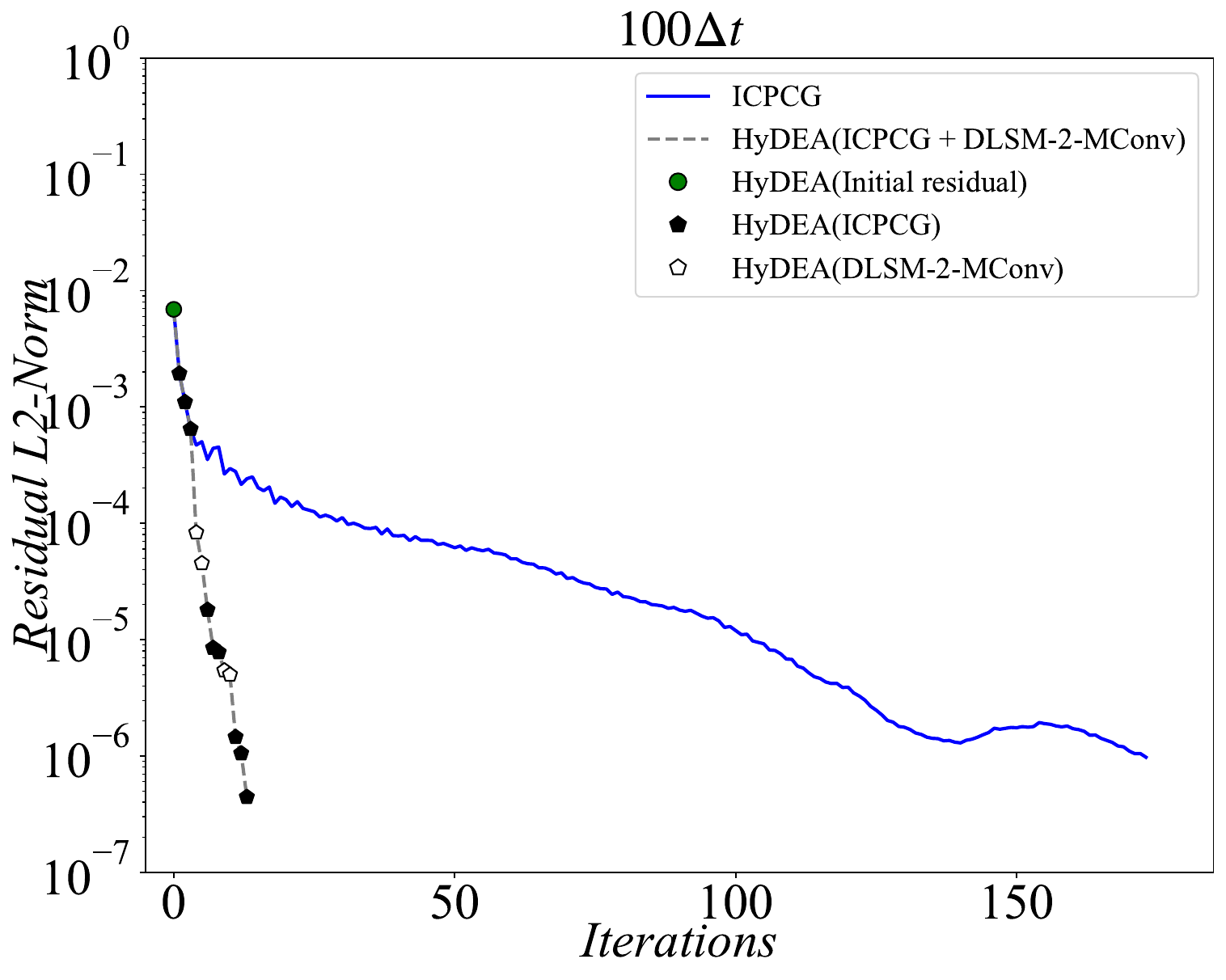}}
  \subfigure[]{
  \label{Case2_suboff_Rline_10000steps_20}
  \includegraphics[scale=0.21]{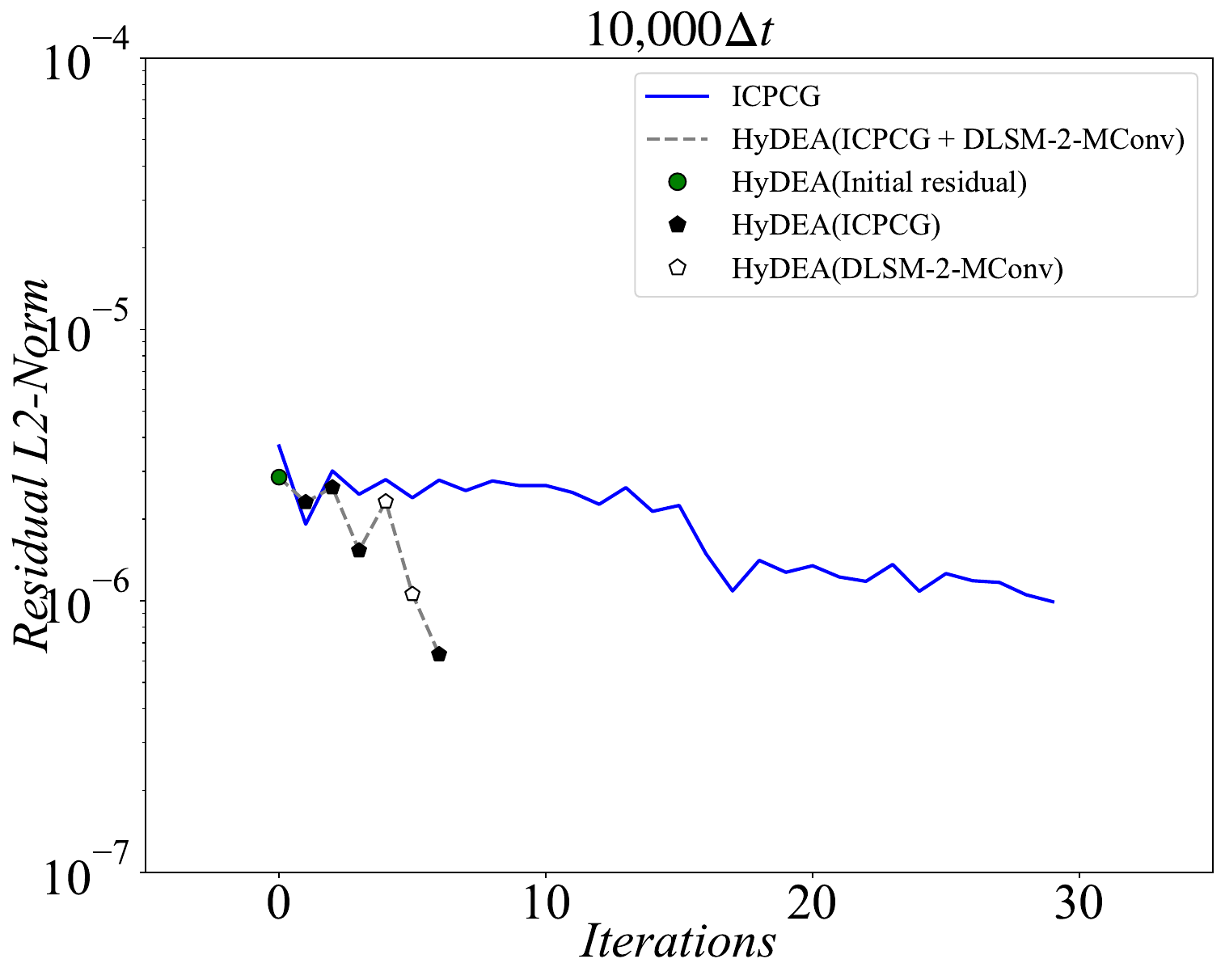}} 
  \caption{Iterative residuals of solving the PPE for 2D flow past the DARPA SUBOFF profile. (a) $10th$ time step. (b) $100th$ time step. (c) $10{,}000th$ time step.}\label{Case2_suboff_Rline_3+2}
\end{figure}

To avoid extensive comparisons, we report only the results of HyDEA~(ICPCG + DLSM-2-MConv), together with the corresponding ICPCG method. Fig.~\ref{Case2_suboff_Rline_3+2} depicts the iterative residuals of solving the PPE at the $10th$, $100th$, and $10{,}000th$ time steps. The results demonstrate that HyDEA~(ICPCG + DLSM-2-MConv) requires significantly fewer iterations to reach the predefined tolerance compared to the standalone ICPCG method. Specifically, HyDEA~(ICPCG + DLSM-2-MConv) takes less than $3$ hybrid rounds (totaling 12 iterations) at $t=10\Delta t$, less than $3$ rounds (13 iterations) at $t=100\Delta t$, and less than $2$ rounds~(6 iterations) at $t=10{,}000\Delta t$. This robust convergence behavior firmly establishes the excellent generalizability of HyDEA, particularly its capability to maintain high efficiency when adapting to complex internal geometries within non-uniform Cartesian grid configurations.

\begin{figure}[htbp]
\centering
  \includegraphics[scale=0.2]{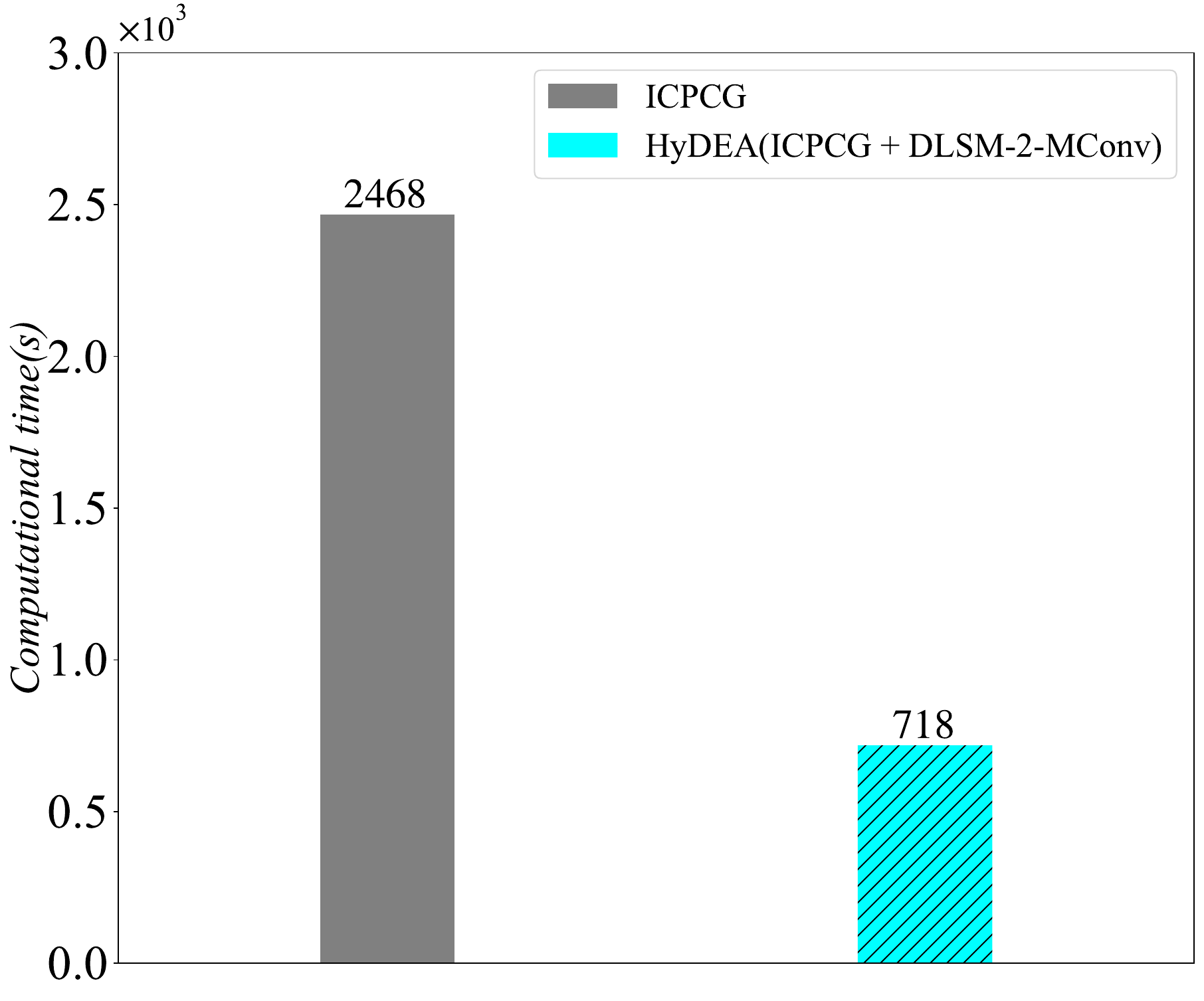}
  \caption{Computational time of the PPE solution over $20{,}000\Delta t$ for 2D flow past the DARPA SUBOFF profile.}\label{Case2_suboff_timecomapre}
\end{figure}

Fig.~\ref{Case2_suboff_timecomapre} illustrates the computational time spent solving the PPE over $20{,}000$ consecutive time steps using HyDEA~(ICPCG + DLSM-2-MConv) and the standalone ICPCG method. The results clearly demonstrate that HyDEA significantly reducing the computational time, providing a remarkable acceleration for the entire simulation process.

\begin{figure}[htbp] 
 \centering  
  \subfigure[]{
  \label{Vorticity_ICPCG_10000step_suboff_30_Case2}
  \includegraphics[scale=0.22]{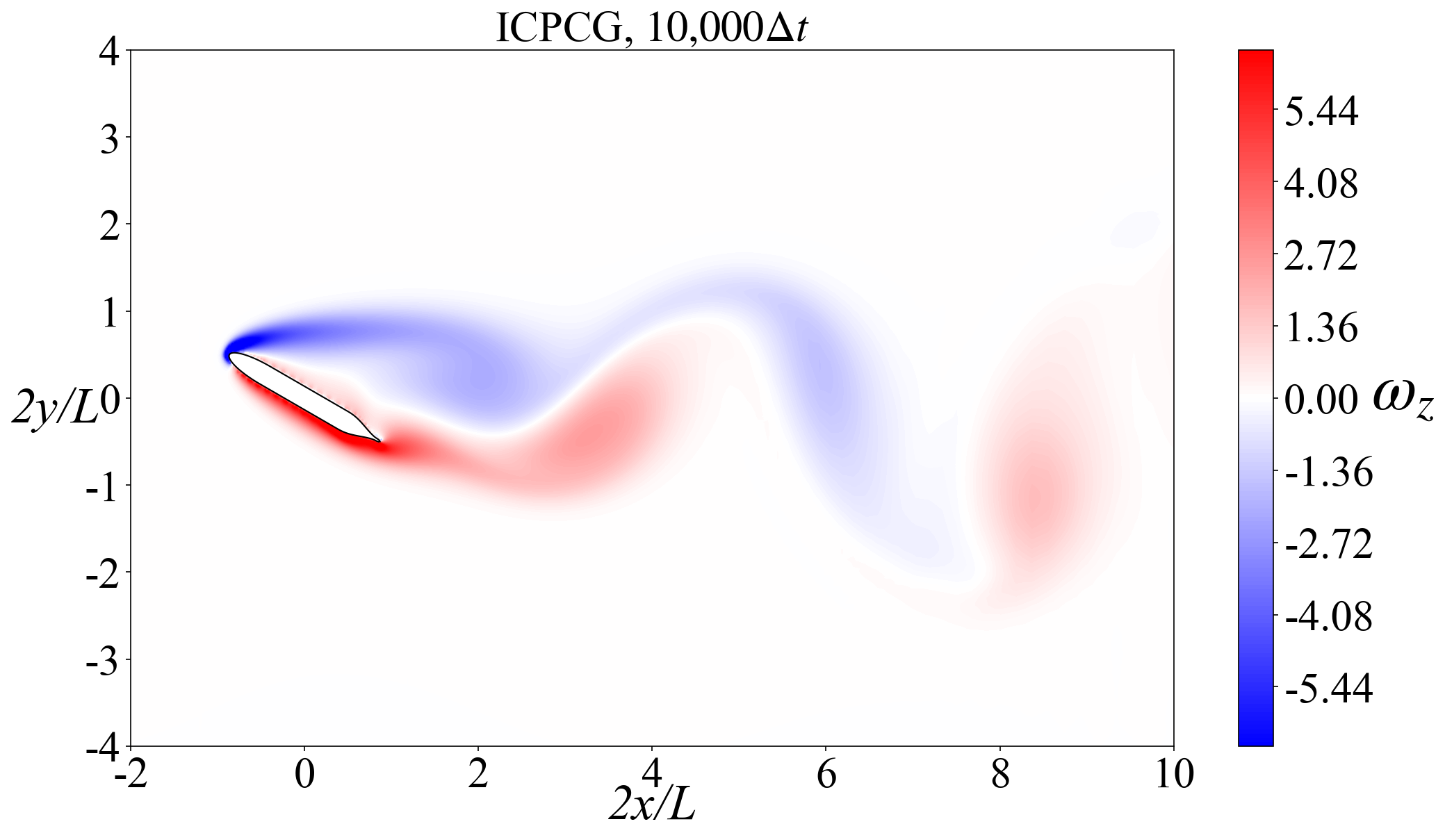}}
  \subfigure[]{
  \label{Vorticity_HyDEA_10000step_suboff_30_Case2}
  \includegraphics[scale=0.22]{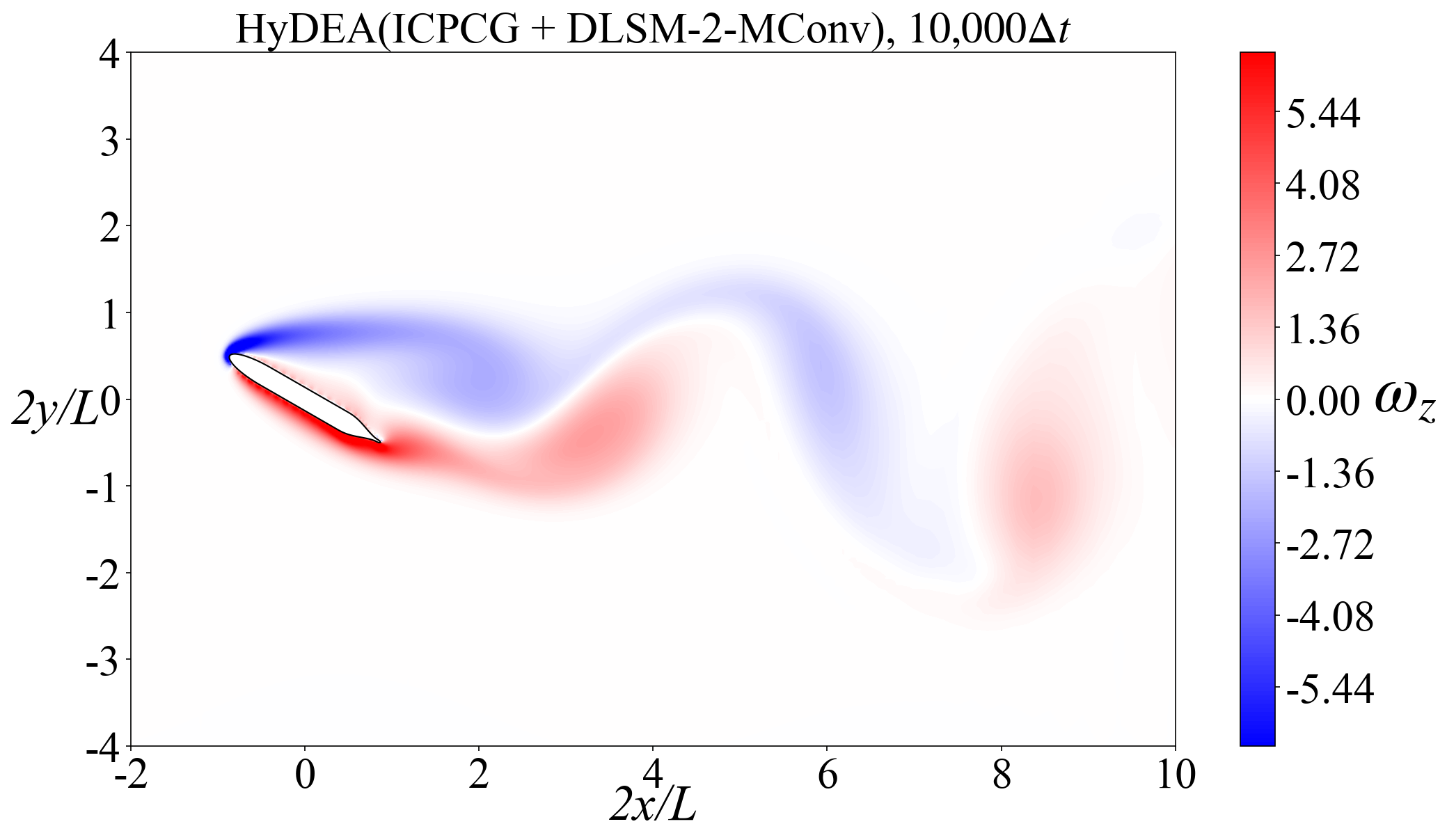}}
  \subfigure[]{
  \label{Vorticity_ICPCG_12000step_suboff_30_Case2}
  \includegraphics[scale=0.22]{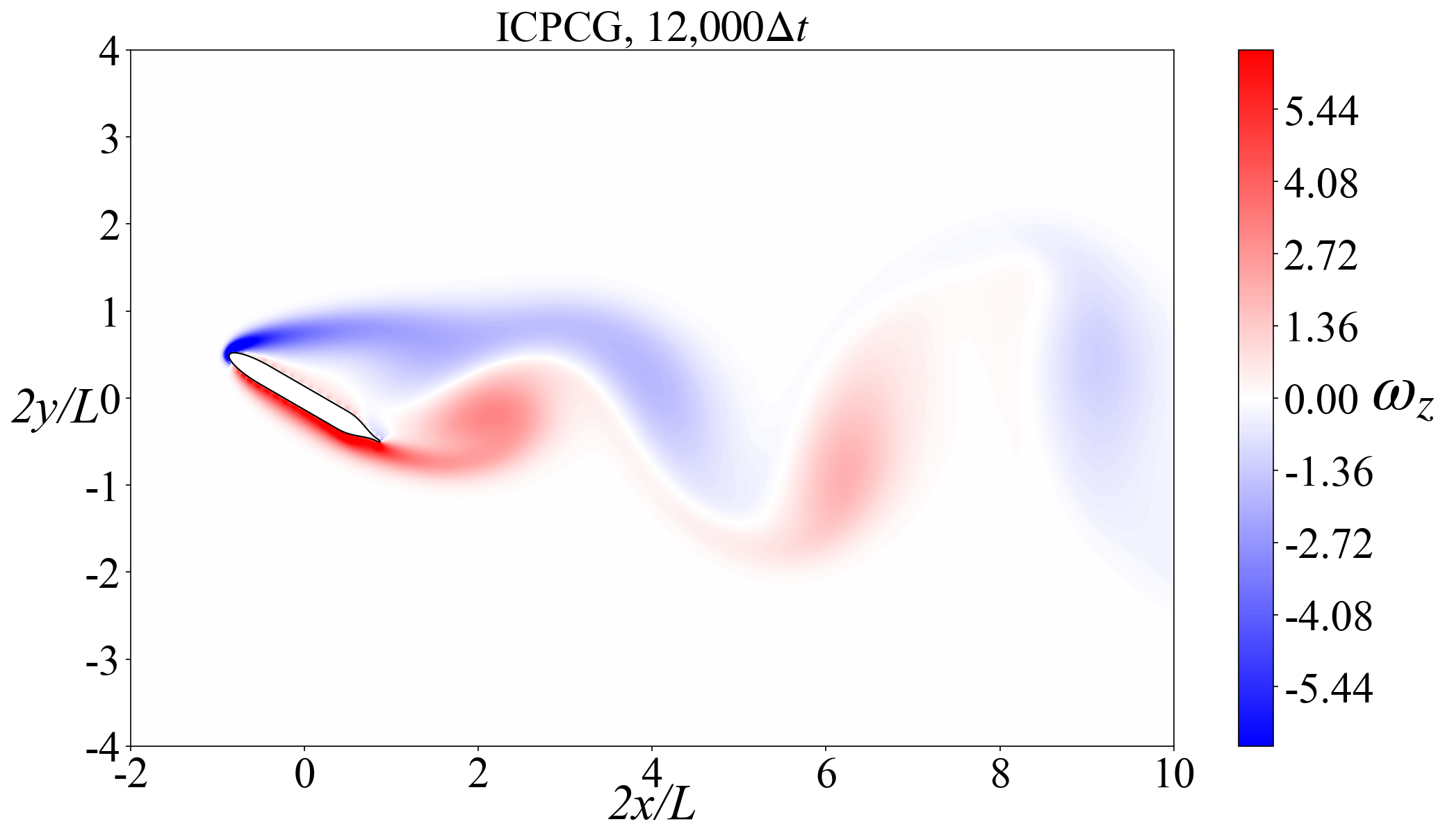}}
  \subfigure[]{
  \label{Vorticity_HyDEA_12000step_suboff_30_Case2}
  \includegraphics[scale=0.22]{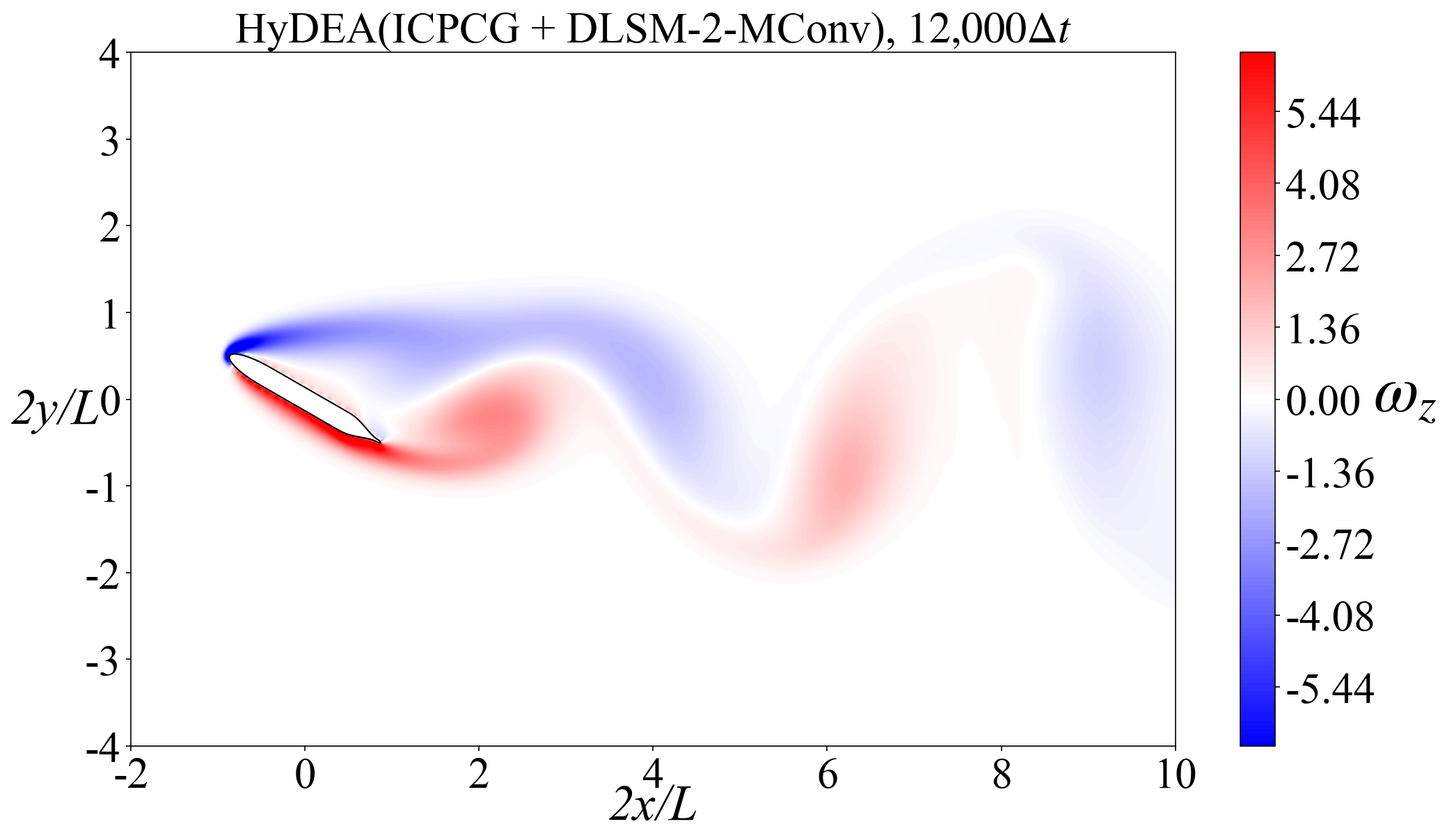}}
  \subfigure[]{
  \label{Vorticity_ICPCG_14000step_suboff_30_Case2}
  \includegraphics[scale=0.22]{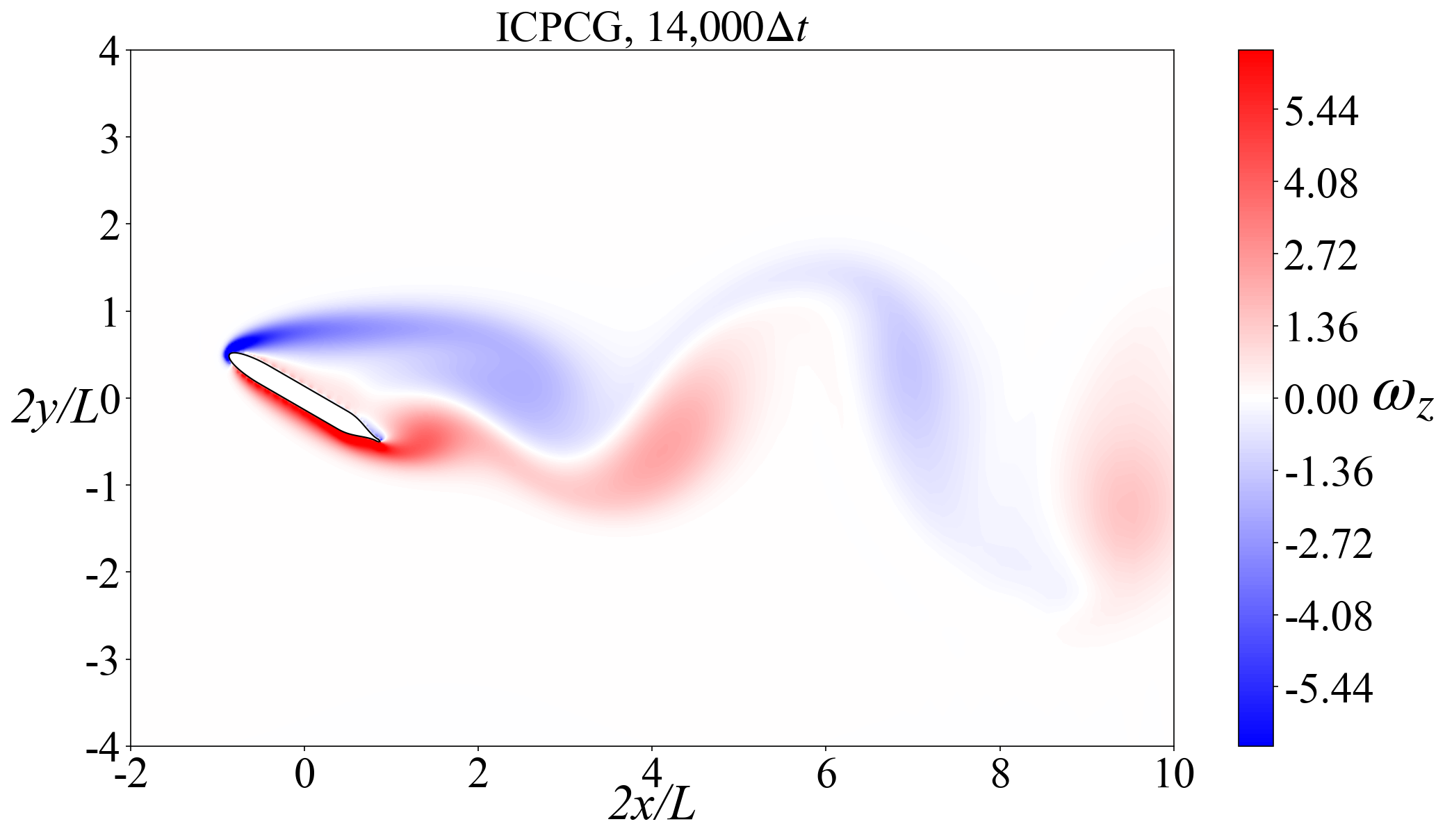}}
  \subfigure[]{
  \label{Vorticity_HyDEA_14000step_suboff_30_Case2}
  \includegraphics[scale=0.22]{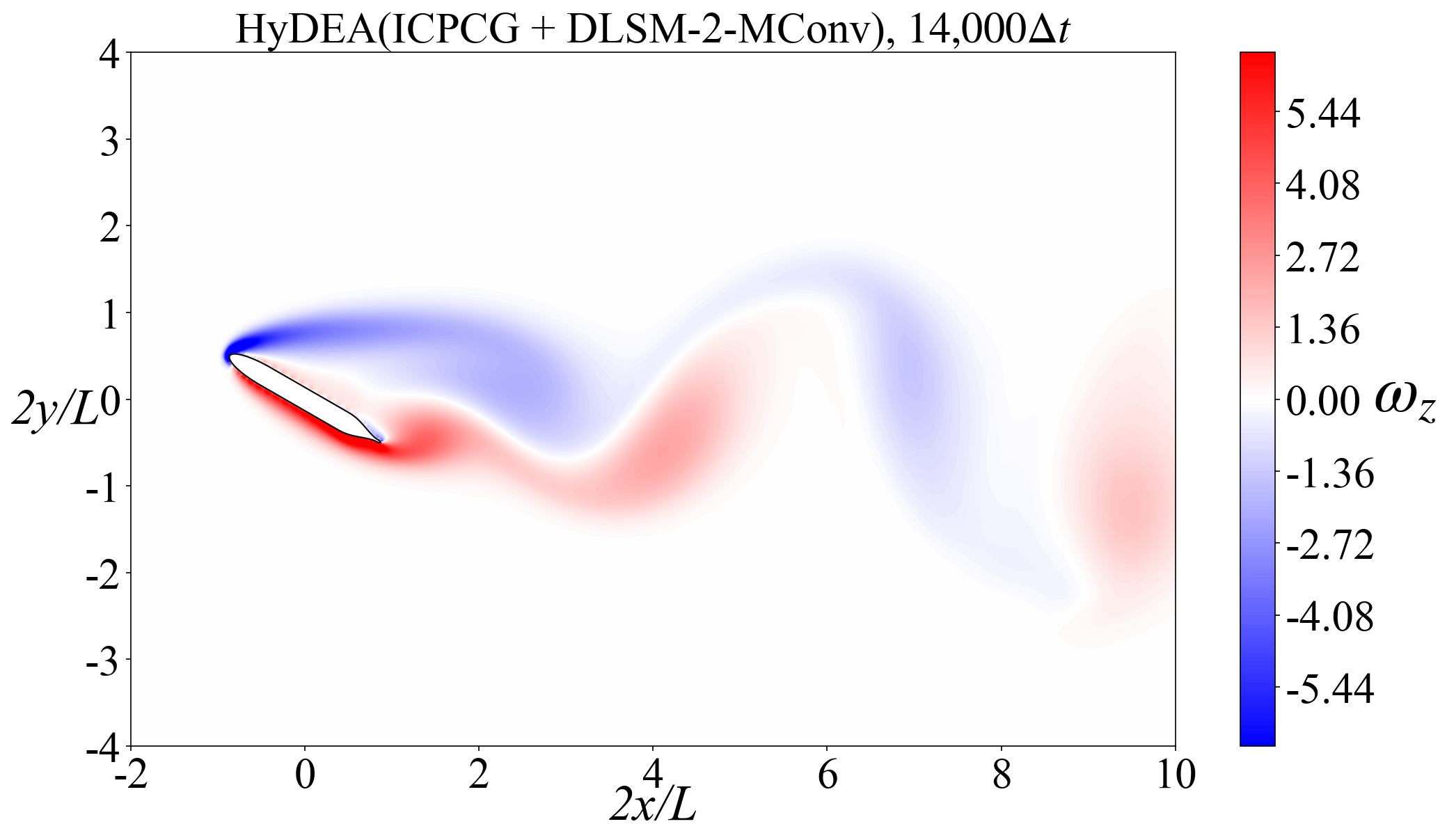}}
 \caption{Vorticity fields for 2D flow past the DARPA SUBOFF profile by ICPCG and HyDEA~(ICPCG + DLSM-2-MConv). (a) $10{,}000th$ time step. (b) $12{,}000th$ time step. (c) $14{,}000th$ time step.}
 \label{Case2_flowfield_suboff}
\end{figure}

Moreover, Fig.~\ref{Case2_flowfield_suboff} displays the vorticity fields at the $10{,}000th$, $12{,}000th$ and $14{,}000th$ time steps. These visualizations confirm that HyDEA faithfully captures the temporal evolution of the flow field.

\subsubsection{2D flow past an inline oscillating cylinder}
\label{flow past Oscylinder}

\begin{figure}[htbp]
\centering
  \includegraphics[scale=0.35]{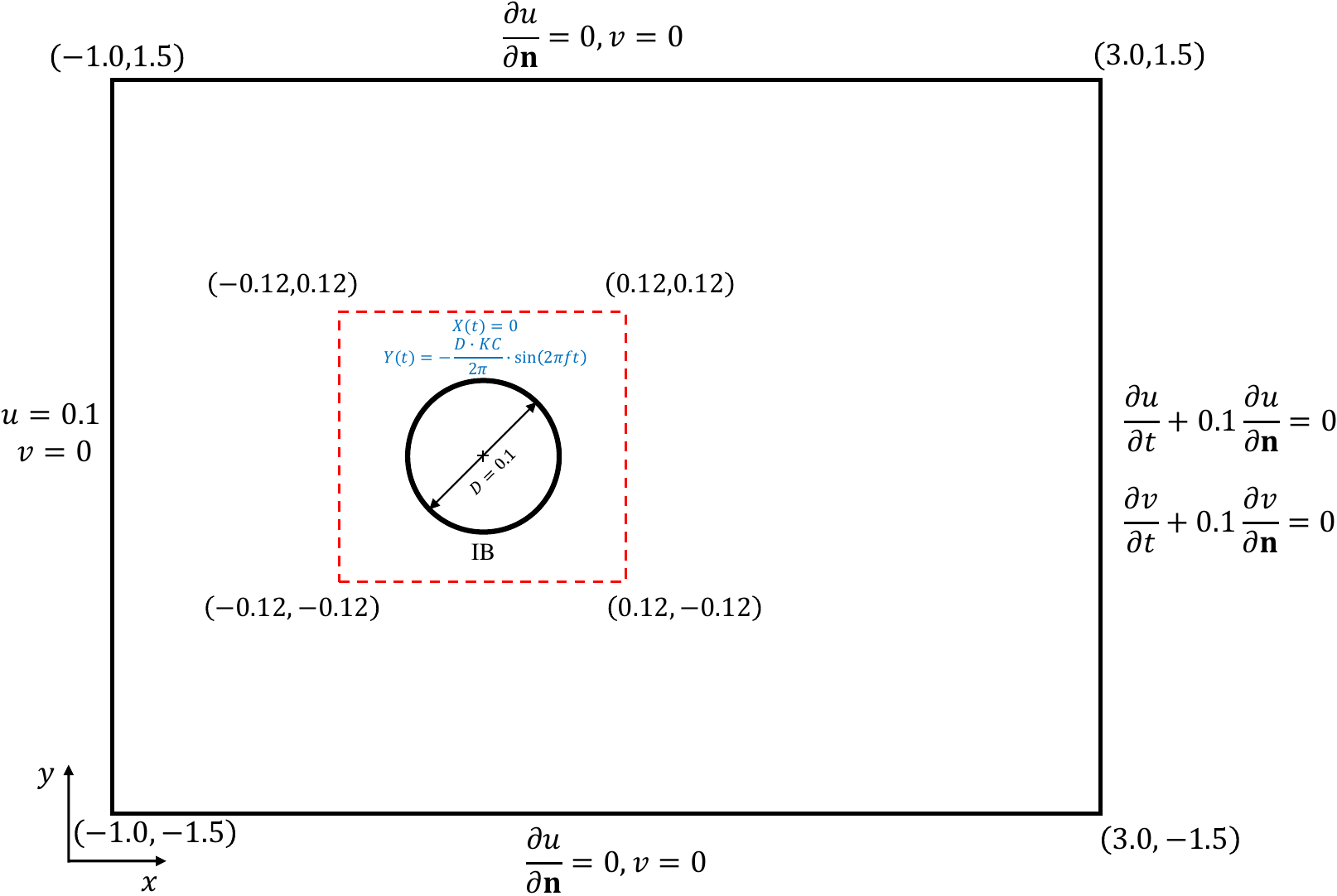}
  \caption{Schematic diagram of 2D flow past an inline oscillating cylinder.}\label{Case2_FAOsCylinder_configurations}
\end{figure}

This section extends the validation of HyDEA to moving boundary problems by simulating flow past an inline oscillating cylinder. While maintaining the identical numerical configurations from Section~\ref{Re100Cylinder}, we introduce an actively moving circular cylinder, as illustrated in Fig.~\ref{Case2_FAOsCylinder_configurations}, and adopt a time step size of $\Delta t=0.005$. The temporal displacement of this cylinder is governed by:
\begin{eqnarray}
\label{Move trajectory of cylinder x}
 X(t) &=& 0 ,
\\
\label{Move trajectory of cylinder y}
  Y(t) &=& -\frac{D\cdot \mathit{KC}}{2\pi}\cdot \sin(2 \pi ft),
\end{eqnarray}
where $(X(t),Y(t))$ denotes the instantaneous center coordinates of the cylinder at time $t$. The geometric and kinematic parameters are specified as follows: cylinder diameter $D=0.1$, oscillation frequency $f=0.1$, velocity amplitude in the y-direction $V_{m}=0.03$, and Keulegan-Carpenter number $\mathit{KC}=V_{m}/fD=3$. \textit{It should be highlighted that both the network architecture and weights are strictly identical to those employed in Section~\ref{Re100Cylinder}}. We set $Num_{\mathrm{CG-type}} = 3$ and $Num_{\mathrm{DLSM}} = 2$, and the iterative process is terminated when the residual $L2$-norm falls below $\epsilon=10^{-6}$.

\begin{figure}[htbp] 
 \centering  
  \subfigure[]{
  \label{Case2_FAOsCylinder_Rline_10steps_20}
  \includegraphics[scale=0.21]{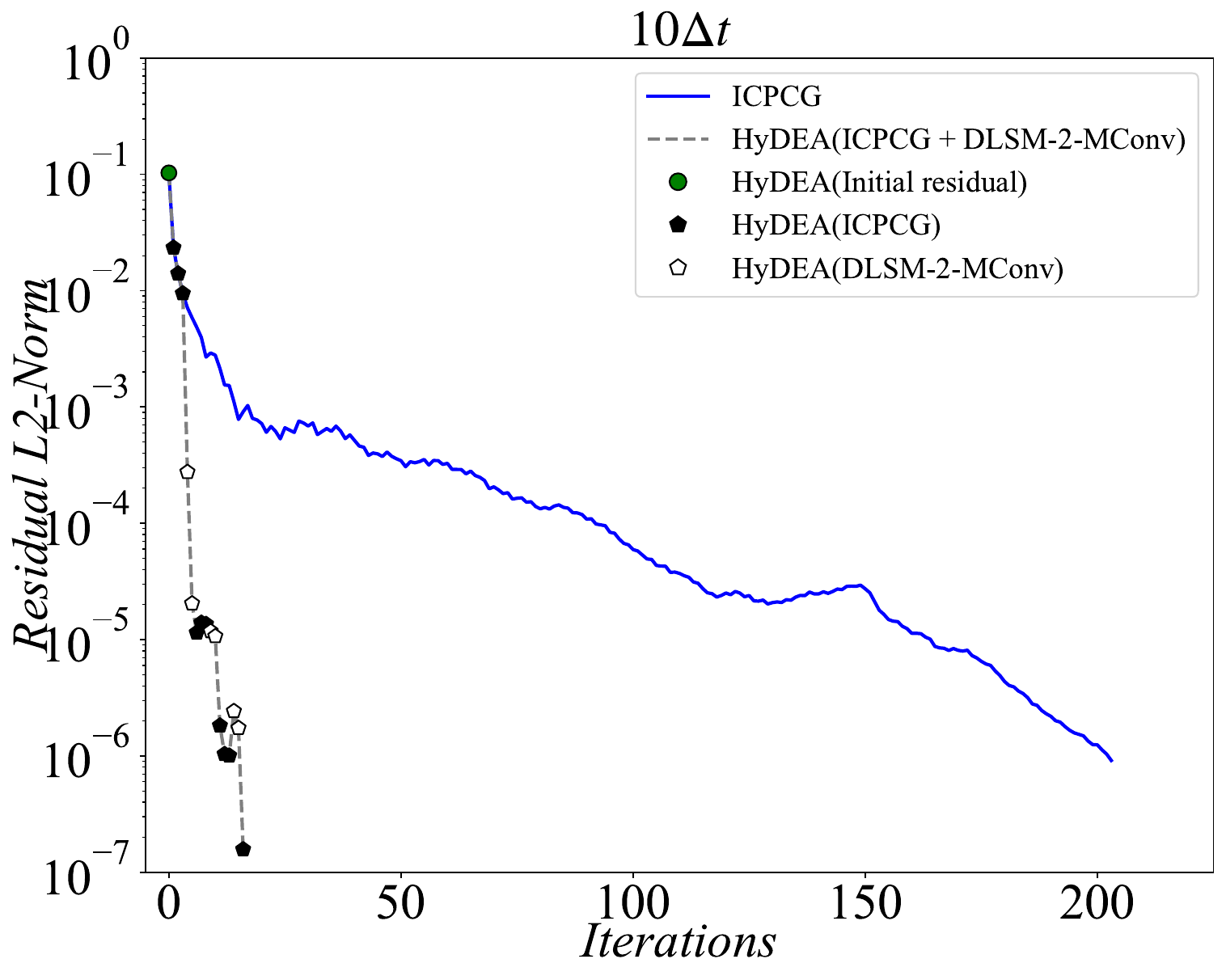}}
  \subfigure[]{
  \label{Case2_FAOsCylinder_Rline_100steps_20}
  \includegraphics[scale=0.21]{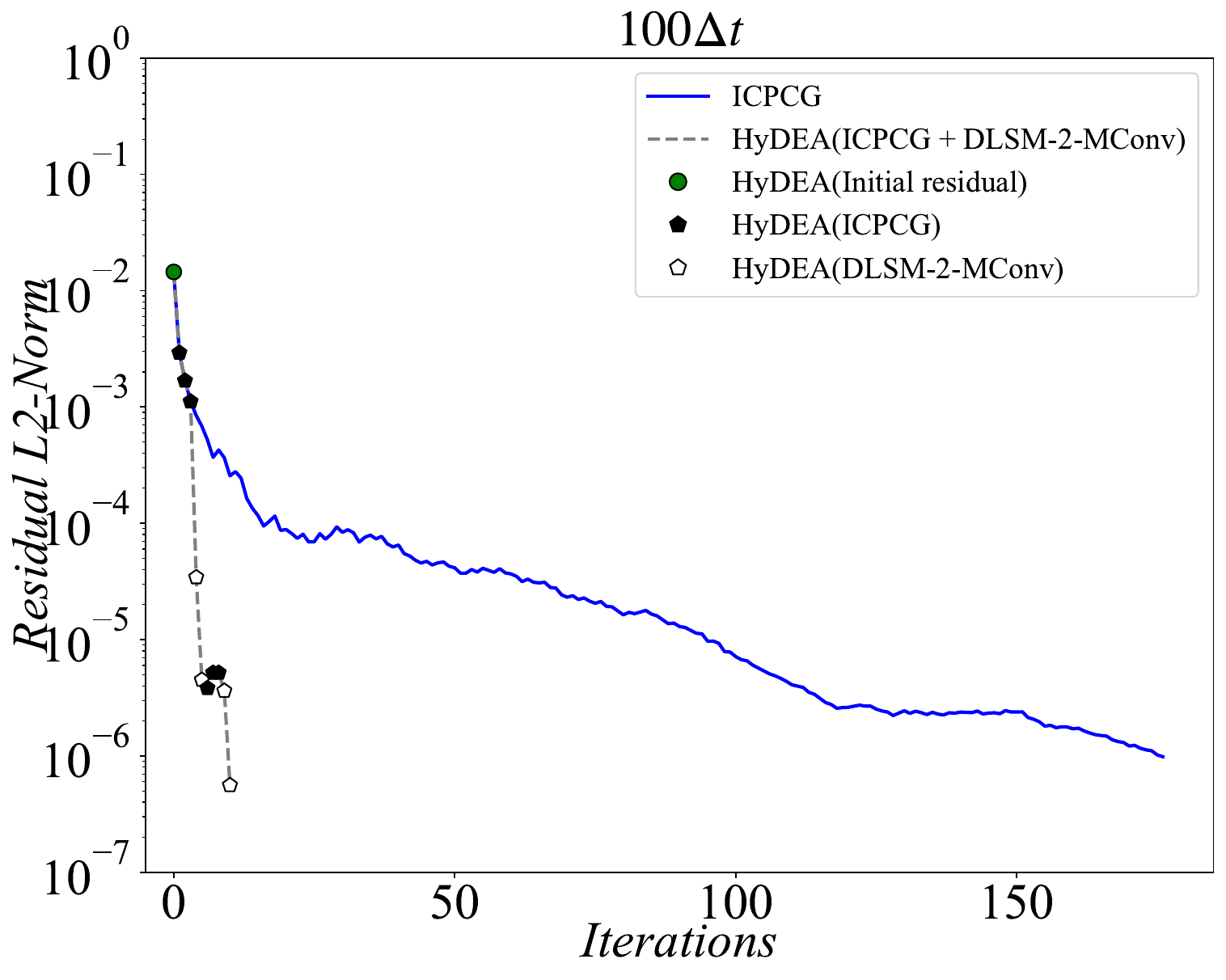}}
  \subfigure[]{
  \label{Case2_FAOsCylinder_Rline_10000steps_20}
  \includegraphics[scale=0.21]{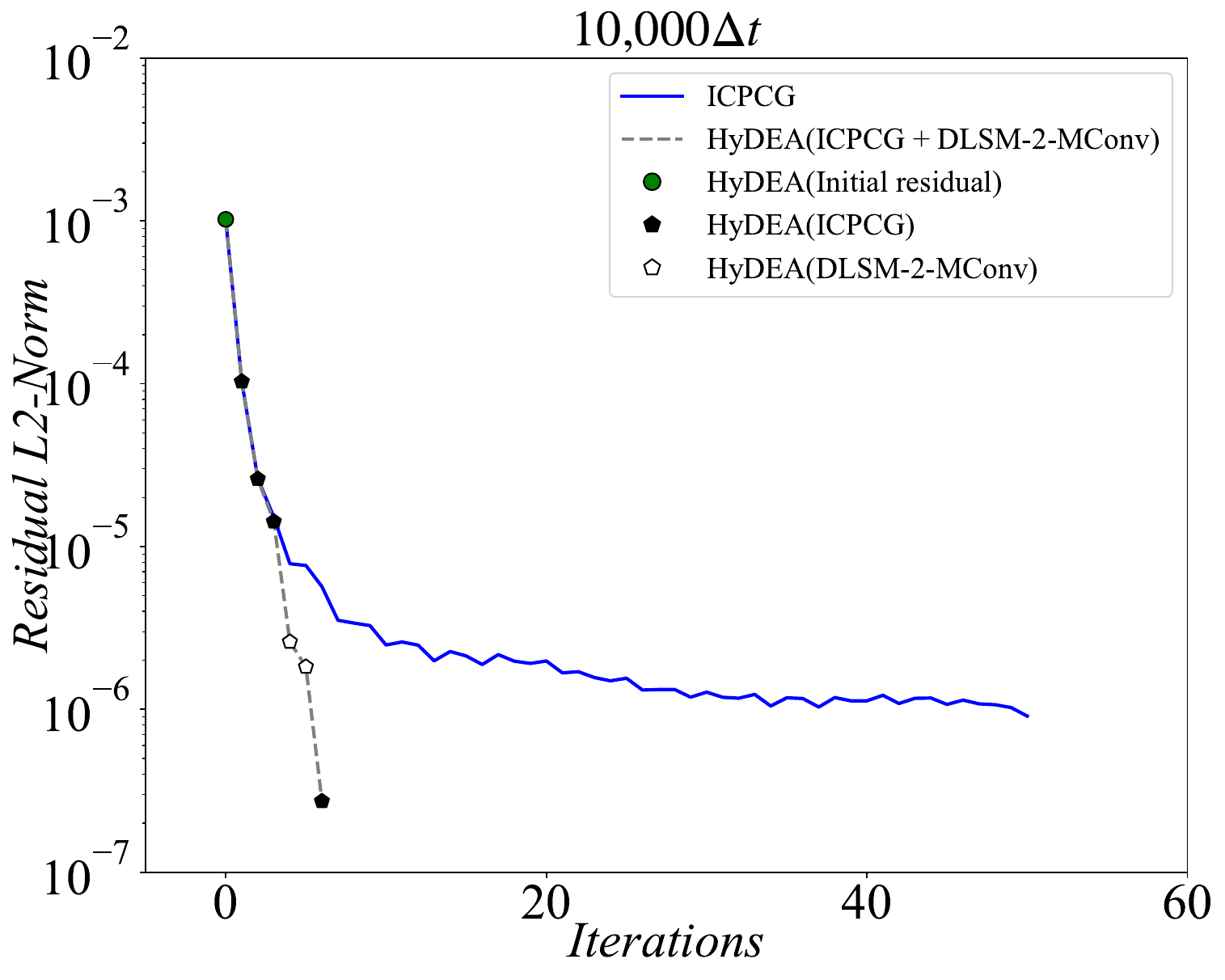}} 
  \caption{Iterative residuals of solving the PPE for 2D flow past an inline oscillating cylinder. (a) $10th$ time step. (b) $100th$ time step. (c) $10{,}000th$ time step.}\label{Case2_FAROsCylinder_Rline_3+2}
\end{figure}

To avoid extensive comparisons, we report only the results of HyDEA~(ICPCG + DLSM-2-MConv), together with the corresponding ICPCG method. Fig.~\ref{Case2_FAROsCylinder_Rline_3+2} presents the iterative residuals of solving the PPE at the $10th$, $100th$, and $10{,}000th$ time steps. The results demonstrate that HyDEA~(ICPCG + DLSM-2-MConv) requires significantly fewer iterations to reach the predefined tolerance compared to the standalone ICPCG method. Specifically, HyDEA~(ICPCG + DLSM-2-MConv) takes less than $4$ hybrid rounds (totaling 16 iterations) at $t=10\Delta t$, $2$ rounds (10 iterations) at $t=100\Delta t$, and less than $2$ rounds (6 iterations) at $t=10{,}000\Delta t$. This robust convergence behavior firmly establishes the excellent generalizability of HyDEA. Most notably, it demonstrates the HyDEA's capability to sustain high computational efficiency when adapting to dynamically evolving internal boundaries within non-uniform Cartesian grid configurations.

\begin{figure}[htbp]
\centering
  \includegraphics[scale=0.2]{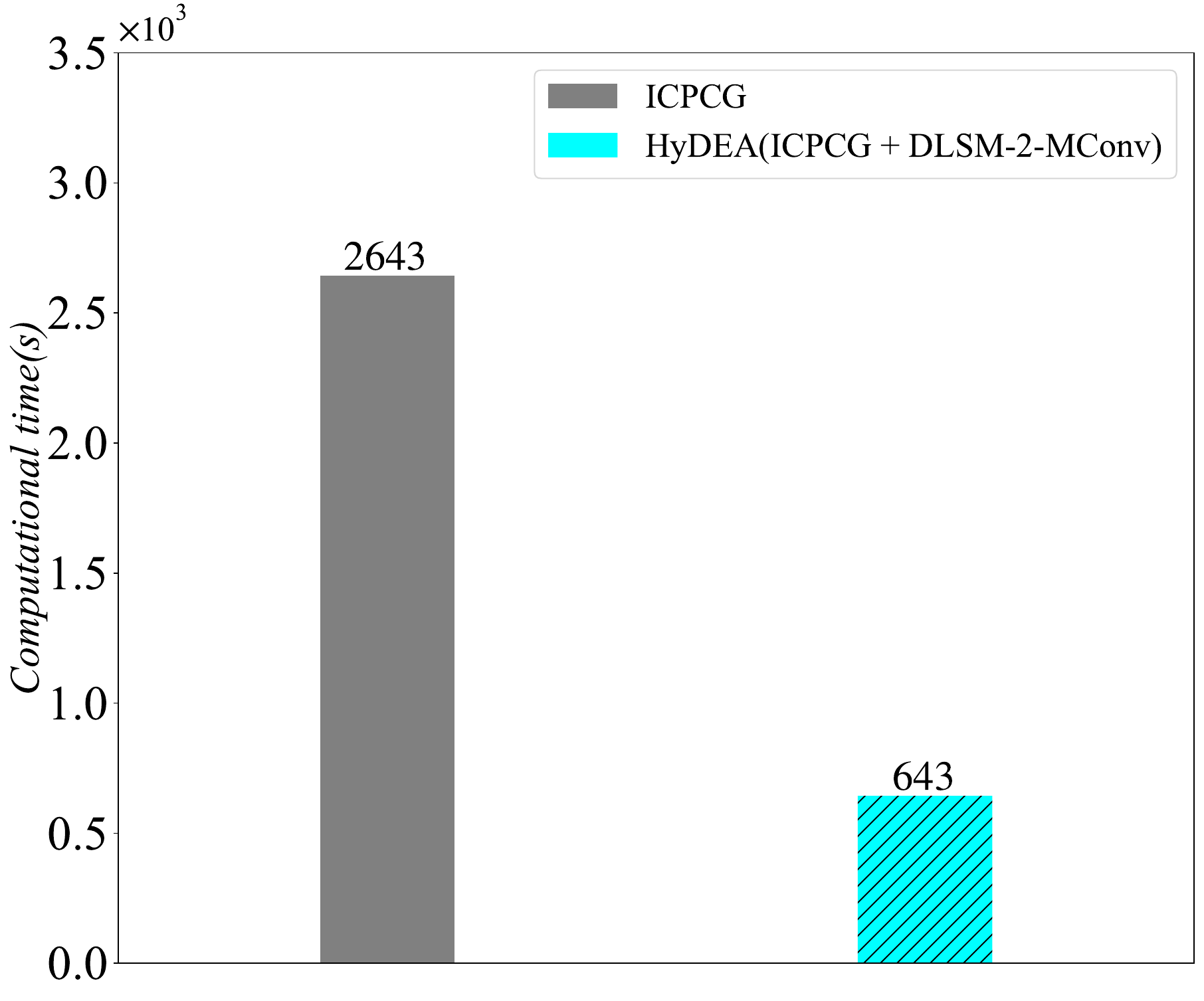}
  \caption{Computational time of the PPE solution over $20{,}000\Delta t$ for 2D flow past an inline oscillating cylinder.}\label{Case2_FAOsCylinder_timecomapre}
\end{figure}

Fig.~\ref{Case2_FAOsCylinder_timecomapre} illustrates the cumulative computational time required to solve the PPE over $20{,}000$ consecutive time steps using both HyDEA~(ICPCG + DLSM-2-MConv) and the standalone ICPCG method. The comparison clearly demonstrates that HyDEA substantially reduces computational time, thereby delivering a significant acceleration to the overall simulation process.

\begin{figure}[htbp] 
 \centering  
  \subfigure[]{
  \label{Vorticity_ICPCG_10000step_FAOsCylinder_Case2}
  \includegraphics[scale=0.22]{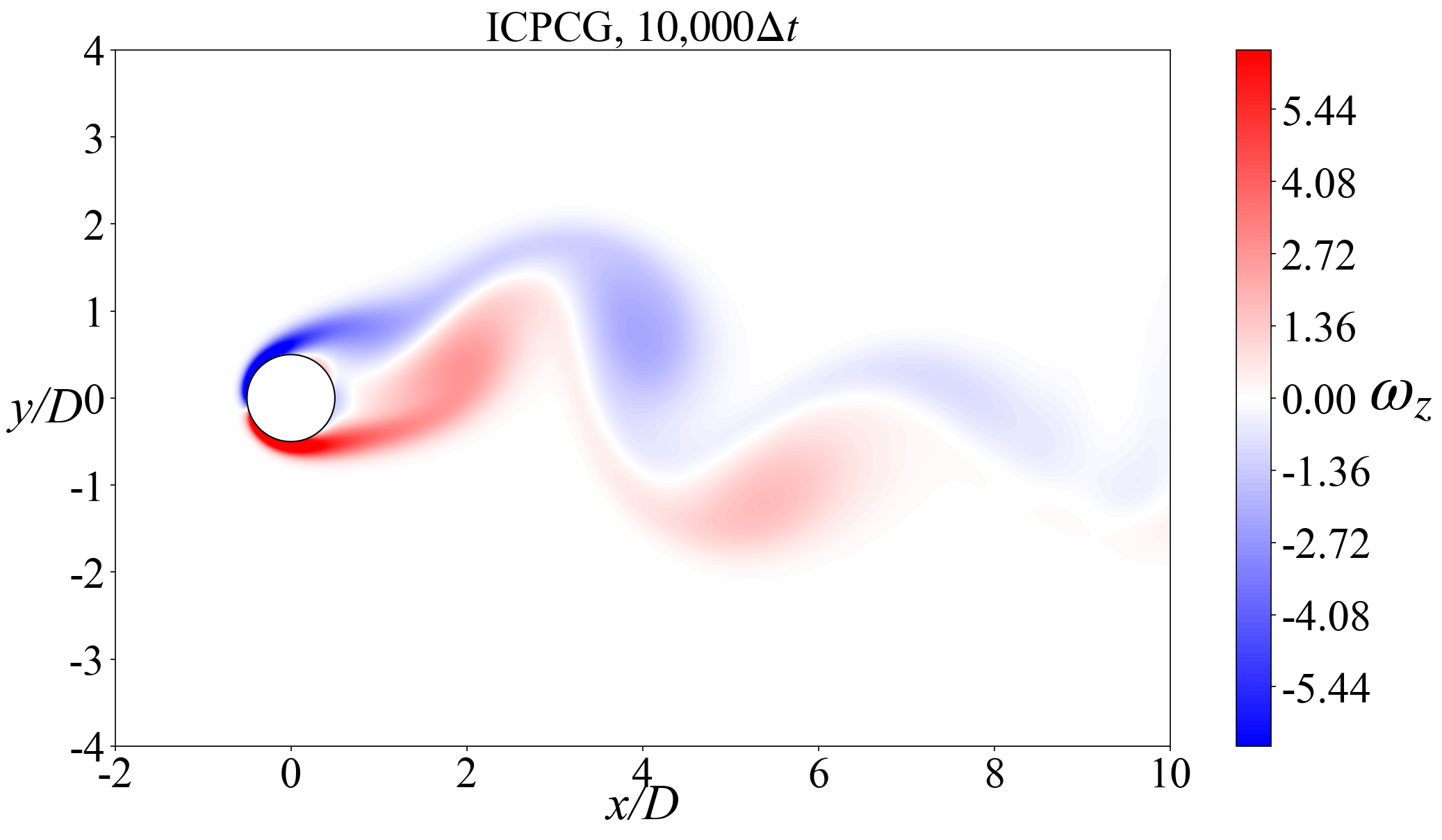}}
  \subfigure[]{
  \label{Vorticity_HyDEA_10000step_FAOsCylinder_Case2}
  \includegraphics[scale=0.22]{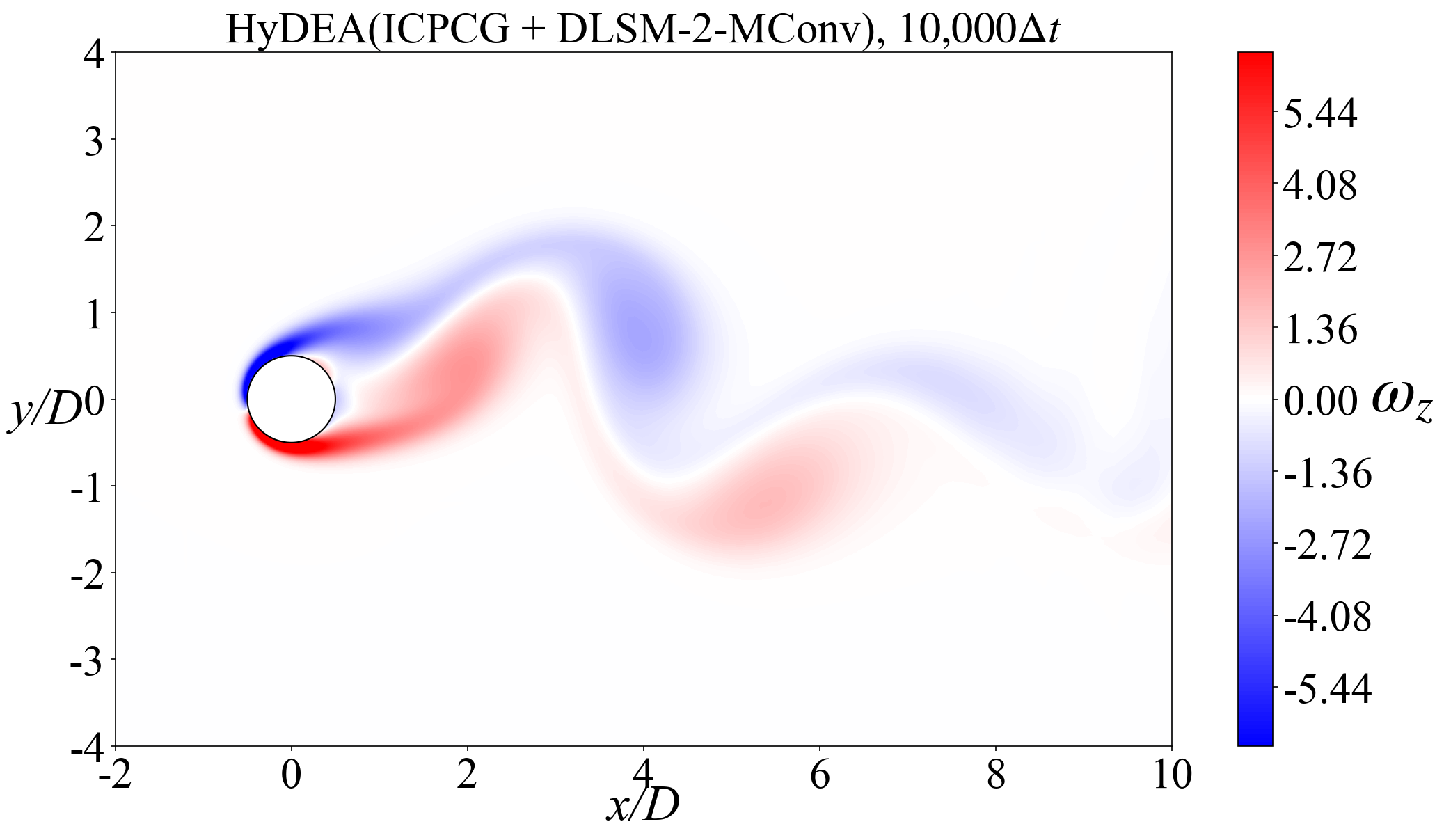}}
  \subfigure[]{
  \label{Vorticity_ICPCG_11500step_FAOsCylinder_Case2}
  \includegraphics[scale=0.22]{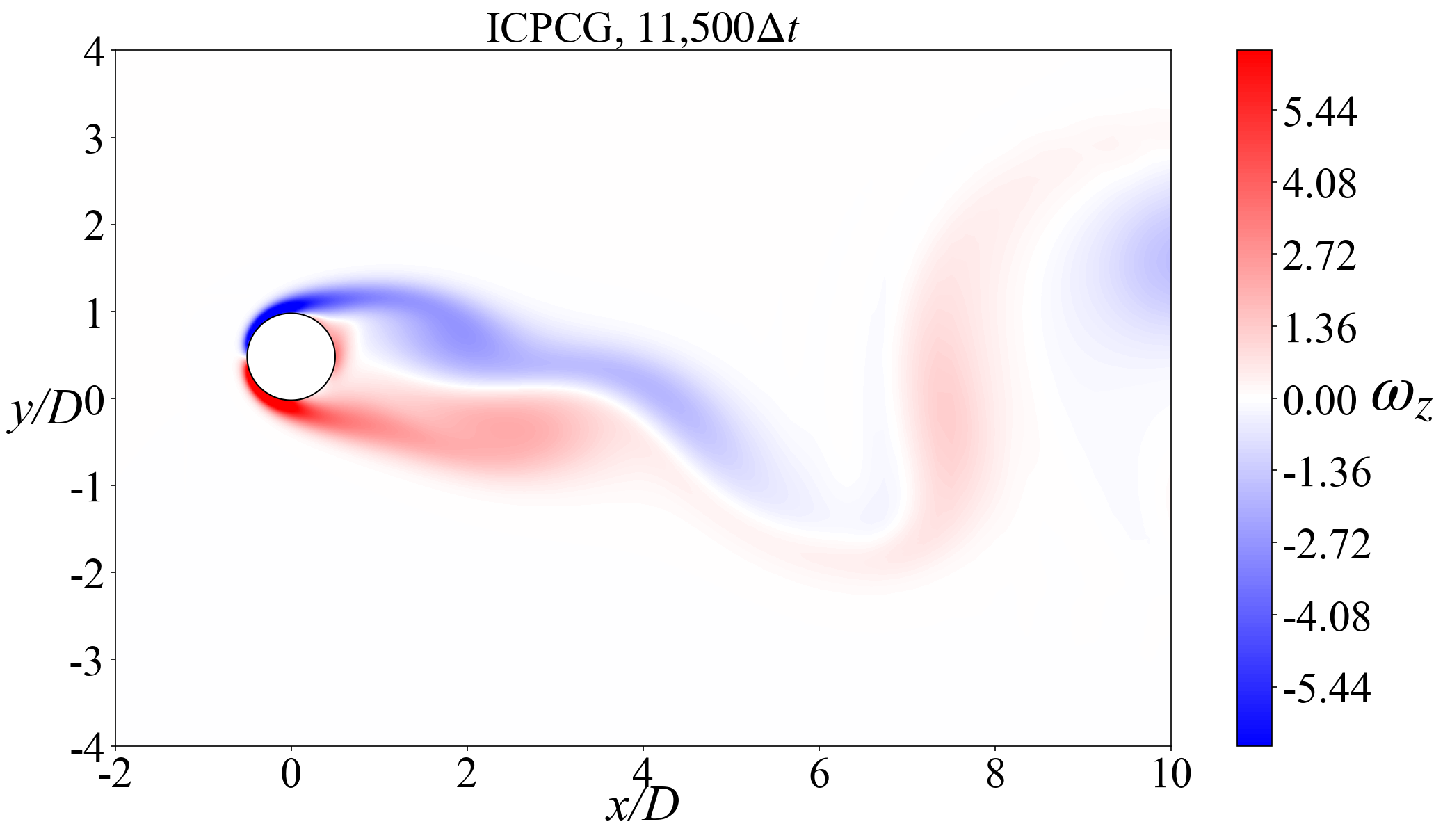}}
  \subfigure[]{
  \label{Vorticity_HyDEA_11500step_FAOsCylinder_Case2}
  \includegraphics[scale=0.22]{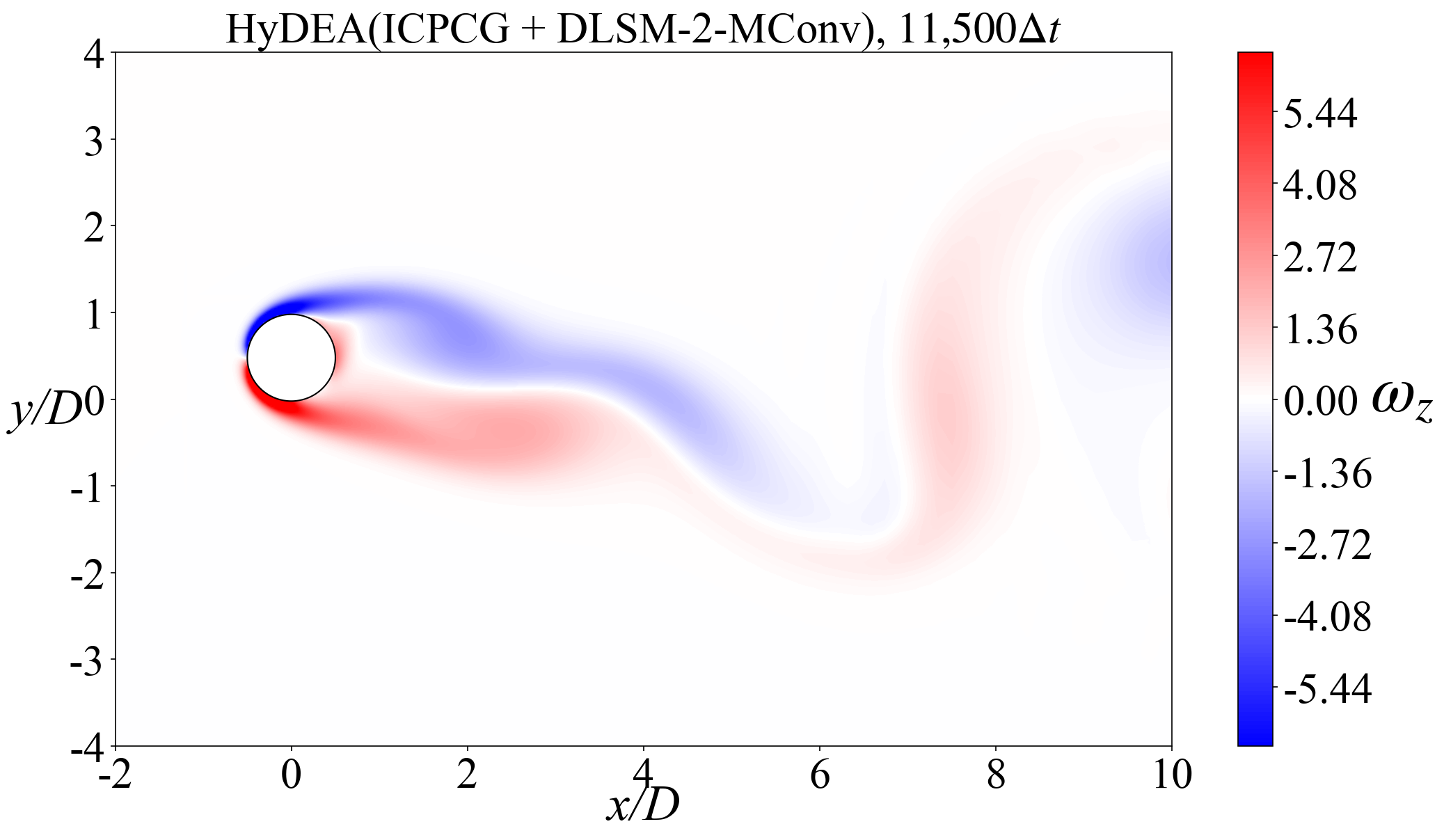}}
  \subfigure[]{
  \label{Vorticity_ICPCG_14500step_FAOsCylinder_Case2}
  \includegraphics[scale=0.22]{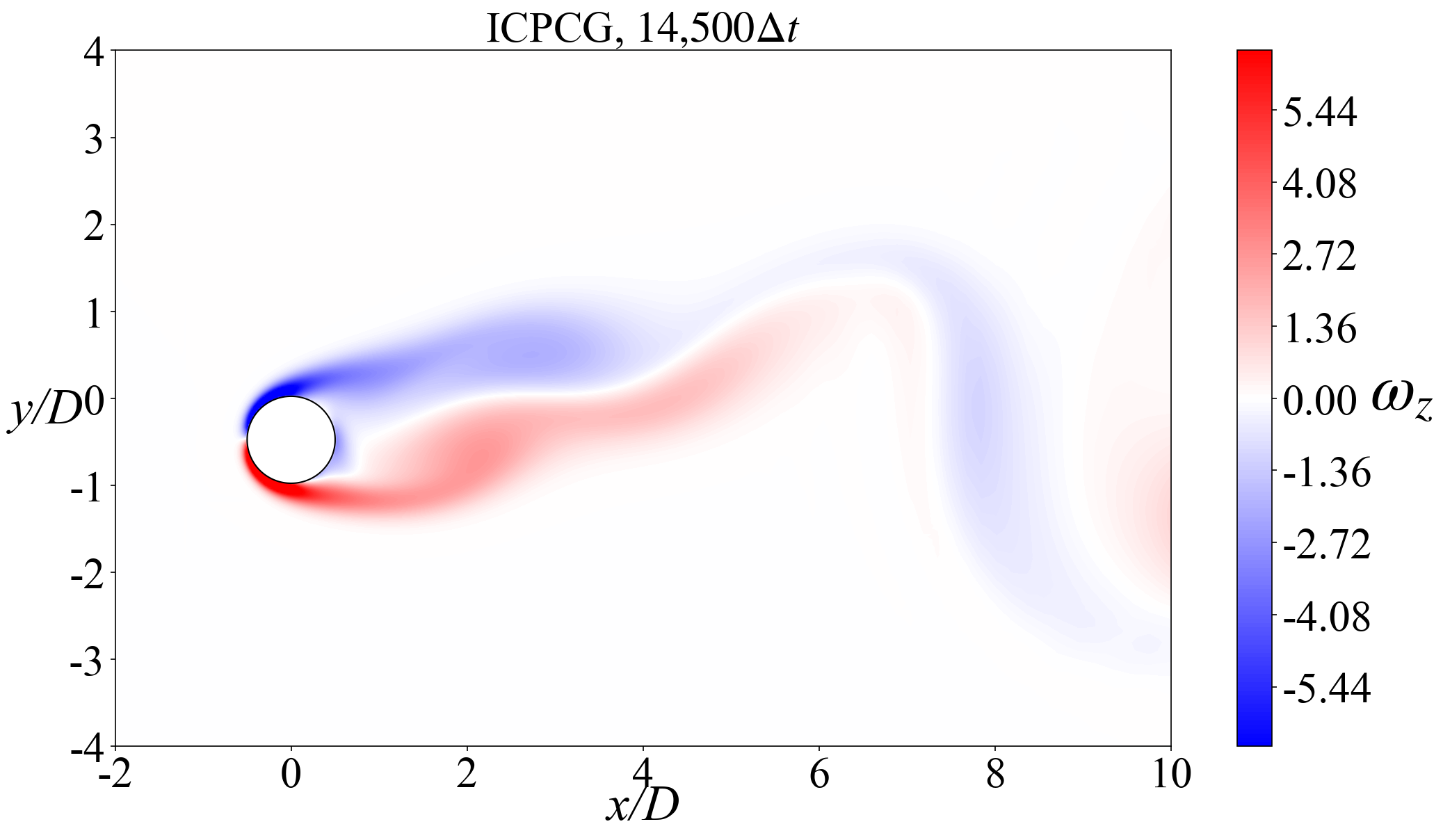}}
  \subfigure[]{
  \label{Vorticity_HyDEA_14500step_FAOsCylinder_Case2}
  \includegraphics[scale=0.22]{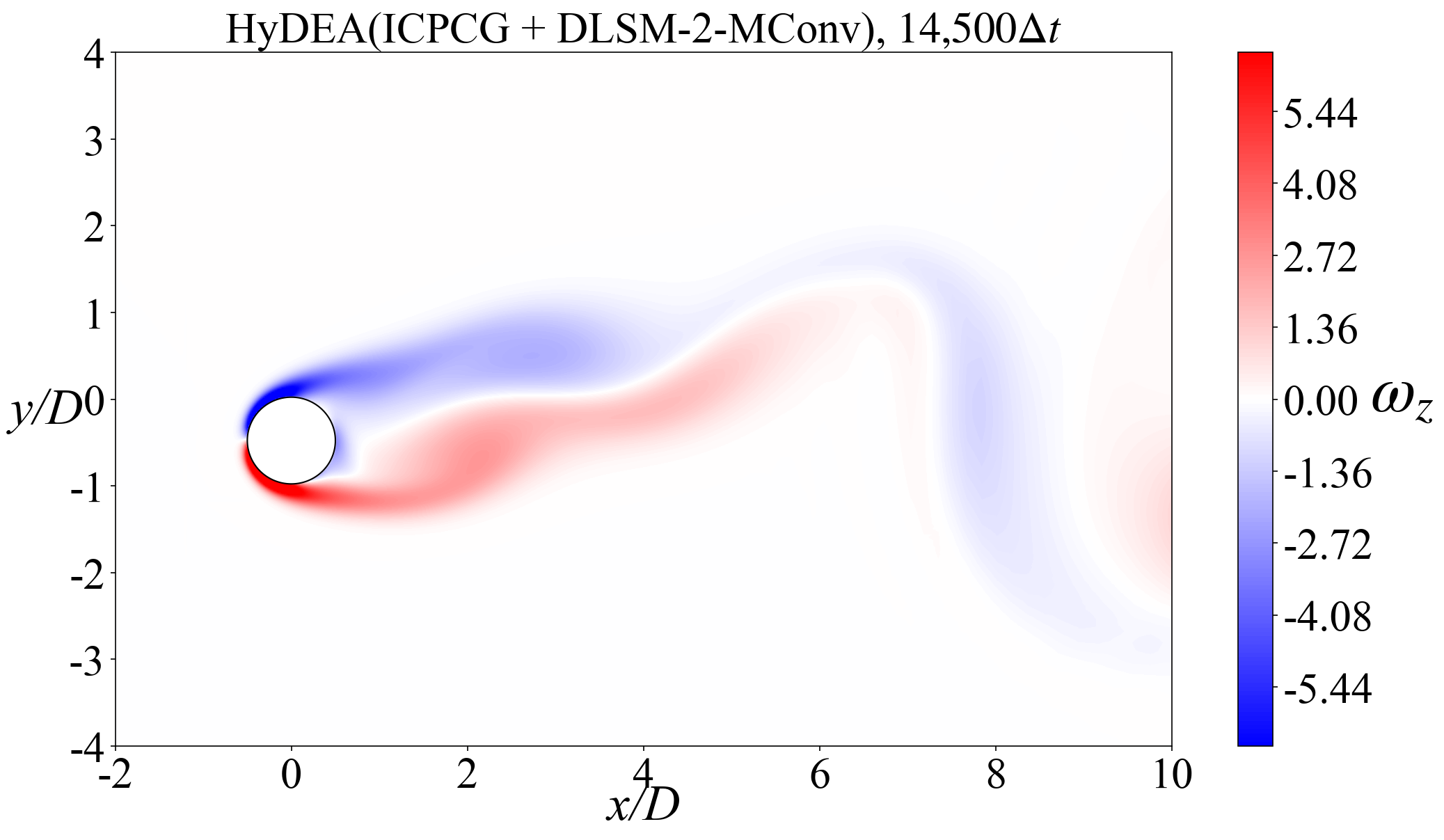}}
 \caption{Vorticity fields for 2D flow past an inline oscillating cylinder by ICPCG and HyDEA~(ICPCG + DLSM-2-MConv). (a) $10{,}000th$ time step. (b) $11{,}500th$ time step. (c) $14{,}500th$ time step.}
 \label{Case2_flowfield_FAOsCylinder}
\end{figure}

Furthermore, the vorticity fields at the $10{,}000th$, $11{,}500th$ and $14{,}500th$ time steps are presented in Fig.~\ref{Case2_flowfield_FAOsCylinder}, which clearly demonstrate that the temporal evolution of the flow field is accurately calculated.

\subsection{Case 3: Flapping elliptical wing}
\label{Flapwing}

\begin{figure}[htbp] 
 \centering  
  \subfigure[]{
  \label{Case2_domain}
  \includegraphics[scale=0.45]{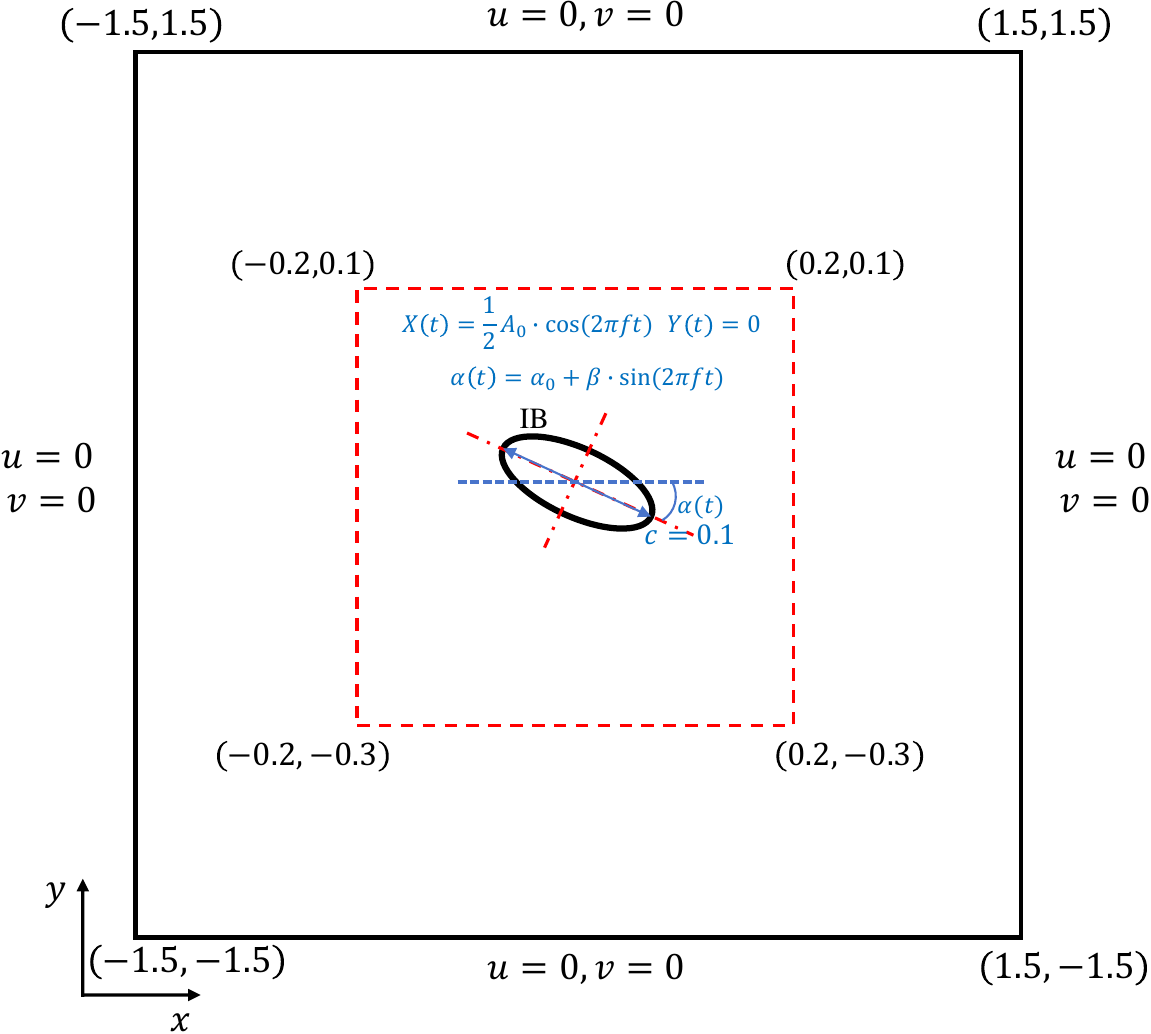}}
  \subfigure[]{
  \label{Case2_grid}
  \includegraphics[scale=0.29]{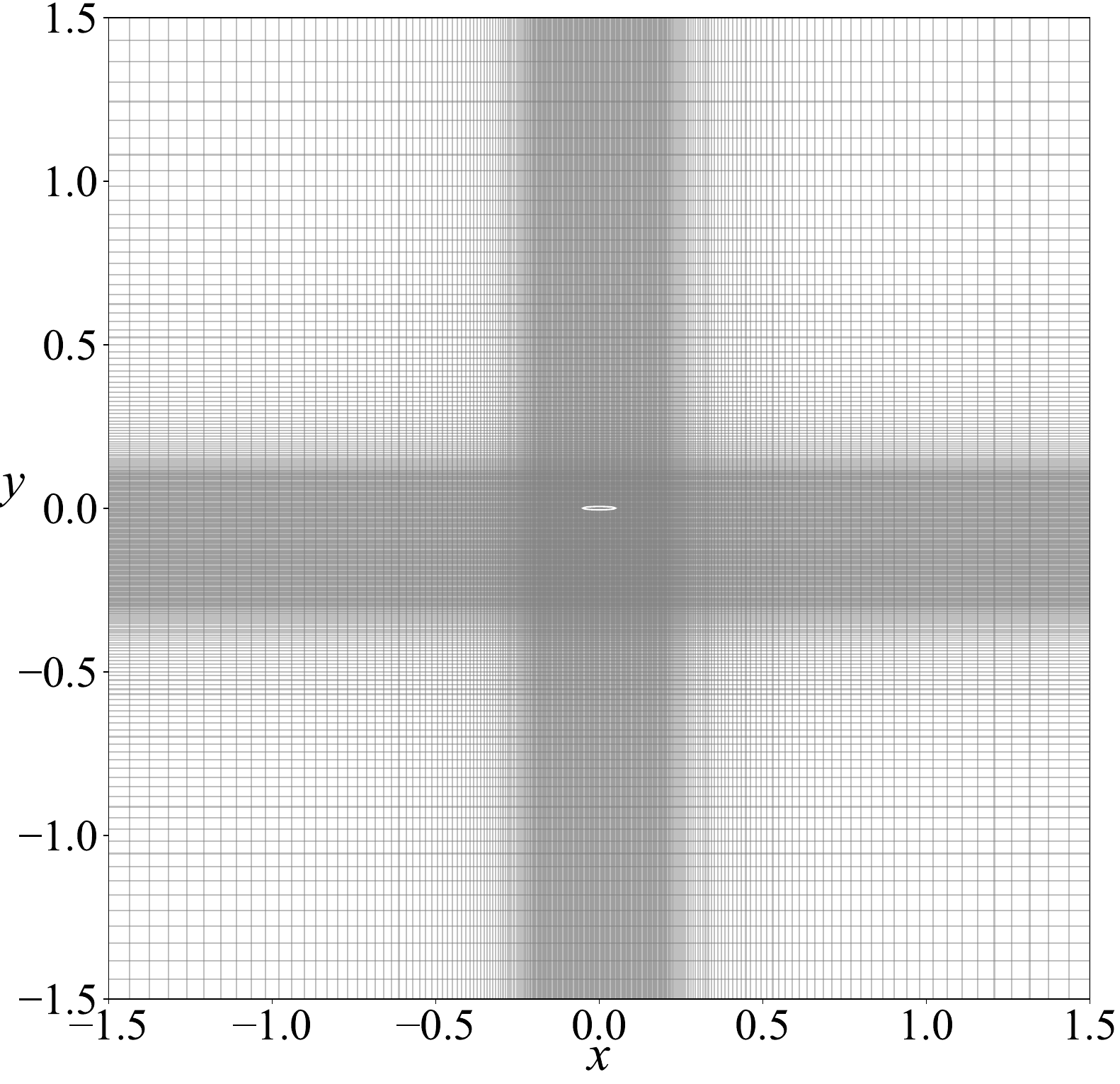}}
 \caption{Schematic diagram of 2D flapping elliptical wing at $Re=75$. (a) Flow boundary conditions and geometric configuration. (b) Computational grid.}
 \label{Case3Domain_Flapwing_Grid}
\end{figure}

The flow boundary conditions, geometric configuration, and computational grid of the flow are illustrated in Fig.~\ref{Case3Domain_Flapwing_Grid}. The computational domain is discretized using a non-uniform Cartesian grid, featuring a locally refined uniform grid~($\Delta x=\Delta y=0.0025$) within a specified square region around the elliptical wing of aspect ratio 10. Beyond this refined region, the grid undergoes progressive coarsening, resulting in $\Delta_{\max}/\Delta_{\min} \approx 28$. The computational grid consists of $87{,}320$ cells. $\Delta t=0.005$ and the chord length of the elliptical wing is $c=0.1$. 

The trajectory of the elliptical wing is governed by
\begin{eqnarray}
\label{Move trajectory of wing x}
 X(t) &=& \frac{1}{2}A_{0}\cdot\cos(2 \pi ft),
\\
\label{Move trajectory of wing y}
  Y(t) &=& 0,
\\
\label{angle of wing}
  \alpha(t) &=& \alpha_{0}+\beta\cdot \sin(2 \pi ft),
\end{eqnarray}
where $(X(t),Y(t))$ represents the instantaneous center coordinates of the elliptical wing at time $t$, $\alpha(t)$ denotes the angular orientation relative to the positive $x$-axis, and $f=0.25$ is flapping frequency. The initial angle $\alpha_{0}=\pi/2$, the translational amplitude $A_{0}=2.8c$, the rotational amplitude $\beta=\pi/4$. Moreover, with $\nu=0.0002932$ and $Re=U_{\max}c/\nu=A_{0} \pi f c/\nu=75$.

We report only the results of HyDEA~(ICPCG + DLSM-3-MConv), specifically configured with $Num_{\mathrm{CG-type}}=3$ and $Num_{\mathrm{DLSM}}=2$. The iteration terminates when the residual $L2$-norm falls below $\epsilon=10^{-6}$.

\begin{figure}[htbp] 
 \centering  
  \subfigure[]{
  \label{Case3_10_Rline}
  \includegraphics[scale=0.21]{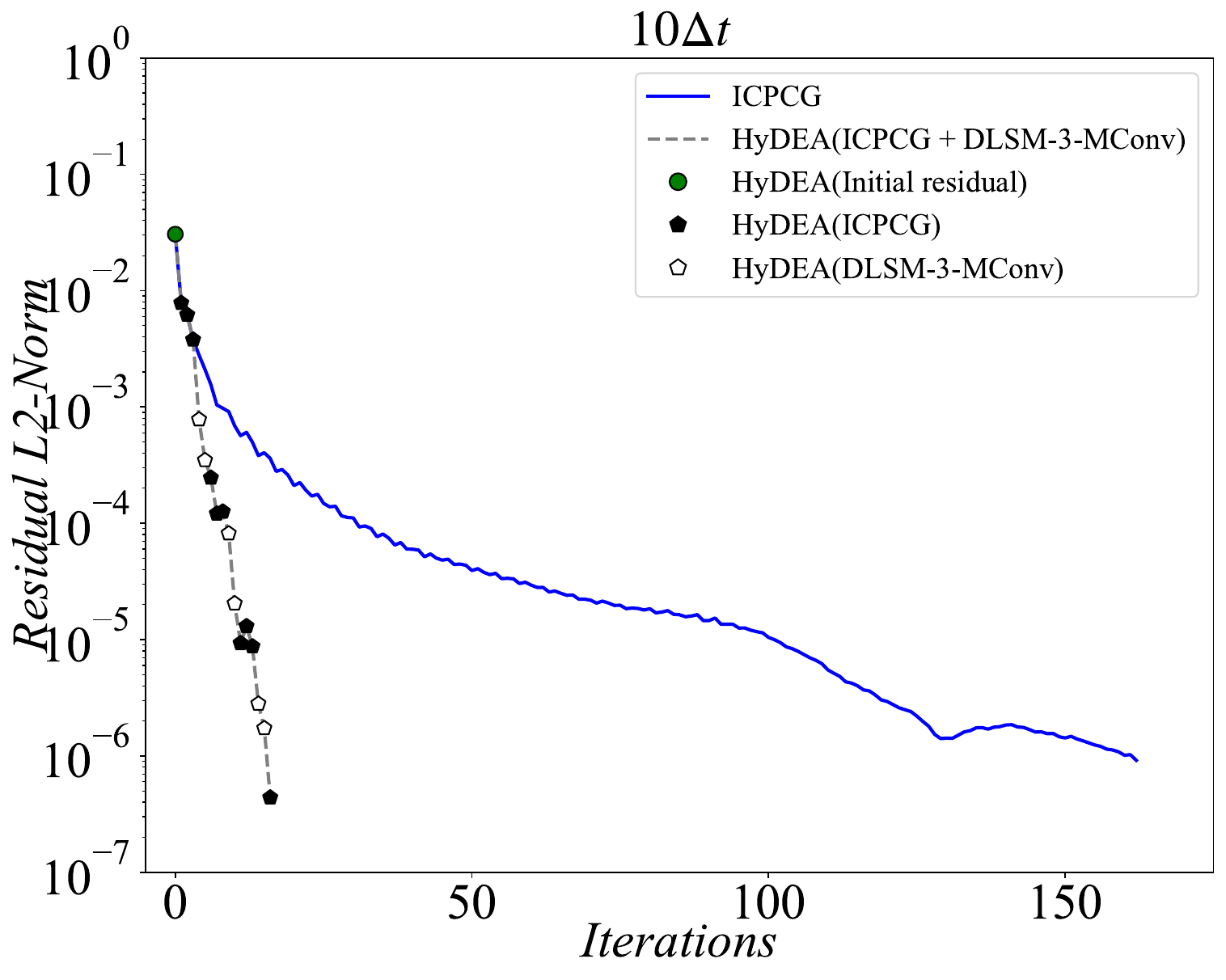}}
  \subfigure[]{
  \label{Case3_100_Rline}
  \includegraphics[scale=0.21]{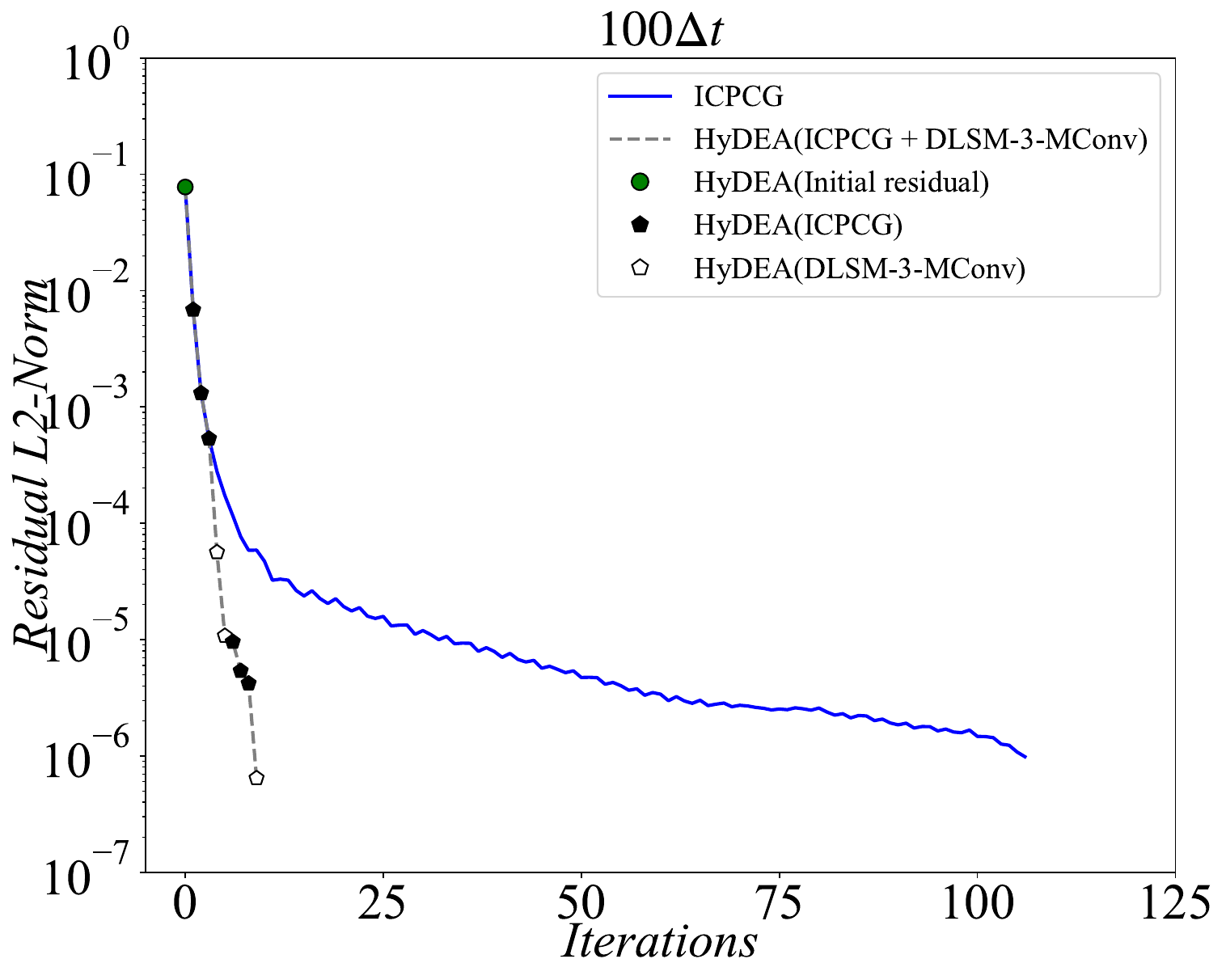}}
  \subfigure[]{
  \label{Case3_3000_Rline}
  \includegraphics[scale=0.21]{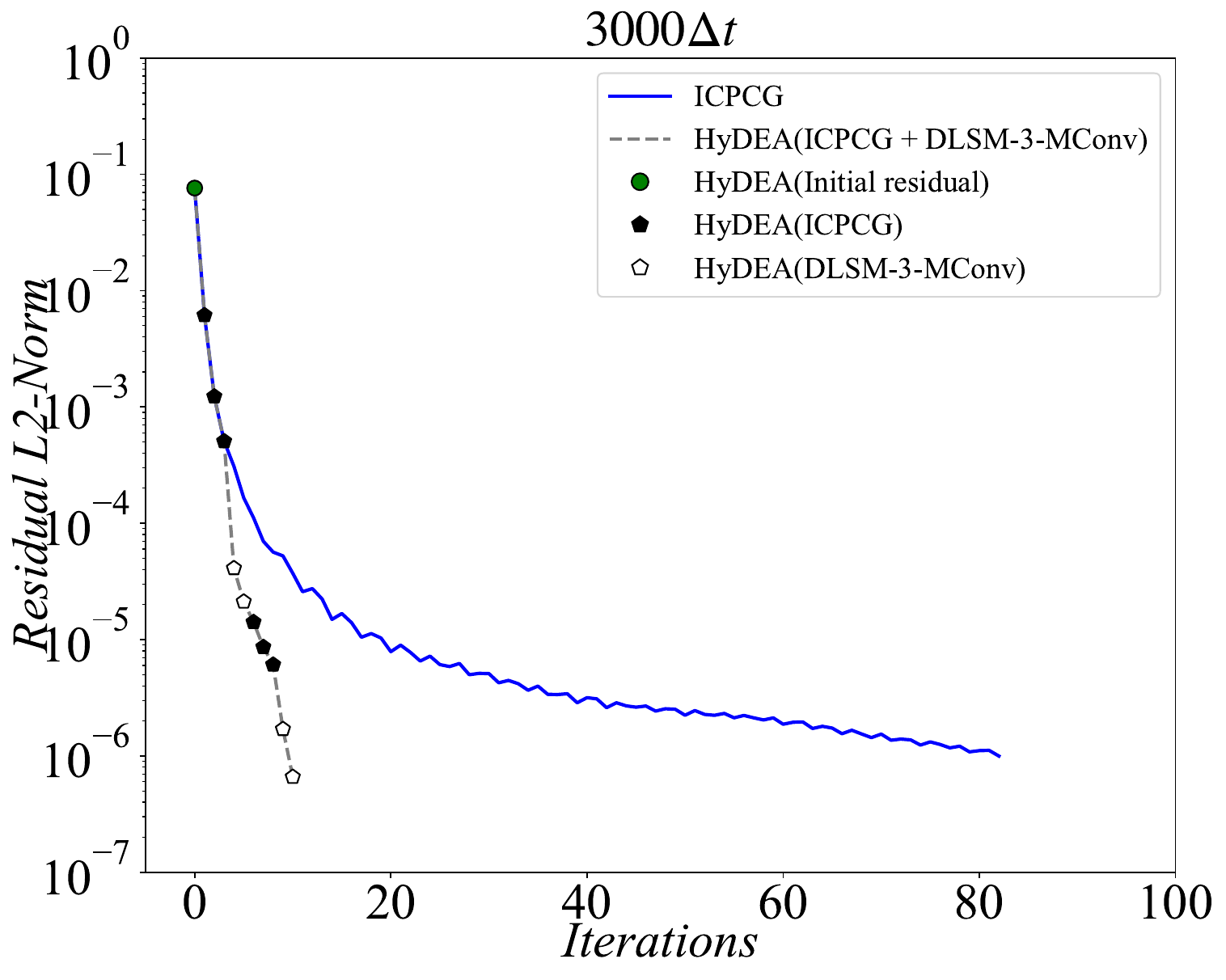}} 
  \caption{Iterative residuals of solving the PPE for 2D flapping elliptical wing at $Re=75$. (a) $10th$ time step. (b) $100th$ time step. (c) $3000th$ time step.}\label{Case3_Rline_3+2}
\end{figure}

The iterative residuals of solving the PPE at the $10th$, $1000th$ and $3000th$ time steps are presented in Fig.~\ref{Case3_Rline_3+2}. As evident from the results, HyDEA~(ICPCG + DLSM-3-MConv) requires significantly fewer iterations than the standalone ICPCG method to meet the predefined tolerance.

\begin{figure}[htbp]
\centering
  \includegraphics[scale=0.2]{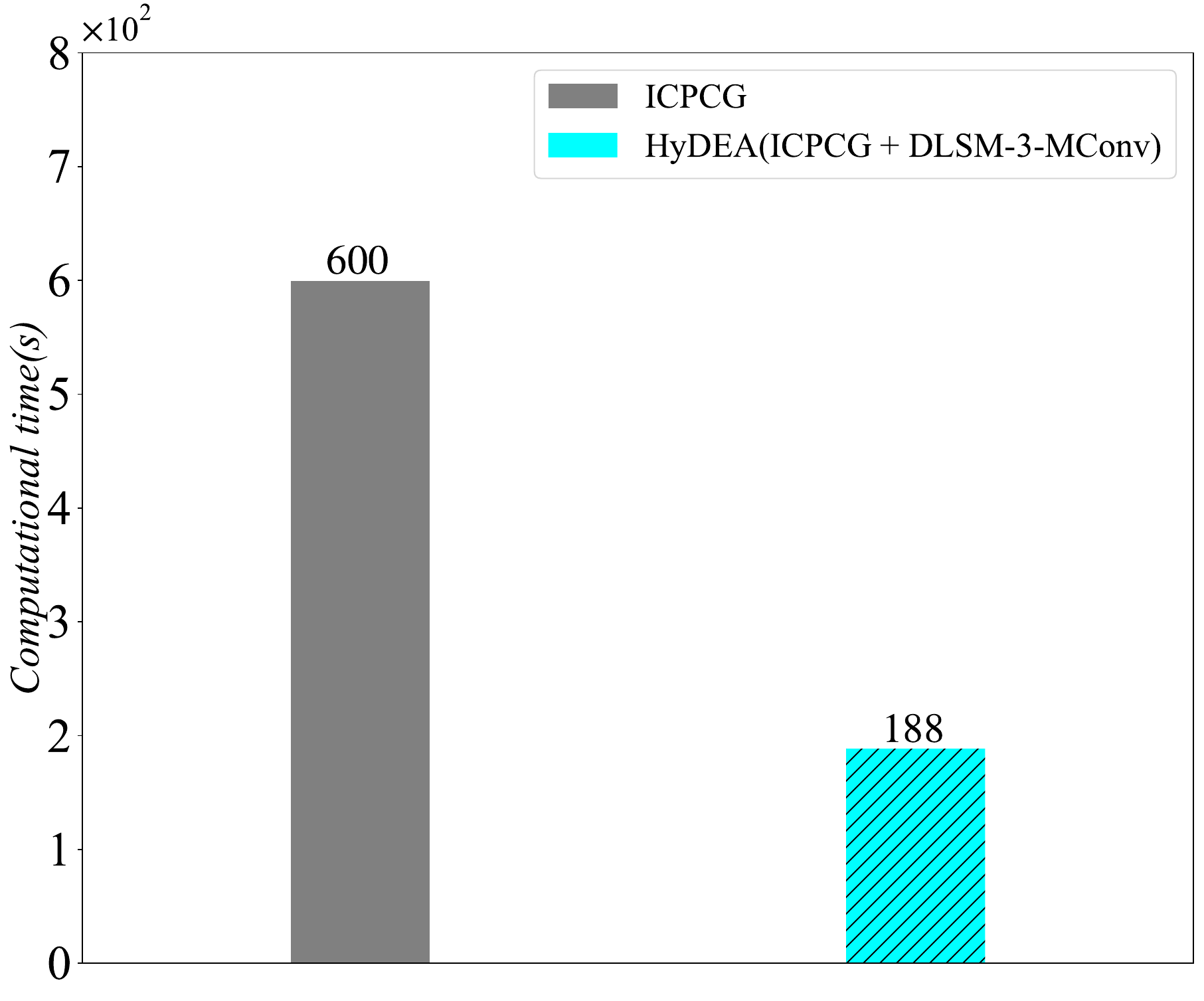}
  \caption{Computational time of the PPE solution over $3200\Delta t$ for 2D flapping elliptical wing at $Re=75$.}\label{Case3_Flapwing_timecomapre}
\end{figure}

Fig.~\ref{Case3_Flapwing_timecomapre} illustrates the computational time required for the PPE solution over 3200 consecutive time steps using HyDEA~(ICPCG + DLSM-3-MConv) and the standalone ICPCG method. The results clearly indicate that HyDEA significantly reducing the computational time, providing a considerable acceleration of the overall computational process.

\begin{figure}[htbp] 
 \centering  
  \subfigure[]{
  \label{Case3_2600_ICPCG}
  \includegraphics[scale=0.124]{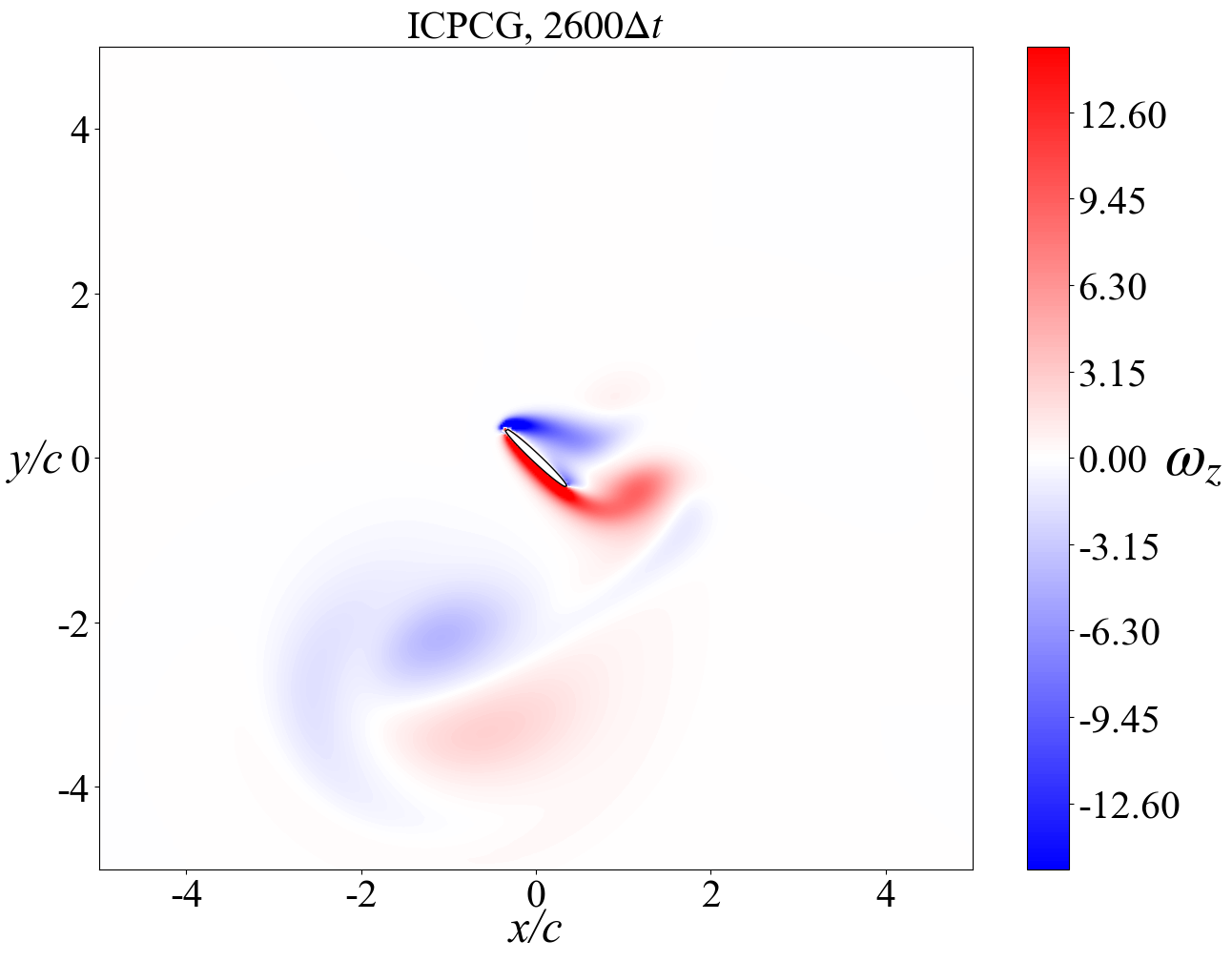}}
  \subfigure[]{
  \label{Case3_2600_HyDEA}
  \includegraphics[scale=0.124]{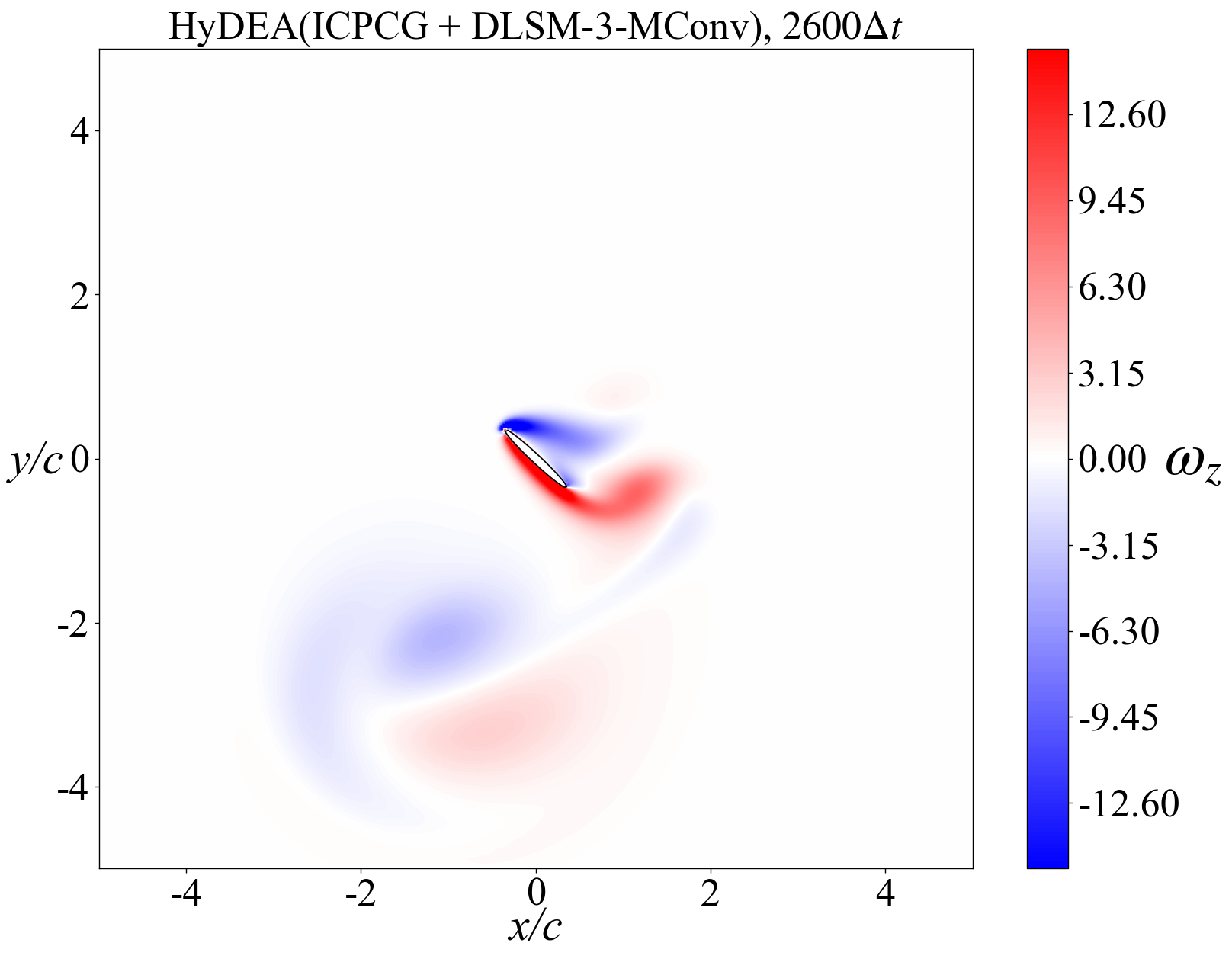}}
  \subfigure[]{
  \label{Case3_2800_ICPCG}
  \includegraphics[scale=0.124]{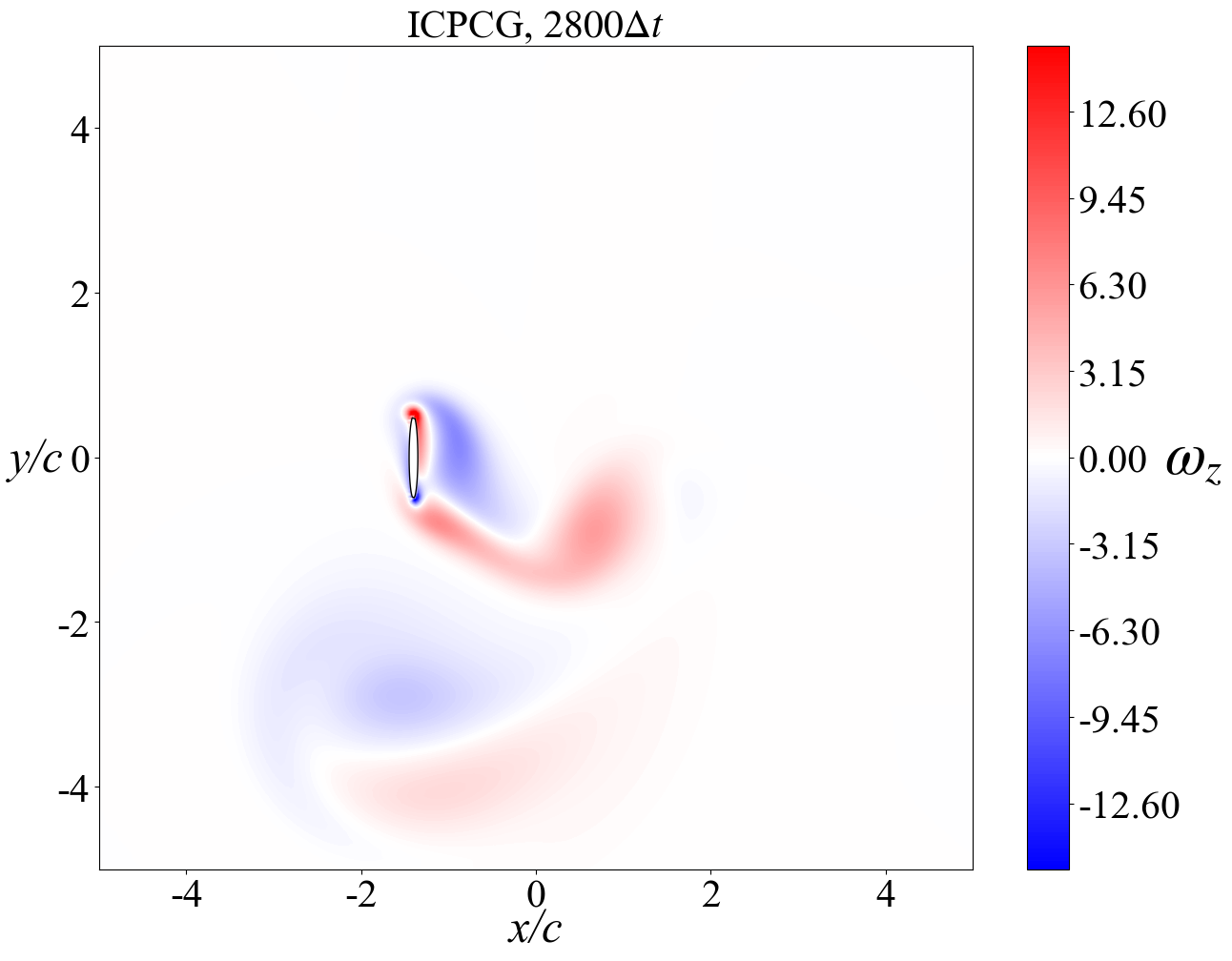}}
  \subfigure[]{
  \label{Case3_2800_HyDEA}
  \includegraphics[scale=0.124]{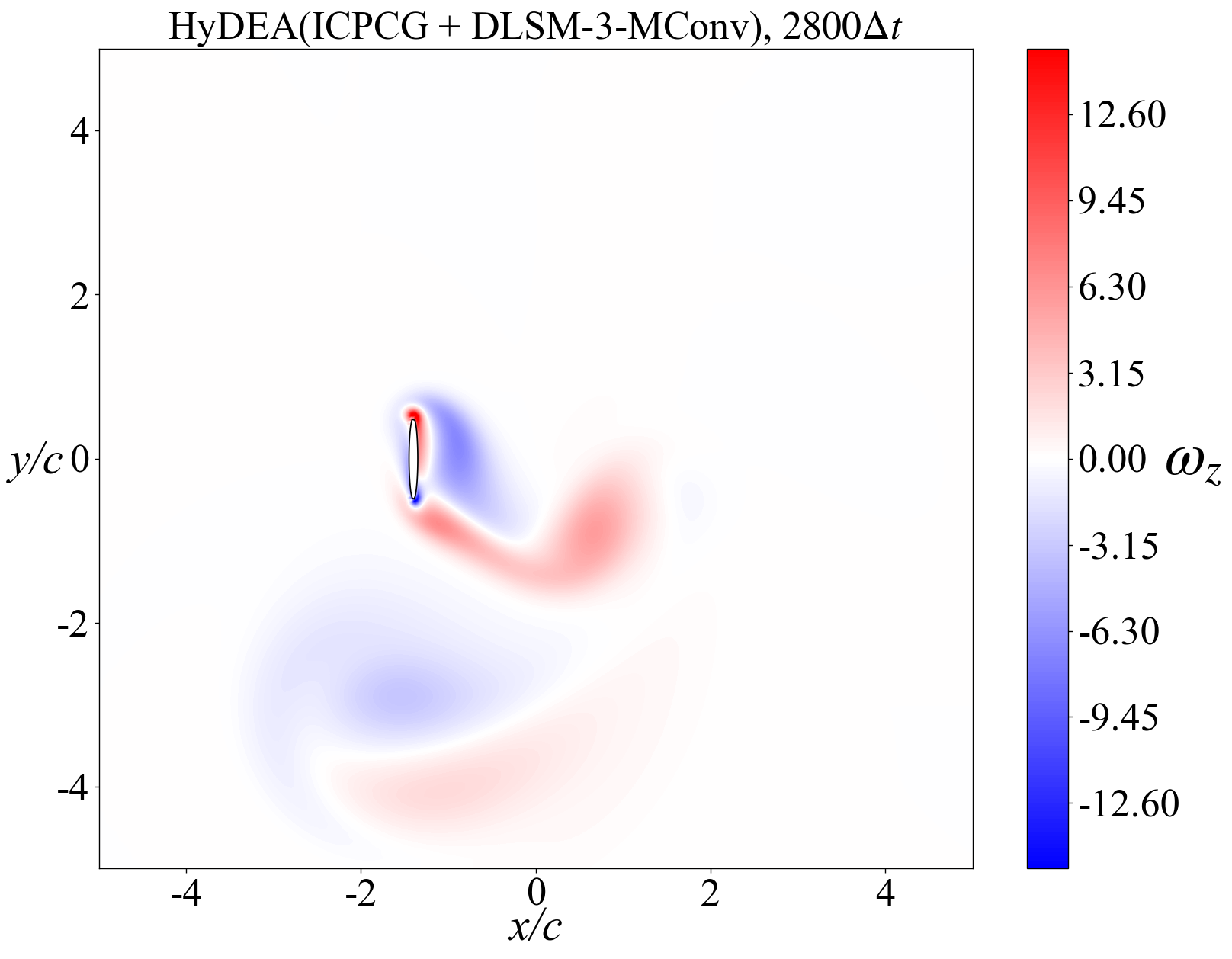}}
  \subfigure[]{
  \label{Case3_3000_ICPCG}
  \includegraphics[scale=0.124]{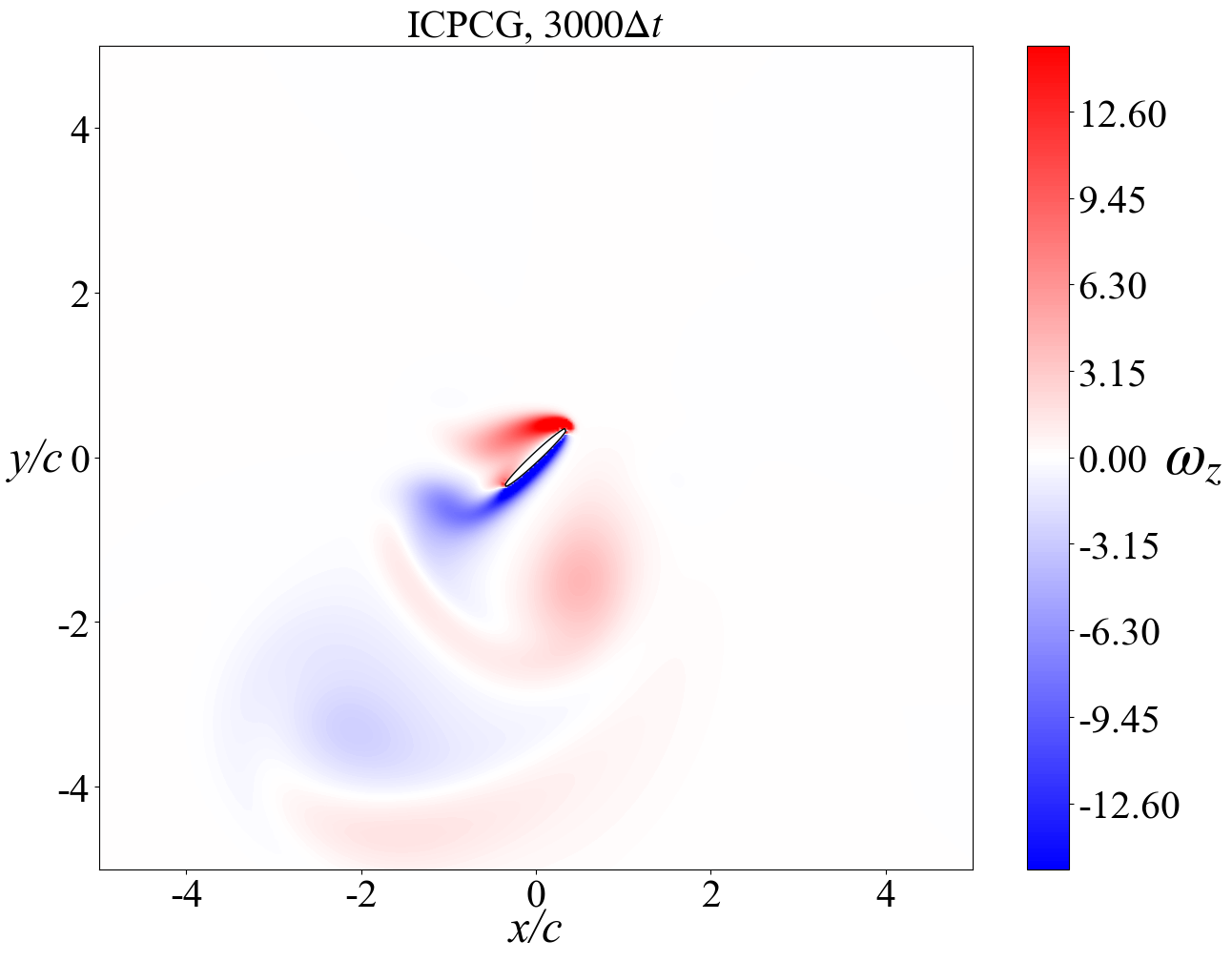}}
  \subfigure[]{
  \label{Case3_3000_HyDEA}
  \includegraphics[scale=0.124]{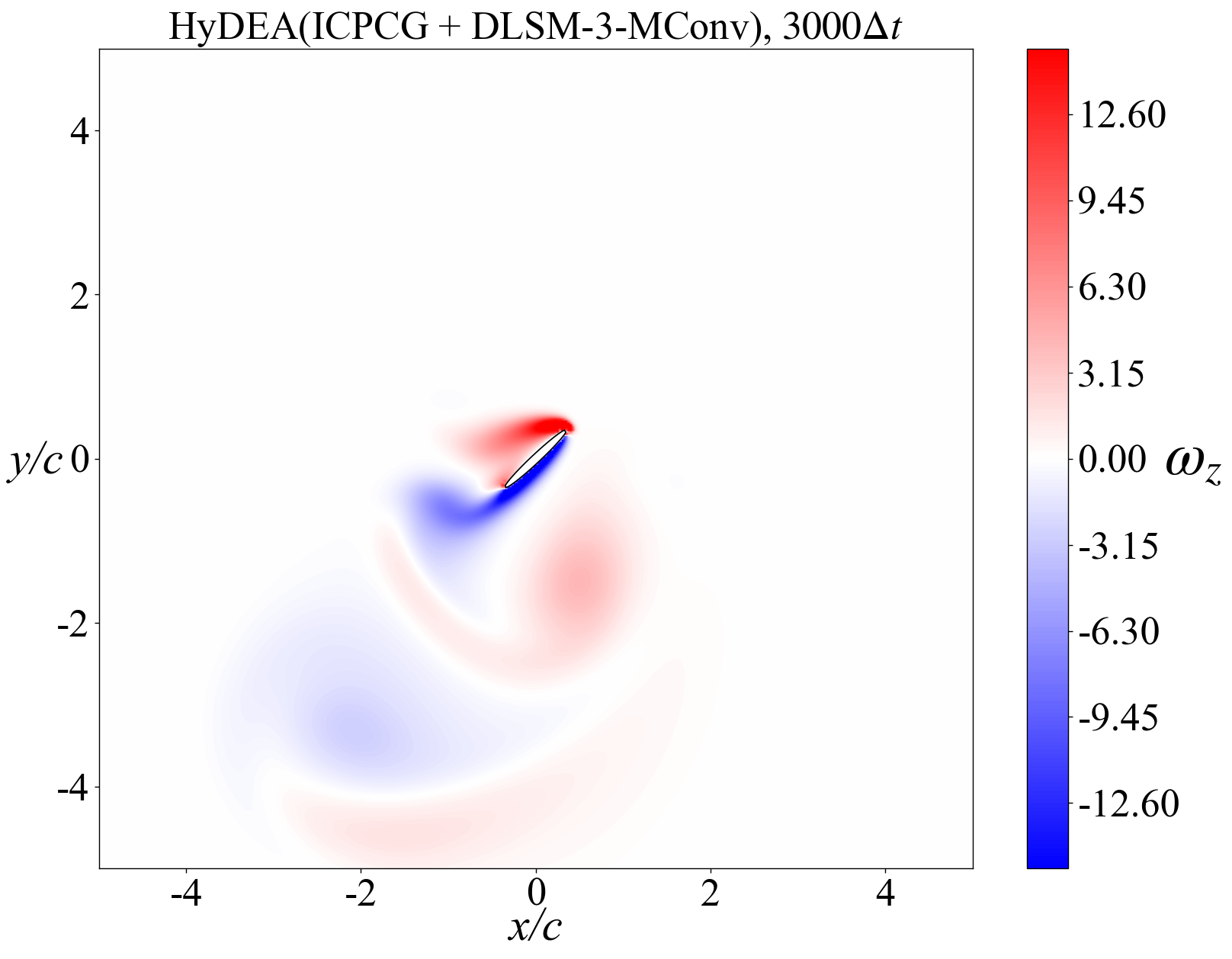}}
  \subfigure[]{
  \label{Case3_3200_ICPCG}
  \includegraphics[scale=0.124]{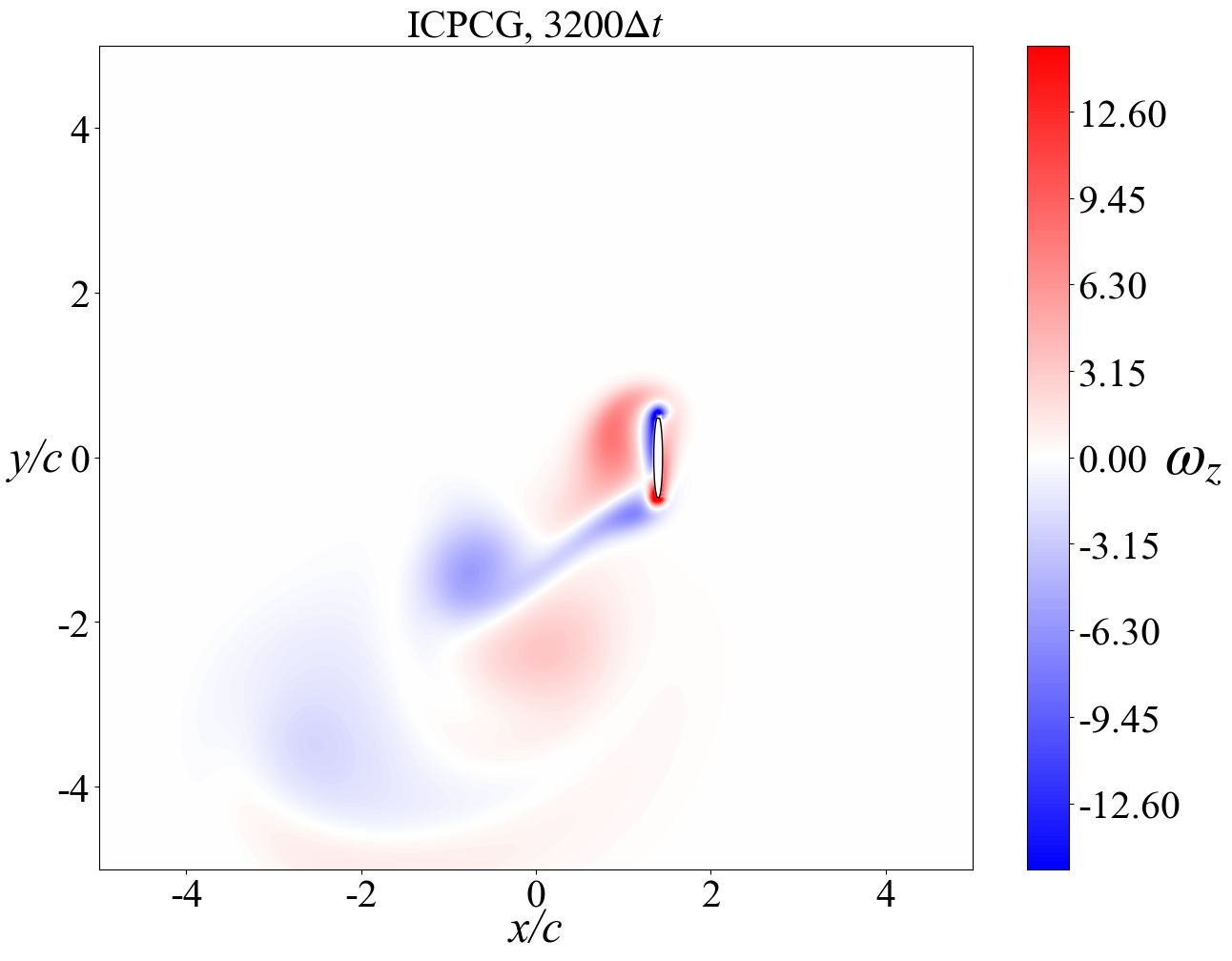}}
  \subfigure[]{
  \label{Case3_3200_HyDEA}
  \includegraphics[scale=0.124]{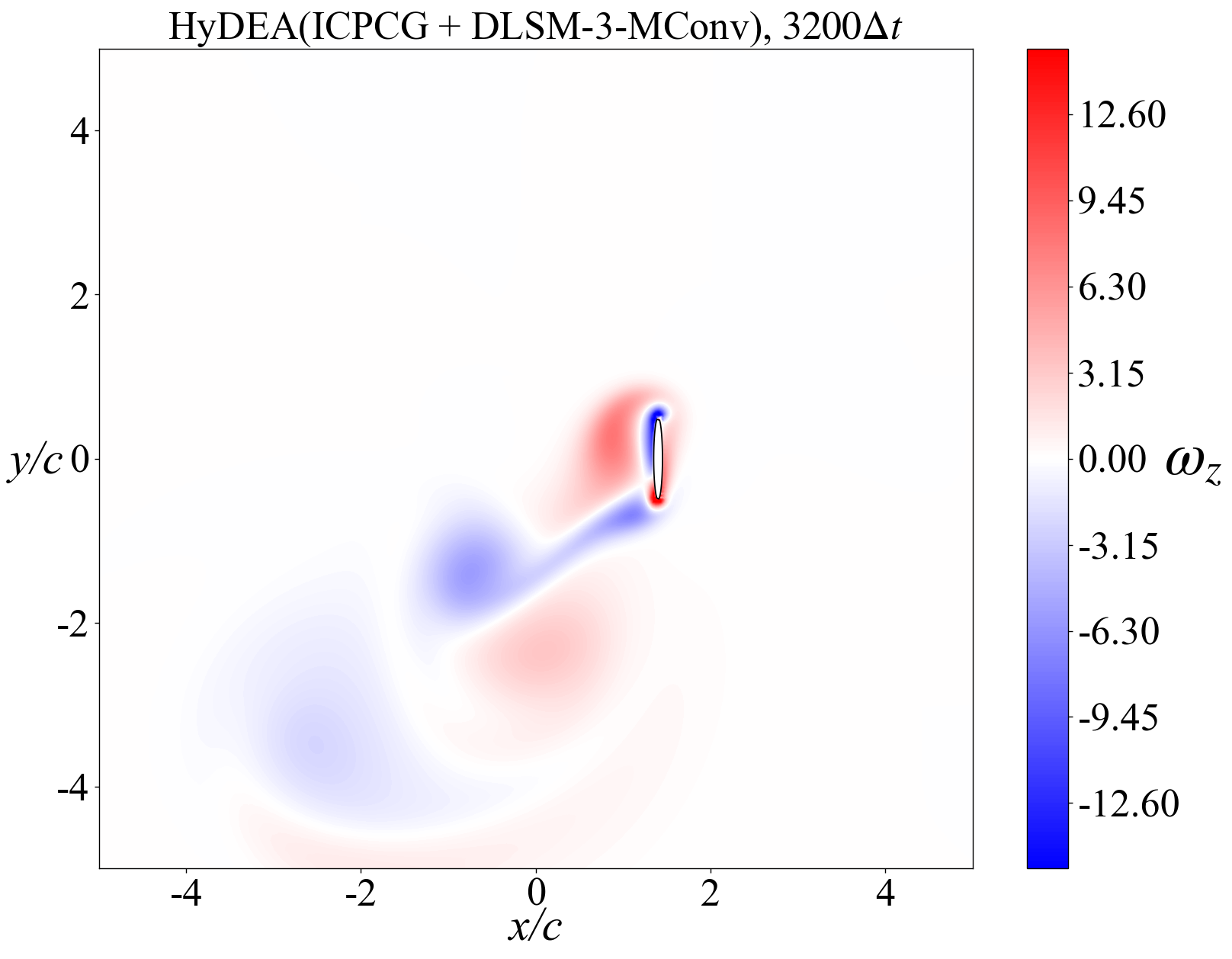}}
 \caption{Vorticity fields for 2D flapping elliptical wing at $Re=75$ by ICPCG and HyDEA~(ICPCG + DLSM-3-MConv). (a)-(b) $2600th$ time step. (c)-(d) $2800th$ time step. (e)-(f) $3000th$ time step. (g)-(h) $3200th$ time step.}
 \label{Case3_flowfield}
\end{figure}

\begin{figure}[htbp]
\centering
  \includegraphics[scale=0.62]{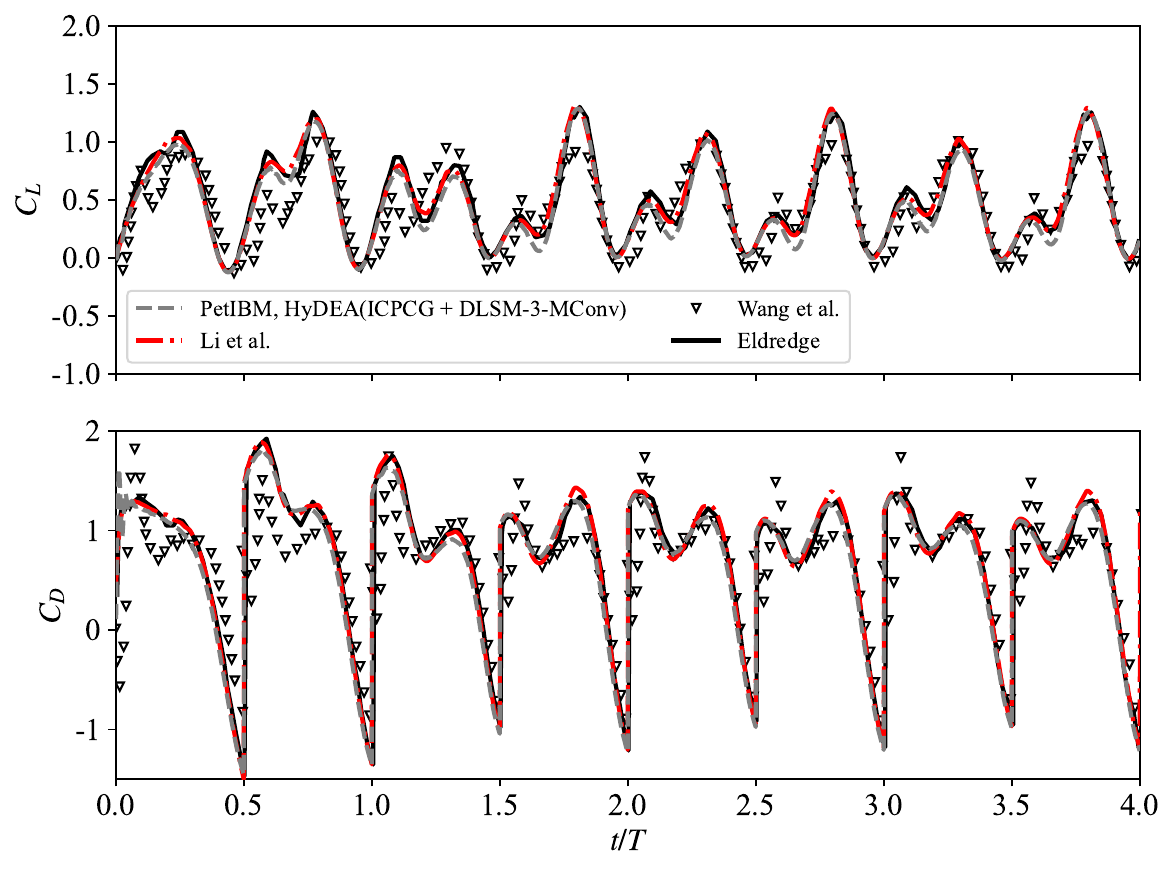}
  \caption{The temporal evolution of the $C_L$ and $C_{D}$ for 2D flapping elliptical wing at $Re=75$. $T=\frac{1}{f}$ represents the flapping period.}\label{CL_CD_wing_timehistory}
\end{figure}

The vorticity fields at the $2600th$, $2800th$, $3000th$ and $3200th$ time steps are presented in Fig.~\ref{Case3_flowfield}, which clearly demonstrate that the temporal evolution of the flow field is calculated accurately. Fig.~\ref{CL_CD_wing_timehistory} further compares the temporal evolution of the lift coefficient~($C_{L}$) and drag coefficient~($C_{D}$) of the elliptical wing with the results reported by Wang et al.~\cite{wang2004unsteadyWing}, Eldredge~\cite{eldredge2007numericalWing} and Li et al.~\cite{li2015effectsWing}. These results demonstrate that HyDEA~(ICPCG + DLSM-3-MConv) for solving the PPE achieves excellent accuracy in predicting these global coefficients of 2D flapping elliptical wing at $Re=75$.

\section{Conclusion}
\label{conclusion}

In this work, we extend the HyDEA framework to accelerate the iterative solution of the pressure Poisson equation~(PPE) for incompressible flows on non-uniform Cartesian grids. Recognizing that standard convolution operators are inherently ill-suited for processing spatially varying resolutions, we propose a multi-level distance vector map construction strategy that evaluates discrete local grid spacings corresponding to each hierarchical level of the U-Net architecture within the branch network and explicitly fuses this grid-spacing information with the input features prior to the convolution operations.

Benchmark results underscore the superior generalizability and computational performance of the grid-spacing-aware HyDEA:
\begin{itemize}
\item Reduces the number of iterations required to solve the PPE across diverse flows on non-uniform Cartesian grids, compared with standalone CG-type methods.
\item Outperforms the standard convolution-based HyDEA significantly on non-uniform Cartesian grid configurations where substantial variations in grid spacing exist across the domain.
\item Generalizes to diverse obstacle geometries in flow simulations with \textit{fixed neural network weights}.
\end{itemize}

Although the grid-spacing-aware HyDEA demonstrates excellent performance, it currently faces inherent limitations regarding computational efficiency and generalizability across arbitrary grids. The existing architecture is exclusively tailored for non-uniform Cartesian grid discretizations and necessitates retraining for new grid configurations. Recognizing the significantly broader application value of unstructured grids, adapting HyDEA to process entirely unstructured spatial data would necessitate advanced network architectures, such as graph neural networks or Transformer. While this would predictably introduce substantial computational overhead, it nonetheless represents a compelling direction for further investigation. Consequently, future efforts will be directed toward enhancing code efficiency, refining the neural network architecture to accommodate unstructured spatial data, optimizing training strategies, and improving generalization to unseen grid configurations. Ultimately, these advancements aim to facilitate large-scale three-dimensional flow simulations across diverse and complex grid topologies.

\section*{Acknowledgments}

The authors appreciate support from Research fund of National Key Laboratory of Hydrodynamics (Project No. NKLH2025KF05) of China Ship Scientific Research Center and National Key R\&D Program of China (2022YFA1203200).

\appendix

\section{Sensitivity analysis of $Num_{\mathrm{CG-type}}$ and $Num_{\mathrm{DLSM}}$ on HyDEA's performance}
\label{appendixA}

This section performs a parametric sensitivity analysis to evaluate the impact of two critical parameters $Num_{\mathrm{CG-type}}$ and $Num_{\mathrm{DLSM}}$ on the performance of HyDEA. The investigation is conducted based on the flow configuration detailed in Section~\ref{Re100Cylinder}, and more specifically:
\begin{itemize}
 \item 2D flow past a circular cylinder at $Re=100$. The computational grid consists of $107{,}016$ cells, and $\Delta_{\max}/\Delta_{\min} \approx 41$.
\end{itemize}

\begin{figure}[htbp]
\centering
  \includegraphics[scale=0.24]{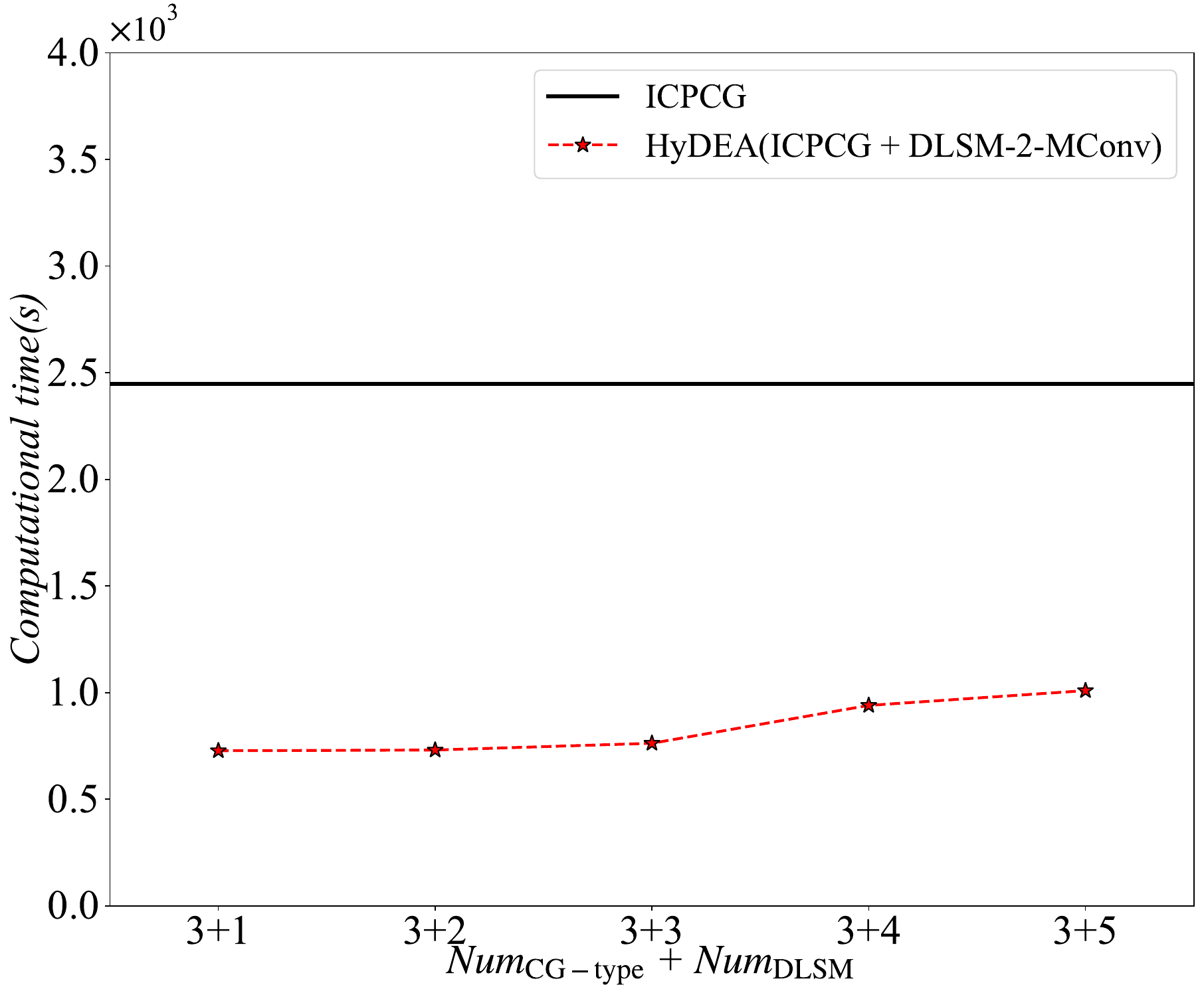}
  \caption{Computational time required to solve the PPE over $20{,}000\Delta t$ in the simulation of 2D flow past a circular cylinder at $Re=100$, using different $Num_{\mathrm{DLSM}}$ values.}\label{Case2_iterationNum_3addM}
\end{figure}

For the sake of conciseness and to avoid redundant benchmarking, the present discussion is restricted to the performance of HyDEA~(ICPCG + DLSM-2-MConv). First, $Num_{\mathrm{CG-type}}$ is fixed at 3, and a parametric study is performed by testing the $Num_{\mathrm{DLSM}}$ values of 1 to 5. Fig.~\ref{Case2_iterationNum_3addM} depicts the computational time over $20{,}000$ consecutive time steps required to solve the PPE. The results indicate that the computational efficiency remains largely consistent across different values of $Num_{\mathrm{DLSM}}$. Although a slight increase in computational time is observed as $Num_{\mathrm{DLSM}}$ increases, all hybrid configurations consistently deliver acceleration compared to the standalone ICPCG method.

\begin{figure}[htbp]
\centering
  \includegraphics[scale=0.24]{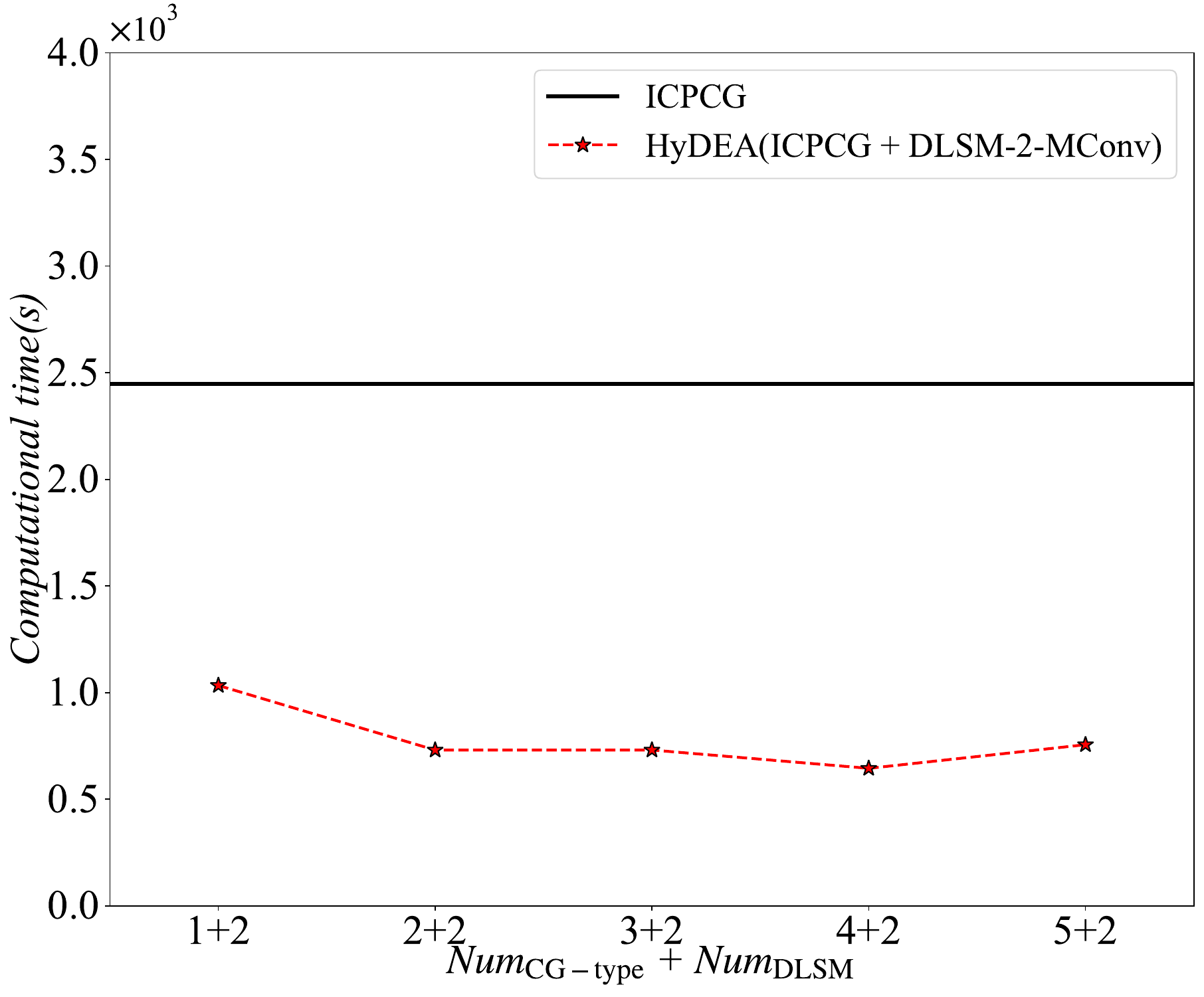}
  \caption{Computational time required to solve the PPE over $20{,}000\Delta t$ in the simulation of 2D flow past a circular cylinder at $Re=100$, using different $Num_{\mathrm{CG-type}}$ values.}\label{Case2_iterationNum_Madd2}
\end{figure}

Subsequently, $Num_{\mathrm{DLSM}}$ is fixed at 2, and $Num_{\mathrm{CG-type}}$ is varied from 1 to 5 to investigate its effect on the performance of HyDEA, with the results depicted in Fig.~\ref{Case2_iterationNum_Madd2}. The results indicate that the computational efficiency remains comparable across varying $Num_{\mathrm{CG-type}}$ settings. Furthermore, all tested configurations consistently deliver acceleration compared to the standalone ICPCG method.

\bibliographystyle{elsarticle-num} 
\bibliography{main.bib}





\end{document}